\documentclass[journal ]{new-aiaa}
\usepackage[utf8]{inputenc}
\usepackage{textcomp}

\usepackage{graphicx}
\usepackage{subcaption}
\usepackage{amsmath}
\usepackage[version=4]{mhchem}
\usepackage{siunitx}
\usepackage{longtable,tabularx}
\title{Comprehensive Study on the Slat Noise of 30P30N High-Lift Airfoil Basd on High-Order Wall-Resolved Large-Eddy Simulation}

\author{Keli Zhang\footnote{Ph.D. Canditate, School of Aeronautic Science and Engineering, zkl70@buaa.edu.cn.}, Shizhi Lin\footnote{Ph.D. Canditate, School of Aeronautic Science and Engineering, LinSZ@buaa.edu.c.} and Peiqing Liu\footnote{Professor, School of Aeronautic Science and Engineering, lpq@buaa.edu.cn (Corresponding Author).}}
\affil{Beihang University, Beijing, 100191, People's Republic of China}

\author{Keli Zhang\footnote{CTO Assistant (Internship), Solver R\&D and Development Center.}, Shihao Liu\footnote{Senior CFD Algorithm Engineer, Solver R\&D and Development Center, liush@rankyee.com.} and Kai Liu\footnote{CTO, kai@dimaxer.com.}}
\affil{Rankyee Technology Beijing, Beijing, 100097, People's Republic of China}

\begin{document}

\maketitle

\begin{abstract}
This study presents wall-resolved large-eddy simulations (WRLES) of a high-lift airfoil, based on high-order flux reconstruction (FR) commercial software Dimaxer, which runs on consumer level GPUs. A series of independence tests are conducted, including various Ffowcs Williams-Hawkings sampling surfaces, different mesh densities, simulations at $4^{th}$ and $5^{th}$ order accuracies, and varying spanwise lengths, to establish best practice for predicting slat noise through high-order WRLES. The results show excellent agreement with experimental data while requiring significantly fewer computational resources than traditional second-order methods. An investigation on the effects of Reynolds number (Re) is performed by scaling the airfoil size, with Reynolds numbers ranging from $8.55\times 10^5$ to a real aircraft level of $1.71\times 10^7$. By applying simple scaling through Strouhal number (St), spanwise correction, and distance from the receiver, the far-field noise spectra for different Reynolds numbers can be coincided. Additionally, simulations are performed at four angles of attack: $3^\circ$, $5.5^\circ$, $9.5^\circ$, and $14^\circ$. The results indicate that higher angles of attack lead to a less intense feedback loop, resulting in lower tonal noise frequencies and reduced noise amplitude. The maximum noise reduction observed is over 14dB when comparing $14^\circ$ to $3^\circ$. Furthermore, an improved formula is proposed to enhance the prediction of slat noise tonal frequencies and to better elucidate the mechanism behind tonal noise generation.
\end{abstract}

\section*{Nomenclature}

{\renewcommand\arraystretch{1.0}
\noindent\begin{longtable*}{@{}l @{\quad=\quad} l@{}}
$\alpha$  & FR coefficient \\
$c$  & speed of sound \\
$C_D$& drag coefficient \\
$C_L$& lift coefficient \\
$C_{\omega z}$& normalized spanwise vorticity\\
$C_p$& pressure coefficient \\
$C_w$& model coefficient of wall-adaptive local eddy-viscosity model\\
$c_s$& stowed chord length \\
$c_{slat}$& slat chord length \\
$\Box$& $\frac{1}{c^2}\frac{\partial^2}{\partial t^2}-\nabla^2$, d'Alembert operator \\
$\delta$& Dirac delta \\
$\delta_{ij}$& Kronecker delta \\
$\Delta$& $\text{cell volume}^{1/3}$ \\
$\Delta t$& time step \\
$\boldsymbol{F}$& flux matrix\\
$\widetilde{\boldsymbol{F}}$& common flux matrix\\
$f$& frequency\\
$\gamma$& ratio of specific heat\\
$H$& Heaviside function\\
$\boldsymbol{J}$& Jacobian matrix\\
$K$& bulk modulus/polynomial degree\\
$\boldsymbol{\xi}$& computational domain coordinate vector\\
$L_a$& distance between the slat cusp and the impingement point of the slat shear layer\\
$L_a^{\prime}$& distance between the slat cusp and the slat trailing edge\\
$L_a^{\prime\prime}$& distance between the slat cusp and the acoustic wave generation point\\
$L_v$& curvilinear length of the slat shear layer (from the slat cusp to the impingement point)\\
$L_v^{\prime}$& curvilinear length of a stream line which extend from the slat cusp to the slat trailing edge\\
$L_v^{\prime\prime}$& curvilinear length along the slat shear layer from the slat cusp to the acoustic wave generation point\\
$M$& Mach number of the moving surface\\
$M_i$& Mach number component of the moving surface\\
$M_0$& inflow Mach number\\
$\mu$& viscosity\\
$N$& number of elements\\
$N_f$& number of flux points on the interface\\
$N_s$& number of element interfaces\\
$N_{qp}^t$& number of time quadrature point\\
$N_z$& number of spanwise elements\\
$n$& mode number\\
$\boldsymbol{n}$& unit outward normal of the surface\\
$\omega_z$& spanwise vorticity\\
$p$& pressure\\
$p^{\prime}$& sound pressure\\
$P_{ij}$& stress tensor\\
$\phi$& $(1,u,v,w)$, kinetic preserving variables\\
$Q$& $Q$ criterion\\
$R$& effective acoustic distance between source and observer\\
$r$& distance between observer and source\\
$\hat{r}_i$& unit radiation vector component\\
$Re$& Reynolds number\\
$\rho$& density\\
$S_{ij}$& strain rate tensor\\
$S_{ij}^d$& filtered velocity gradient tensor\\
$s_t$& slat trailing edge thickness\\
$T_h$& the time a vortex takes from generation at the slat cusp to emit an acoustic wave upon interacting with the slat trailing edge\\
$T_{ij}$& Lighthill tensor stress\\
$\tau$& source time\\
$\tau_{ij}$& viscous stress tensor\\
$\tau_{qp}$& non-dimensional 1D quadrature point value\\
$t$& time/observer time\\
$t_{qp}$& physical time corresponding to $\tau_{qp}$\\
$\boldsymbol{U}$& conservation variable matrix in Cartesian domain\\
$\widehat{\boldsymbol{U}}$& conservation variable matrix in computational domain\\
$U_0$& incoming flow velocity magnitude\\
$U_{0i}$& incoming flow velocity component\\
$U_v$& average convection velocity magnitude along $L_v$\\
$U_v^{\prime}$& average convection velocity magnitude along $L_v^{\prime}$\\
$U_v^{\prime\prime}$& average convection velocity magnitude along $L_v^{\prime\prime}$\\
$u_i$& velocity component of fluid\\
$u$& streamwise component of velocity\\
$v$& vertical component of velocity\\
$\boldsymbol{v}$& initial value matrix\\
$v_i$& velocity component of the moving surface\\
$w$& spanwise component of velocity\\
$w_{qp}^{t}$& nondimensional quadrature weight on time quadrature point qp\\
$\boldsymbol{x}$& Cartesian domain coordinate vector/coordinate vector of observer\\
$x_i$& Cartesian domain coordinate component\\
$y_{mesh}^{+}$& nondimensional mesh resolution in the wall-normal direction\\
$y^{+}$& $\frac{y_{mesh}^{+}}{K+1}$, nondimensional mesh resolution in the wall-normal direction based on high-order method\\
\multicolumn{2}{@{}l}{Subscripts}\\
0 & far-field value\\
L & loading component\\
ret & retarded time\\
sgs & subgrid-scale\\
T & thickness component\\
\end{longtable*}}

\section{Introduction}
\lettrine{S}{ince} the development of high bypass ratio turbofan engines, slat, flap, and landing gears have emerged as significant sources of noise in commercial aircraft. This is particularly true during landing, when the engines are running at low-power state, and the noise produced by the slat can rival that of the engines \cite{dobrzynski2010almost}. Consequently, it is crucial to accurately predict the noise generated by such device using numerical methods to facilitate the development of quieter commercial aircraft that meet the growing demands of noise regulations \cite{jimenez2011probabilistic,european_commission_flightpath_2011}.

\begin{figure}[hbt!]
    \centering
    \includegraphics[width=.35\textwidth]{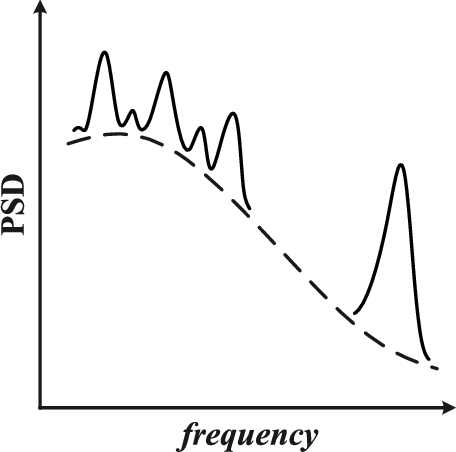}
    \caption{\label{fig:slat_spectrum} Schematic of the slat noise spectrum of a wind tunnel reduced-scale model: -{}-, broadband noise; ---, tonal noise peaks; -.-, high-frequency hump.}
\end{figure}

Three basic noise components of wind tunnel reduced-scale models have been identified in the power spectral density (PSD) spectrum \cite{murayama2014experimental,murayama2018experimental} as illustrated in Fig.~\ref{fig:slat_spectrum}. The first component is a broadband noise that ranges from 0.8 to 7 kHz. On top of that are multiple tonal noise peaks occurring at frequencies between 1 kHz and 10 kHz, along with a hump that rises above 20 kHz.

As proposed by \citet{roger2000low}, the low-frequency tonal noise peaks are associated with cavity modes in the slat cove area, i.e. a feedback loop mechanism. \citet{terracol2016investigation} derived a tailored equation to predict the frequency of the tonal noise peaks based on Rossiter's original formula, and they successfully predicted the tonal noise frequencies of a slat under $4^\circ$ angle of attack (AoA). Their equation considers a feedback loop that cycles as follows: (i) initially, a vortex is generated at the slat cusp; (ii) this vortex convects along the shear layer, which surrounds the main vortex core inside the slat cove; (iii) the vortex impinges the slat surface, producing an acoustic wave; (iv) the acoustic wave propagates through the slat cove; (v) upon reaching the slat cusp, it triggers the formation of a new vortex. In addition, \citet{souza2019dynamics} performed a spectral proper orthogonal decomposition analysis on the flow field of a slat under various AoAs and slat deflection angels, concluding that the interaction between the shear layer disturbance and the slat trailing edge (TE) after the impingement is responsible for generating the acoustic wave.

Numerical simulation done by \citet{terracol2016investigation} and wind tunnel experiments \cite{takeda2001unsteady,choudhari2002aeroacoustic} identified the high-frequency vortex shedding from the slat TE as the phenomenon responsible for the high-frequency hump. As for the broadband noise, the mechanism behind has not yet settled.

To effectively resolve these phenomena, at least a simulation level of detached eddy simulation (DES) is necessary \cite{chen2021noise,terracol2015aeroacoustic}, in conjunction with the Ffowcs Williams-Hawkings (FW-H) acoustic analogy \cite{ffowcs1969sound} to obtain the far-field noise signal. However, this kind of simulation requires very small time step size, leading to significant amount of computing resources, not to mention performing large-eddy simulation (LES). For example, \citet{terracol2020wall} costed 4096 CPU cores and 6 million CPU hours to obtain wall-resolved LES (WRLES) flow field on the LEISA2 high-lift airfoil under $Re=1.23\times 10^6$, \citet{asada2024aeroacoustics} utilized 55296 CPU cores to achieve wall-modeled LES (WMLES) on the 30P30N high-lift airfoil at $Re=1.71\times 10^6$. When considering a Reynolds number (Re) of real aircraft, i.e. above $10^7$, the computational demands of conventional solvers will grow exponentially because of additional constraints on mesh element size and time step.

However, with the help of Dimaxer, which runs on the latest Nvidia Ada Lovelace GPUs and utilizes the space time expansion of high-order kinetic preserving flux reconstruction (STE-KEP-FR) scheme \cite{liu2019ultra,lu2019grid}, conducting high-order WRLES on a high-lift airfoil becomes quite affordable, even at Reynolds number of real aircraft. This paper discusses the high-order WRLES of the flow around the typical 30P30N high-lift airfoil with focus on the noise generated by the slat. The primary objectives are: 
\begin{enumerate}
\item to provide best practice toward slat noise prediction with high-order WRLES and FW-H acoustic analogy;
\item to explore how the Reynolds number affects slat noise; 
\item to enhance the comprehension of AoA effects on the slat noise emissions.
\end{enumerate}

A total of nine meshes are generated, featuring various mesh densities, spanwise lengths, and airfoil chord lengths. These cases are selected from the experiments of \citet{murayama2014experimental,murayama2018experimental} to facilitate better comparisons between simulations and experimental results under different AoAs.

The structure of the paper is as follows: Sec.~\ref{sec:NM} provides a brief overview of the numerical methods. Sec.~\ref{sec:SD} outlines the geometry, mesh specifics, and the simulation procedure. Then the results of different FW-H sampling surfaces, meshes, simulation orders, Reynolds numbers, slat TE thicknesses and AoAs are presented and discussed accordingly in Sec.~\ref{sec:RaD}. Finally, the main conclusions are summarized in Sec.~\ref{sec:C}.

\section{\label{sec:NM}Numerical Methods}
\subsection{Governing Equations for Weakly Compressible Flow}
The spatially filtered continuity and momentum equations of the Navier-Stokes equations can be written as:
\begin{equation}
    \label{eq:continuity_equation}
    \frac{\partial\rho}{\partial t}+\frac{\partial(\rho u_i)}{\partial x_i}=0
\end{equation}
\begin{equation}
    \label{eq:momentum_equation}
    \frac{\partial(\rho u_i)}{\partial t}+\frac{\partial(\rho u_iu_j)}{\partial x_j}+\frac{\partial p}{\partial x_i}=\frac{\partial\tau_{ij}}{\partial x_j}
\end{equation}
with viscous stress tensor defined as:
\begin{equation}
    \label{eq:tau_ij}
    \tau_{ij}=(\mu+\mu_{sgs})(S_{ij}-\frac{1}{3}S_{kk}\delta_{ij})
\end{equation}
\begin{equation}
    \label{eq:S_ij}
    S_{ij}=\frac{1}{2}\biggl(\frac{\partial u_i}{\partial x_j}+\frac{\partial u_j}{\partial x_i}\biggr)
\end{equation}

Considering weakly compressible flow, $p$ is only a function of $\rho$, the barotropic gas state equation \cite{bernard2014low} can be used:
\begin{equation}
    \label{eq:state_equation}
    p(\rho)=p_0K\left[\left(\frac\rho{\rho_0}\right)^\gamma-1\right]+p_0
\end{equation}
\begin{equation}
    \label{eq:K}
    K=\frac{\rho_0c_0^2}{\gamma p_0}
\end{equation}
with $\gamma$ is set to 1 and $c_0$ is assigned by the user.

Above all, viscosity is considered as constant.

\subsection{Subgrid-scale Model}
To close the governing equations, wall-adaptive local eddy-viscosity (WALE) model \cite{kim2020assessment} is deployed:
\begin{equation}
    \label{eq:WALE}
    \mu_{sgs}=\rho(C_w\Delta)^2\frac{(S_{ij}^dS_{ij}^d)^{3/2}}{(S_{ij}S_{ij})^{5/2}+(S_{ij}^dS_{ij}^d)^{5/4}}
\end{equation}
with ${S}_{ij}^d$ defined as:
\begin{equation}
    \label{eq:S_ij^d}
    {S}_{ij}^d=\frac12\left(\frac{\partial u_i}{\partial x_k}\frac{\partial u_k}{\partial x_j}+\frac{\partial u_j}{\partial x_k}\frac{\partial u_k}{\partial x_i}\right)-\frac13\delta_{ij}\frac{\partial u_l}{\partial x_k}\frac{\partial u_k}{\partial x_l}
\end{equation}
and the model coefficient is set to be $C_w = 0.2$ in the current study.

\subsection{\label{subsec:STE-KEP-FR}STE-KEP-FR Scheme}
STE-KEP-FR \cite{liu2019ultra,lu2019grid} has been developed to combine the advantages of simple differential form high-order FR scheme \cite{huynh2007flux}, one-step time integration method that enables time-accurate local time stepping \cite{gassner2011explicit}, and enhanced nonlinear stability by adopting the split form kinetic preserving method to address the convective term \cite{abe2018stable}. To briefly review the STE-KEP-FR approach, first rewrite Eqs.~(\ref{eq:continuity_equation}) and (\ref{eq:momentum_equation}) in conservative form:
\begin{equation}
    \label{eq:conservative_form}
    \frac{\partial\boldsymbol{U}}{\partial t}+ \nabla^{\boldsymbol{x}}\cdot\boldsymbol{F}^{\boldsymbol{x}} = 0
\end{equation}
with
\begin{equation}
    \label{eq:U_F}
    \boldsymbol{U}=\begin{bmatrix}
        \rho\\
        \rho u\\
        \rho v\\
        \rho w
    \end{bmatrix},
    \quad\boldsymbol{F}^{\boldsymbol{x}}=\begin{bmatrix}
        \rho u_i\\
        \rho u_iu+p\delta_{i1}-\tau_{i1}\\
        \rho u_iv+p\delta_{i2}-\tau_{i2}\\
        \rho u_iw+p\delta_{i3}-\tau_{i3}
    \end{bmatrix}
\end{equation}
and the conversion from Cartesian coordinates to the computational domain can be described as:
\begin{equation}
    \label{eq:J_1}
    \widehat{\boldsymbol{U}}=|\boldsymbol{J}|\boldsymbol{U}
\end{equation}
\begin{equation}
    \label{eq:J_2}
    \boldsymbol{F}^{\xi}=|\boldsymbol{J}|\big(\xi_{x}\boldsymbol{F}^{x} + \xi_{y}\boldsymbol{F}^{y} + \xi_{z}\boldsymbol{F}^{z}\big)
\end{equation}
\begin{equation}
    \label{eq:J_3}
    \boldsymbol{F}^{\eta}=|\boldsymbol{J}|\big(\eta_{x}\boldsymbol{F}^{x} + \eta_{y}\boldsymbol{F}^{y} + \eta_{z}\boldsymbol{F}^{z}\big)
\end{equation}
\begin{equation}
    \label{eq:J_4}
    \boldsymbol{F}^{\zeta}=|\boldsymbol{J}|\big(\zeta_{x}\boldsymbol{F}^{x} + \zeta_{y}\boldsymbol{F}^{y} + \zeta_{z}\boldsymbol{F}^{z}\big)
\end{equation}
with $\boldsymbol{J}$ given by:
\begin{equation}
    \label{eq:J_5}
    \boldsymbol{J}=\begin{bmatrix}  
        x_\xi & x_\eta & x_\zeta \\  
        y_\xi & y_\eta & y_\zeta \\  
        z_\xi & z_\eta & z_\zeta  
        \end{bmatrix}
\end{equation}

Then Eq.~(\ref{eq:conservative_form}) in computational domain can be written as:
\begin{equation} 
    \label{eq:conservative_form_c}
    \frac{\partial\widehat{\boldsymbol{U}}}{\partial t}+\nabla^{\boldsymbol{\xi}}\cdot{\boldsymbol{F}}^{\boldsymbol{\xi}}=0
\end{equation}

\begin{figure}[hbt!]
    \centering
    \captionsetup{justification=raggedright, singlelinecheck=false}
    \subcaptionbox{\label{fig:solution_points_K3}K3}{
    \includegraphics[width = 0.4\textwidth]{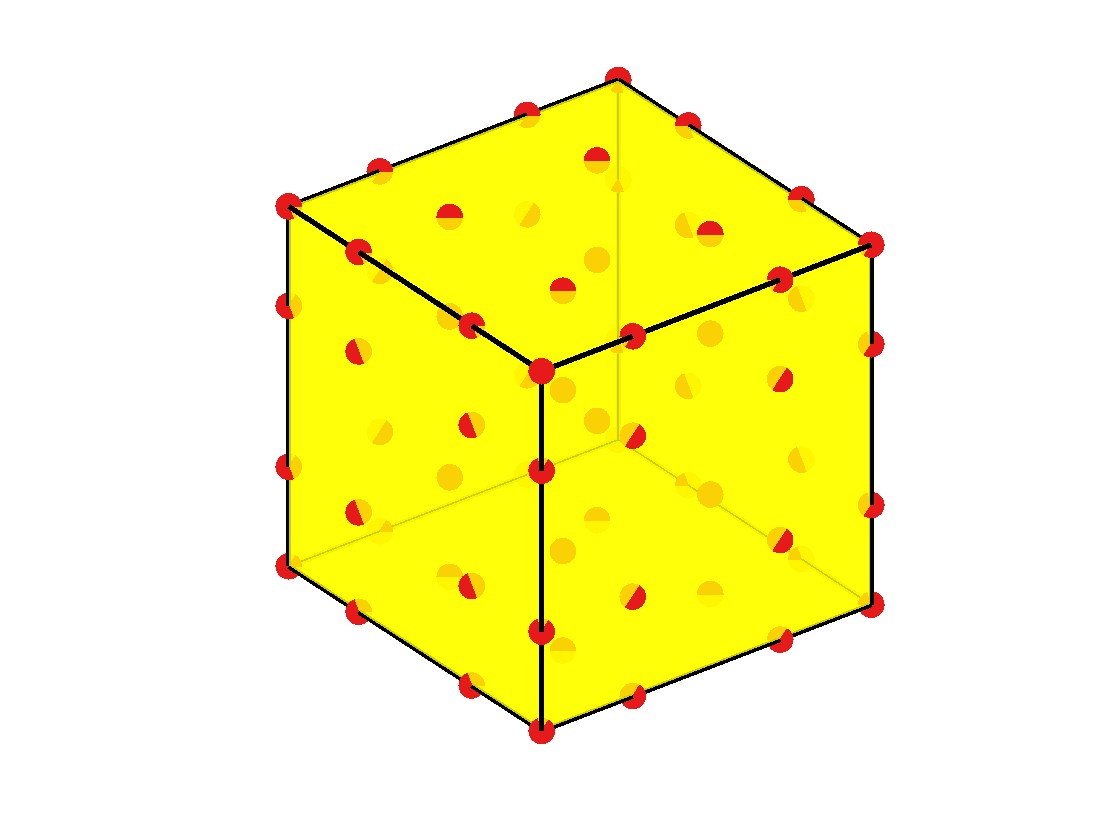}}
    \subcaptionbox{\label{fig:solution_points_K4}K4}{\includegraphics[width = 0.4\textwidth]{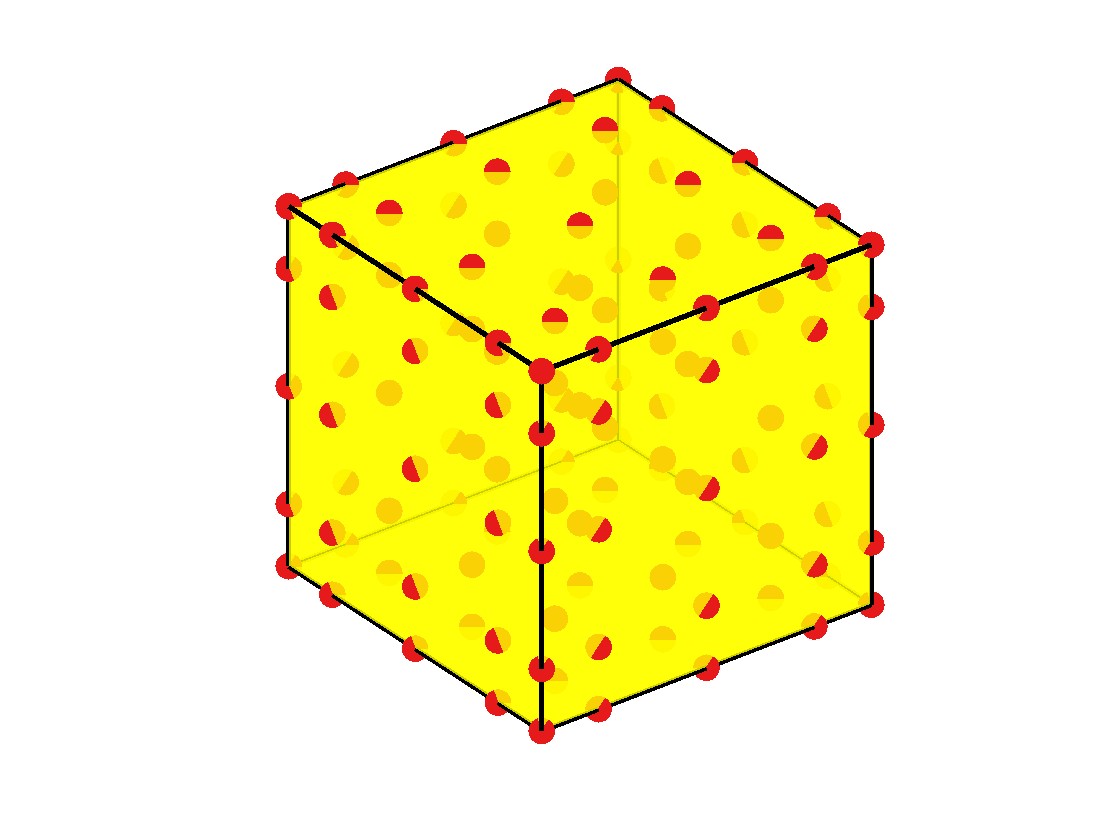}}
    \captionsetup{justification=centering}
    \caption{\label{fig:solution_points} LGL solution points discretization for hexahedral element with K = 3 and 4.}
\end{figure}

Illustration of using Lagrange-Gauss-Legendre (LGL) points as the solution points for hexahedral element with $3^{rd}$ degree polynomial reconstruction ($4^{th}$ order of accuracy) and $4^{th}$ degree polynomial reconstruction ($5^{th}$ order of accuracy) are shown in Fig.~\ref{fig:solution_points}. For $k^{th}$ solution point of $i^{th}$ element, the uniform FR scheme is given as:
\begin{equation} 
    \label{eq:FR_scheme}
    \frac{\partial\hat{\boldsymbol{U}}_{j,k}}{\partial t}+ \left(\nabla^{\boldsymbol{\xi}}\cdot{\boldsymbol{F}}^{\boldsymbol{\xi}}({\boldsymbol{U}}_{j})\right)_{j,k}+ \sum_{s=1}^{N_{s}}\sum_{f=1}^{N_{f}}\alpha_{k,s,f}\big(\widetilde{\boldsymbol{F}}^{\boldsymbol{\xi}}|_{n}- {\boldsymbol{F}}^{\boldsymbol{\xi}}|_{n}\big)_{k,s,f}=0
\end{equation}
where common flux $\widetilde{\boldsymbol{F}}$ generally takes the form of Riemann fluxes for nonlinear inviscid flux (HLLC Riemann solver is used in this study) and central average for viscous component. $\boldsymbol{F}|_{n}=\boldsymbol{F}\cdot{\boldsymbol{n}}$ where $\boldsymbol{n}$ is outer normal unit vector at each flux point. The difference between the common flux and the outer normal projection of local flux is referred to as correction flux, which is used to update the DoF by exchanging information with adjoint elements in a conservative way. The FR coefficient $\alpha$ can be obtained through lift operation \cite{lu2015high} for standard elements.

The following split form $Split_{Fe}$ is used to calculate convective term (without pressure) of divergence part in the governing equations:
\begin{equation} 
    \label{eq:Split_Fe}
    \nabla^{\boldsymbol{\xi}}\cdot{\boldsymbol{F}}^{\boldsymbol{\xi}}(\boldsymbol{U}_{j})= \frac{1}{2}\frac{\partial}{\partial\xi}\Big(\rho\widehat{\boldsymbol{U}}\phi\Big)+\frac{1}{2}\phi \frac{\partial}{\partial\xi}\Big(\rho\widehat{\boldsymbol{U}}\Big)+ \frac{1}{2}\rho\widehat{\boldsymbol{U}}\frac{\partial}{\partial\xi}(\phi)
\end{equation}
which enhances the nonlinear stability by its kinetic energy preserving property. Rewrite Eq.~(\ref{eq:FR_scheme}) as:
\begin{equation} 
    \label{eq:FR_scheme_rewrite}
    \frac{\partial{\boldsymbol{U}}_{j,k}}{\partial t}=\mathbb{R}_{j,k}^{Div}(\boldsymbol{U}_j)+\sum_{s=1}^{N_s}\sum_{f=1}^{N_f}\mathbb{R}_{j,k,s,f}^{Cor}(\boldsymbol{U}_j)
\end{equation}
\begin{equation} 
    \label{eq:R_jk_Div}
    \mathbb{R}_{j,k}^{Div}(\boldsymbol{U}_j)= -\frac{1}{|\boldsymbol{J}|_{j,k}}\left(\nabla^{\boldsymbol{\xi}}\cdot{\boldsymbol{F}}^{\boldsymbol{\xi}}({\boldsymbol{U}}_{i})\right)_{j,k}
\end{equation}
\begin{equation} 
    \label{eq:R_jksf_Corr}
    \mathbb{R}_{j,k,s,f}^{Cor}(\boldsymbol{U}_j)=-\frac{1}{|\boldsymbol{J}|_{j,k}} \alpha_{k,s,f}\big(\widetilde{\boldsymbol{F}}^{\boldsymbol{\xi}}|_{n}(\boldsymbol{U}_j)- {\boldsymbol{F}}^{\boldsymbol{\xi}}|_{n}(\boldsymbol{U}_j)\big)_{k,s,f}
\end{equation}
where $\mathbb{R}^{Div}$ and $\mathbb{R}^{Cor}$ are flux divergence part and the linear combination of correction flux part, respectively.

Integrating Eq.~(\ref{eq:FR_scheme_rewrite}) over $t\in[t^{n},t^{n+1}]$ can yield:
\begin{equation} 
    \label{eq:FR_scheme_rewrite_time}
    \boldsymbol{U}_{j,k}^{n+1}-\boldsymbol{U}_{j,k}^n=\int_{t^n}^{t^{n+1}}\left[\mathbb{R}_{j,k}^{Div}(\boldsymbol{U}_j)+\sum_{s=1}^{N_s}\sum_{f=1}^{N_f}\mathbb{R}_{j,k,s,f}^{Cor}(\boldsymbol{U}_j)\right] dt
\end{equation}

Construct a local space-time prediction $\boldsymbol{v}_j=\boldsymbol{v}(\boldsymbol{x}_j,t)$ for $t\in[t^{n},t^{n+1}]$ by solving the following time-dependent equation:
\begin{equation} 
    \label{eq:v_jk}
    \frac{d\boldsymbol{v}_{j,k}}{dt}= \mathbb{R}_{j,k}^{Div}\big(\boldsymbol{v}(\boldsymbol{x}_j,t)\big)
\end{equation}
with initial value $\boldsymbol{v}(\boldsymbol{x}_j,t=t^n)=\boldsymbol{U}_{j}^n$. The continuous explicit Runge-Kutta (CERK) method is deployed to solve the equation, and integrates it over $t\in[t^{n},t^{n+1}]$ derives:
\begin{equation} 
    \label{eq:v_jk_time}
    \int_{t^n}^{t^{n+1}}\mathbb{R}_{j,k}^{Div}\big(\boldsymbol{v}(\boldsymbol{x}_j,t)\big) dt=\boldsymbol{v}_{j,k}(t^{n+1})-\boldsymbol{U}_{j,k}^n
\end{equation}
taking the space-time polynomial $\boldsymbol{v}(\boldsymbol{x}_j,t)$ as a local predictor, and the combination of correction flux $\mathbb{R}_{j,k,s,f}^{Cor}$ as corrector, substituting Eq.~(\ref{eq:v_jk_time}) into Eq.~(\ref{eq:FR_scheme_rewrite_time}) will result in the STE-KEP-FR scheme:
\begin{equation} 
    \label{eq:STE-KEP-FR}
    \boldsymbol{U}_{j,k}^{n+1}= \boldsymbol{v}_{j,k}(t^{n+1})+ \sum_{s=1}^{N_s}\sum_{f=1}^{N_f}\int_{t^n}^{t^{n+1}}\mathbb{R}_{j,k,s,f}^{Cor}(\boldsymbol{v}_j(t^{n+1}))~dt\end{equation}
which means the time integrations of the correction flux can be carried out at each flux point instead of across the entire element interface. This is a very important feature to support flexible mesh schemes like hanging nodes and overset mesh. The numerical time integration process is executed as follows:
\begin{equation} 
    \label{eq:STE-KEP-FR_numerical}
    \boldsymbol{U}_{j,k}^{n+1}= \boldsymbol{v}_{j,k}(t^{n+1})+ \sum_{s=1}^{N_s}\sum_{f=1}^{N_f}\sum_{qp}^{N_{qp}^{t}} \Delta t\cdot w_{qp}^{t} \mathbb{R}_{j,k,s,f}^{Cor}\left(\boldsymbol{v}_{j}(t_{qp})\right)
\end{equation}
where physical time $t_{qp}=t^{n}+\Delta t\cdot\tau_{qp}$ is corresponding to $\tau_{qp}\in[0,1]$.

\subsection{FW-H Acoustic Analogy}
A modified version of OpenCFD-FWH \cite{zhang2023mpiopenmpmixingparallelopen} (Open Computational Fluid Dynamics code for FW-H acoustic analogy) is utilized to calculate far-field noise, with the ability of parallel I/O to further to accelerate the speed of far-field noise prediction. A brief overview of the far-field noise prediction procedure can begin with the FW-H equation for permeable surface \cite{ffowcs1969sound}:
\begin{equation}
    \label{eq:FWH}
    \Box^2c^2(\rho-\rho_0) = \frac{\partial}{\partial t}\left[Q_n\delta(f)\right] \quad   -\frac{\partial}{\partial x_i}\left[L_i\delta(f)\right]+\frac{\partial}{\partial   x_i\partial x_j}\left[T_{ij}H(f)\right]
\end{equation}
with the moving surface described by $f(\boldsymbol{x},t)=0$ such that ${\boldsymbol{n}}=\triangledown f$ represents the unit outward normal of the surface \cite{farassat2007derivation}. The $Q_n$, $L_i$, and Lighthill tensor stress $T_{ij}$ are defined as follows:
\begin{align}
    Q_n&=Q_i{n}_i=\left[\rho_0v_i+\rho(u_i-v_i)\right]{n}_i\label{eq:Qn}\\  
    L_i&=L_{ij}{n}_j=\left[P_{ij}+\rho u_i(u_j-v_j)\right]{n}_j\label{eq:Li}\\
    T_{ij}&=\rho u_iu_j+P_{ij}+c^2(\rho-\rho_0)\delta_{ij}\label{eq:Tij}
\end{align}
with $P_{ij}$ represents the stress tensor:
\begin{equation}
  P_{ij}=(p-p_{0})\delta_{ij}-\tau_{ij}\label{eq:Pij}
\end{equation}
$\tau_{ij}$ is generally considered as a negligible source of sound 
and is neglected by most other FW-H implementations  \cite{shen2020validation,epikhin2015development,farassat2007derivation}. Therefore, $P_{ij}=(p-p_{0})\delta_{ij}$ is used in this paper.

Omitting the quadrupole term in Eq.~(\ref{eq:FWH}), and following the derivation procedure of Farassat 1A formulation  \cite{farassat2007derivation}, the integral solution of the permeable surface FW-H equation can be obtained:
\begin{align}
    4\pi p_{T}^{\prime}(\boldsymbol{x},t)& =\int_{f=0}\left[\frac{\dot{{Q}_i} {n}_i+{Q_i} \dot{n_i}}{r(1-M_r)^2}\right]_{ret}dS\nonumber\\
    &+\int_{f=0}\left[\frac{Q_n(r\dot{M_r}+c_0(M_r-M^2))}{r^2(1-M_r)^3}\right]_{ret}dS\label{eq:pT}
  \end{align}
  \begin{align}
  4\pi p_L^{\prime}(\boldsymbol{x},t)& =\frac1{c_0}\int_{f=0}\left[\frac{\dot{L}_r}{r(1-M_r)^2}\right]_{ret}dS\nonumber\\
  &+\int_{f=0}\left[\frac{L_{r}-L_{M}}{r^{2}(1-M_{r})^{2}}\right]_{ret}dS\nonumber\\
  &+\frac{1}{c_0}\int_{f=0}\left[\frac{L_r(r\dot M_r+c_0(M_r-M^2))}{r^2(1-M_r)^3}\right]_{ret}dS\label{eq:pL}
  \end{align}
\begin{equation}
    \label{eq:p}
    p'(\boldsymbol{x},t)=p'_T(\boldsymbol{x},t)+p'_L(\boldsymbol{x},t)
\end{equation}
where the superscript "\textperiodcentered{}" indicates differentiation
with respect to the source time $\tau$, and $M_r$, $L_r$, $L_M$ are defined as:
\begin{align}
    M_r&=M_i\hat{r}_i\label{eq:Mr}\\
    L_r&=L_i\hat{r}_i\label{eq:Lr}\\
    L_M&=L_iM_i\label{eq:LM}
\end{align}

The subscript $ret$ in Eqs.~(\ref{eq:pT}) and (\ref{eq:pL}) means the quantities inside the square brackets are determined at 
the retarded time:
\begin{equation}
    \label{eq:tau}
    \tau_{ret}=t-r_{ret}/c
\end{equation}

Despite the quadrupole term in Eq.~(\ref{eq:FWH}) has been excluded, the quadrupole source within the permeable FW-H surface are still considered by Eq.~(\ref{eq:p}) according to \citet{brentner1998analytical}.

Eq.~(\ref{eq:p}) is derived in a coordinate system in which the source is in motion within a stationary medium with observers at-rest in the far-field. In the scenario of a wind tunnel case, where both the source and observers are stationary within a uniform flow with an AoA, the Garrick Triangle\cite{garrick1953theoretical} can be utilized to transform the coordinate system. This transformation results in a significant simplification of the formulation and enhances the computational efficiency of the code.

Consider the mean flow to possess a velocity $U_0$ in the direction of the positive $x_1$ axis.
Equations (\ref{eq:tau}) will be changed to:
\begin{equation}
    \label{eq:tau2}
    \tau_{ret}=t-R/c_0
\end{equation}
where R is the effective acoustic distance between the source and the observer \cite{bres2010ffowcs}:
\begin{align}
  R&=\frac{-M_0d_1+R^*}{\beta^2}\label{eq:R}\\  
  R^*&=\sqrt{d_1^2+\beta^2[d_2^2+d_3^2]}\label{eq:R_s}\\
  \beta&=\sqrt{1-M_0^2}\label{eq:beta}\\
  M_0&=U_0/c_0\label{eq:M0}\\
  d_i&=x_i-y_i\label{eq:di}
\end{align}
and the unit radiation vector component is now changed to:
\begin{align}
  \hat{R}_1&=\frac{-M_0R^*+d_1}{\beta^2R}\label{eq:R1}\\  
  \hat{R}_2&=d_2/R\label{eq:R2}\\
  \hat{R}_3&=d_3/R\label{eq:R3}
\end{align}

Subsequently, consider a mean flow with an AoA in the x-y plane, and its velocity magnitude remains equal to $U_0$. Using the 2D plane coordinate transformation:
\begin{align}
  d_1^{\prime}&=\;\,\,   d_1cos(\mathrm{AoA})+d_2sin(\mathrm{AoA})\label{eq:d1_prime}\\  
  d_2^{\prime}&=        -d_1sin(\mathrm{AoA})+d_2cos(\mathrm{AoA})\label{eq:d2_prime}
\end{align}
and bring them into the $d_1$ and $d_2$ of the Eq.~(\ref{eq:R}), and Eq.~(\ref{eq:R_s}) yields:
\begin{align}
  &\quad R=\frac{-M_1d_1-M_2d_2+R^*}{\beta^2}\label{eq:R_New}\\  
  R^*&=\sqrt{(M_1d_1+M_2d_2)^2+\beta^2[d_1^2+d_2^2+d_3^2]}\label{eq:R_s_New}\\
  &\qquad \; M_1=M_0cos(\mathrm{AoA})\label{eq:M1}\\
  &\qquad \; M_2=M_0sin(\mathrm{AoA})\label{eq:M2}
\end{align}
with the unit radiation vector component changed to:
\begin{align}
  {\hat{R}_1}^{\prime}&=\frac{-M_0R^*+d_1^{\prime}}{\beta^2R}\label{eq:R1_prime}\\  
  {\hat{R}_2}^{\prime}&=d_2^{\prime}/R\label{eq:R2_prime}\\
  \hat{R}_3&=d_3/R\label{eq:R3_New}\\
  \hat{R}_1&={\hat{R}_1}^{\prime}cos(\mathrm{AoA})-{\hat{R}_2}^{\prime}sin(\mathrm{AoA})\label{eq:R1_New}\\  
  \hat{R}_2&={\hat{R}_1}^{\prime}sin(\mathrm{AoA})+{\hat{R}_2}^{\prime}cos(\mathrm{AoA})\label{eq:R2_New}
\end{align}

Then, substituting all instances of the variable $r$ in Eqs.~(\ref{eq:pT}) $\sim$ (\ref{eq:Lr}) by $R$. Furthermore, adjustments must be made to both the velocity of the moving surface and the fluid velocity to account for the change of the coordinate system. Thus, the mean flow velocity is subtracted. After simplification, Eqs.~(\ref{eq:pT}) and (\ref{eq:pL}) for wind tunnel scenarios are derived as follows:
\begin{align}
  4\pi p_{T}^{\prime}(\boldsymbol{x},t)& =\int_{f=0}\left[\frac{\dot{{Q}_i} {n}_i}{R(1-M_R)^2}\right]_{ret}dS\nonumber\\
  &+\int_{f=0}\left[\frac{Q_nc_0(M_R-M^2)}{R^2(1-M_R)^3}\right]_{ret}dS\label{eq:pT_WT}
\end{align}
\begin{align}
4\pi p_L^{\prime}(\boldsymbol{x},t)& =\frac1{c_0}\int_{f=0}\left[\frac{\dot{L}_R}{R(1-M_R)^2}\right]_{ret}dS\nonumber\\
&+\int_{f=0}\left[\frac{L_R-L_M}{R^{2}(1-M_R)^{2}}\right]_{ret}dS\nonumber\\
&+\int_{f=0}\left[\frac{L_R(M_R-M^2)}{R^2(1-M_R)^3}\right]_{ret}dS\label{eq:pL_WT} 
\end{align}
with
\begin{align}
  &Q_i=\left[-\rho_0U_{0i}+\rho{u_i}\right]\label{eq:Qi_New}\\  
  L_i&=[P_{ij}+\rho (u_i-U_{0i})u_j]{n}_j\label{eq:Li_New}\\
  &\qquad \; M_R=M_i\hat{R}_i\label{eq:MR}\\
  &\qquad \; L_R=L_i\hat{R}_i\label{eq:LR}
\end{align}

It is worth noticing that by setting the surface velocity to zero when using the wall surface as the FW-H sampling surface, one can derive the corresponding far-field noise prediction equation for solid surface.

\section{\label{sec:SD}Simulation Details}
\subsection{30P30N Airfoil Geometric Details}
The 30P30N slat noise problem is chosen from the $3^{rd}$ AIAA Workshop on Benchmark Problems for Airframe Noise Computations (BANC-III) \cite{choudhari2015assessment}. This classic high-lift airfoil was created by McDonnell Douglas (now Boeing) in the early 1990s and has been extensively used in research regarding the aeroacoustic characteristics of high-lift devices, especially in the context of slat noise \cite{murayama2014experimental,terracol2015aeroacoustic,choudhari2015assessment,murayama2018experimental,souza2019dynamics,housman2019predictions,himeno2021spod,chen2021noise,asada2024aeroacoustics}.

\begin{figure}[hbt!]
    \centering
    \includegraphics[width=.45\textwidth]{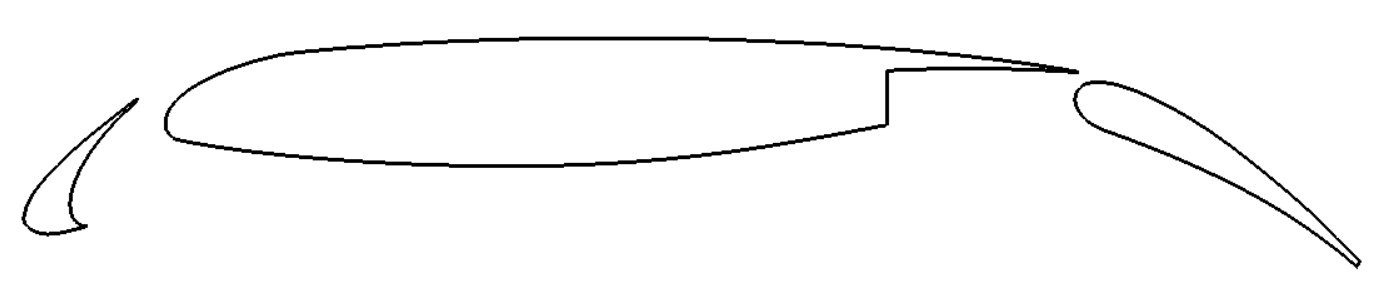}
    \caption{\label{fig:30P30N} Profile of the JAXA modified 30P30N airfoil.}
\end{figure}

The modified 30P30N from Japan Aerospace Exploration Agency (JAXA) \cite{murayama2018experimental} is used as shown in Fig.~\ref{fig:30P30N}. It has a stowed chord length of $c_s=0.4572 m$, and both the slat and flap have deflection angles of $30^\circ$ with corresponding chord lengths of $0.15c_s$ and $0.3c_s$, respectively. Under an inflow Mach number (Ma) of $0.17$, the Reynolds number based on the stowed chord length is equal to $1.71\times 10^6$.

A total of four AoAs, i.e. $3.5^\circ$, $5.5^\circ$, $9.5^\circ$, and $14^\circ$ are simulated for this basic 30P30N airfoil. The model is also scaled by factors of 0.5, 5, and 10 to investigate the Reynolds number effect on the slat noise. Additionally, the 10 times lager model has the slat TE adjusted to match the size of real aircraft, allowing for the study of the high-frequency hump associated with high-frequency vortex shedding at the slat TE.

\subsection{Mesh Details}
A total of nine structure meshes are generated with different densities, spanwise lengths and stowed chord lengths. The specifics of these meshes are detailed in Table~\ref{tab:meshes}. The dgree of freedom (DoF) of these meshes ranges from 6 million to 35 million for the K3 simulation, and they all maintain a consistent $y_{mesh}^+$ value of 10. According to \citet{wang2024towards}, in the K3 simulation, a $y_{mesh}^+=10$ corresponds to an equivalent $y^+$ value of 0.625 or lower when compared to the second-order simulation.
\begin{table}[hbt!]
    \caption{\label{tab:meshes} 30P30N mesh settings}
    \centering
    \begin{tabular}{lcccccccc}
    \hline
    Meshes       &N      & $c_s/m$& Re               & $s_z/c_s$& $N_z$& $s_t/c_{slat}$\\\hline
    Fine         &471,562& 0.4572 & $1.71\times 10^6$& 11.1\%   & 13   & 1.11\%        \\
    Medium       &235,750& 0.4572 & $1.71\times 10^6$& 11.1\%   & 10   & 1.11\%        \\
    Coarse       &117,929& 0.4572 & $1.71\times 10^6$& 11.1\%   & 7    & 1.11\%        \\
    Medium\_4    & 94,300& 0.4572 & $1.71\times 10^6$& 4.44\%   & 4    & 1.11\%        \\
    Medium\_23   &542,225& 0.4572 & $1.71\times 10^6$& 25.6\%   & 10   & 1.11\%        \\
    Medium\_10Re &298,130& 4.5720 & $1.71\times 10^7$& 11.1\%   & 10   & 1.11\%        \\
    Medium\_10Re*&290,330& 4.5720 & $1.71\times 10^7$& 11.1\%   & 10   & 0.28\%        \\
    Medium\_5Re  &274,830& 2.2860 & $8.55\times 10^6$& 11.1\%   & 10   & 1.11\%        \\
    Medium\_0.5Re&247,040& 0.2286 & $8.55\times 10^5$& 11.1\%   & 10   & 1.11\%        \\
\hline
\end{tabular}
\end{table}

\begin{figure}[hbt!]
    \centering
        \captionsetup{justification=raggedright, singlelinecheck=false}
        \subcaptionbox{\label{fig:global_view}Global view of the Medium mesh}{
        \includegraphics[width = 0.45\textwidth]{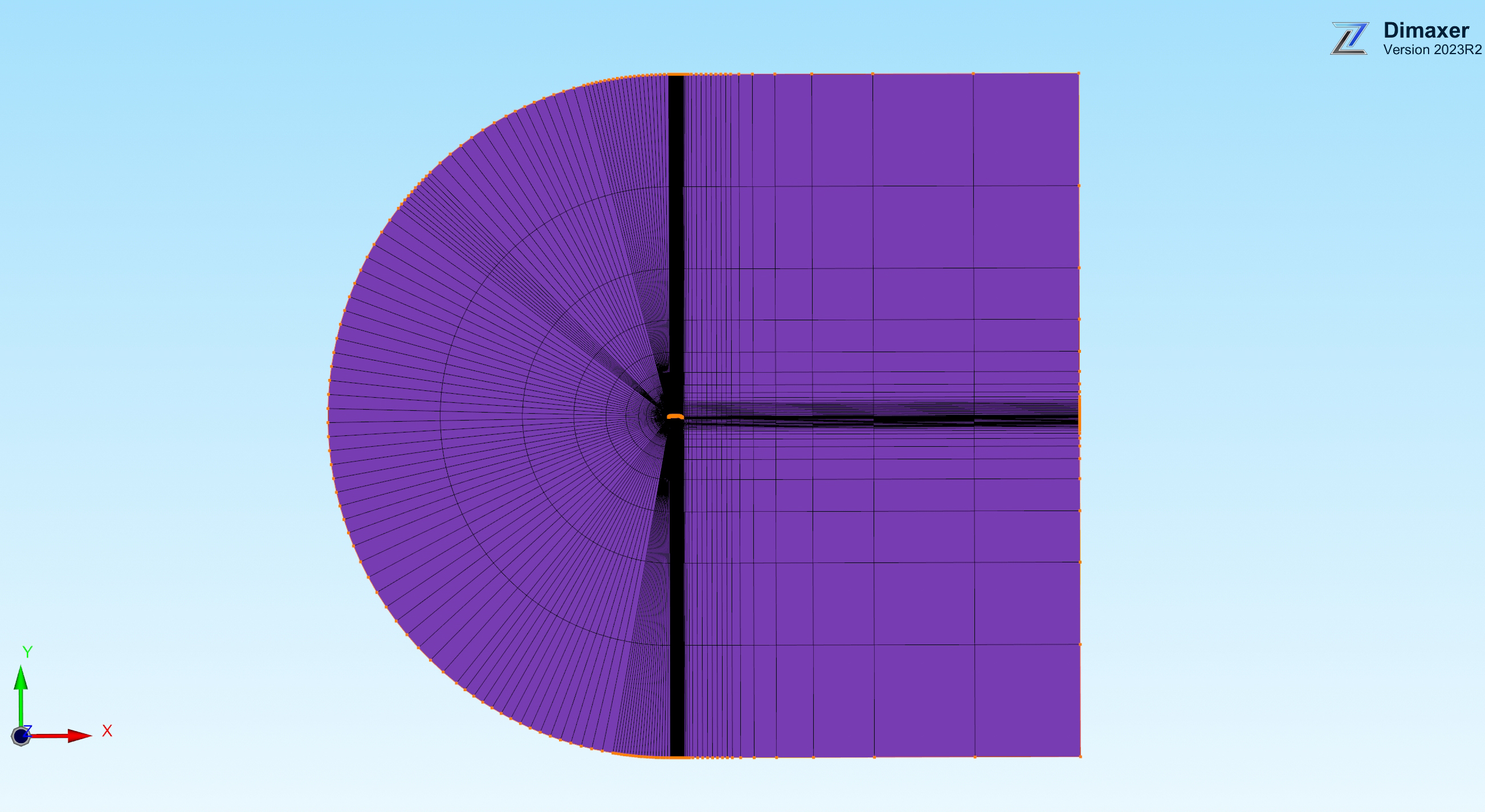}}
        \quad
        \subcaptionbox{\label{fig:Fine}Fine mesh}{\includegraphics[width = 0.45\textwidth]{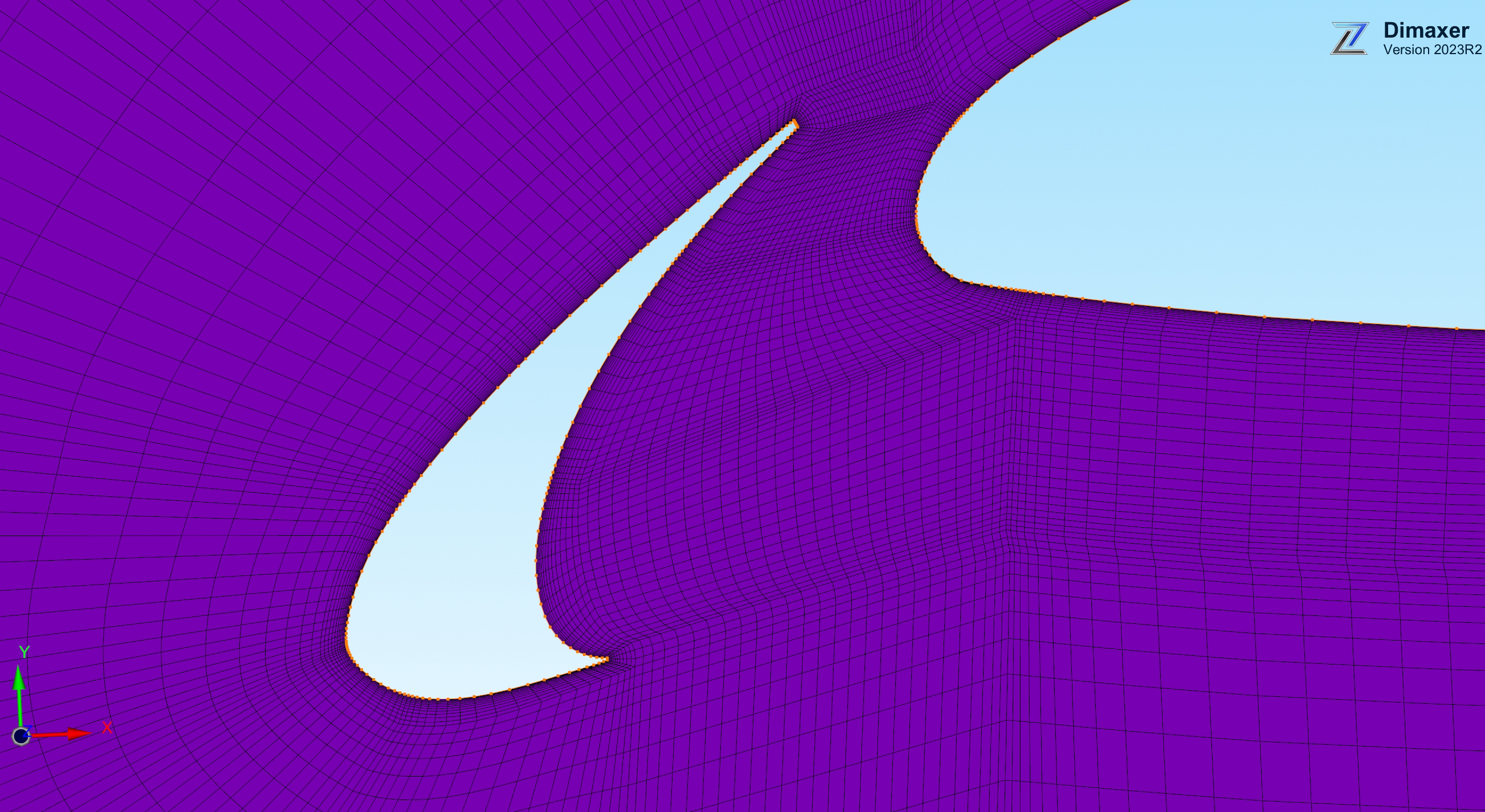}}
        \quad
        \subcaptionbox{\label{fig:Medium}Medium mesh}{\includegraphics[width = 0.45\textwidth]{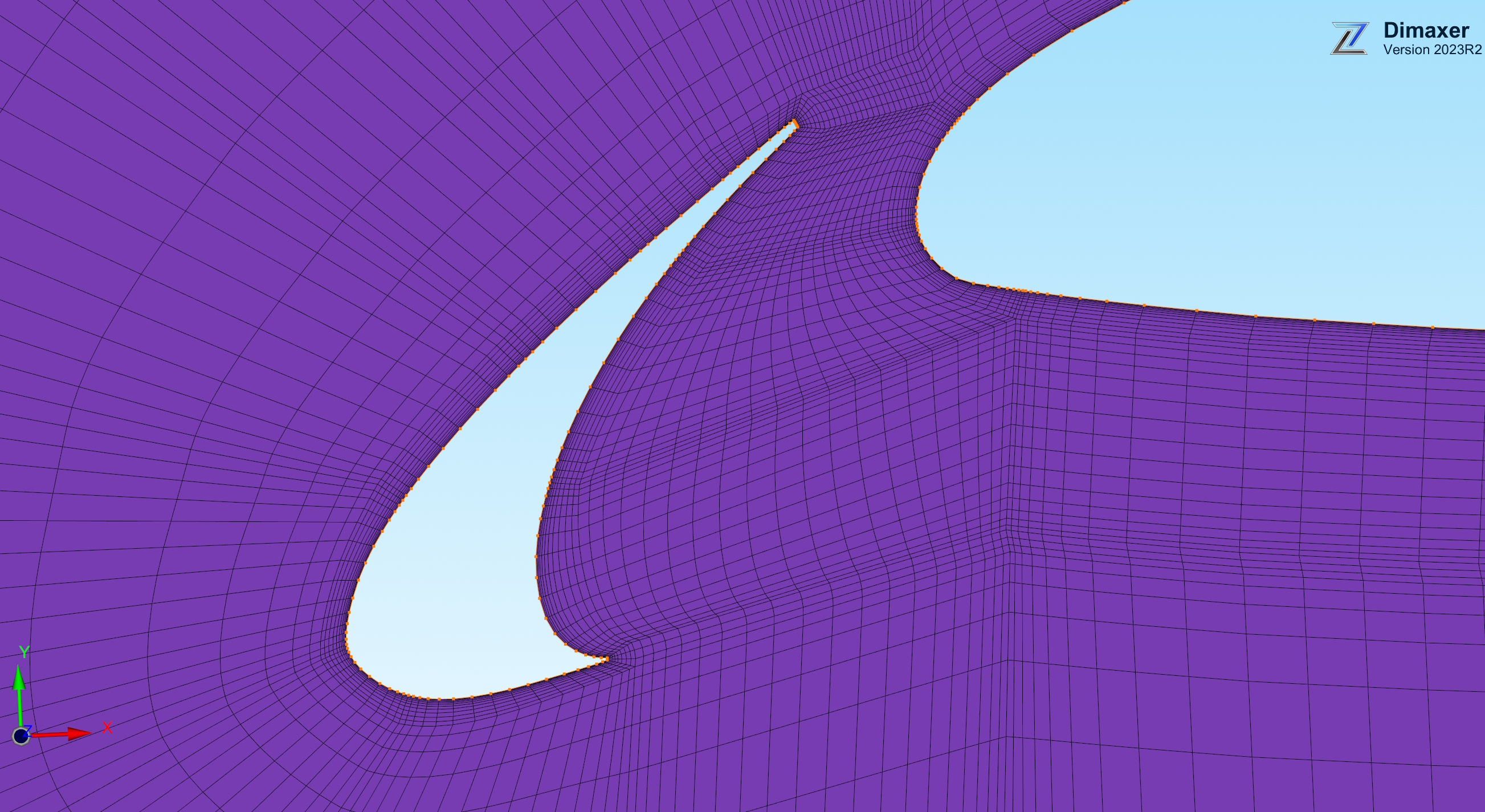}} 
        \quad
        \subcaptionbox{\label{fig:Coarse}Coarse mesh}{\includegraphics[width = 0.45\textwidth]{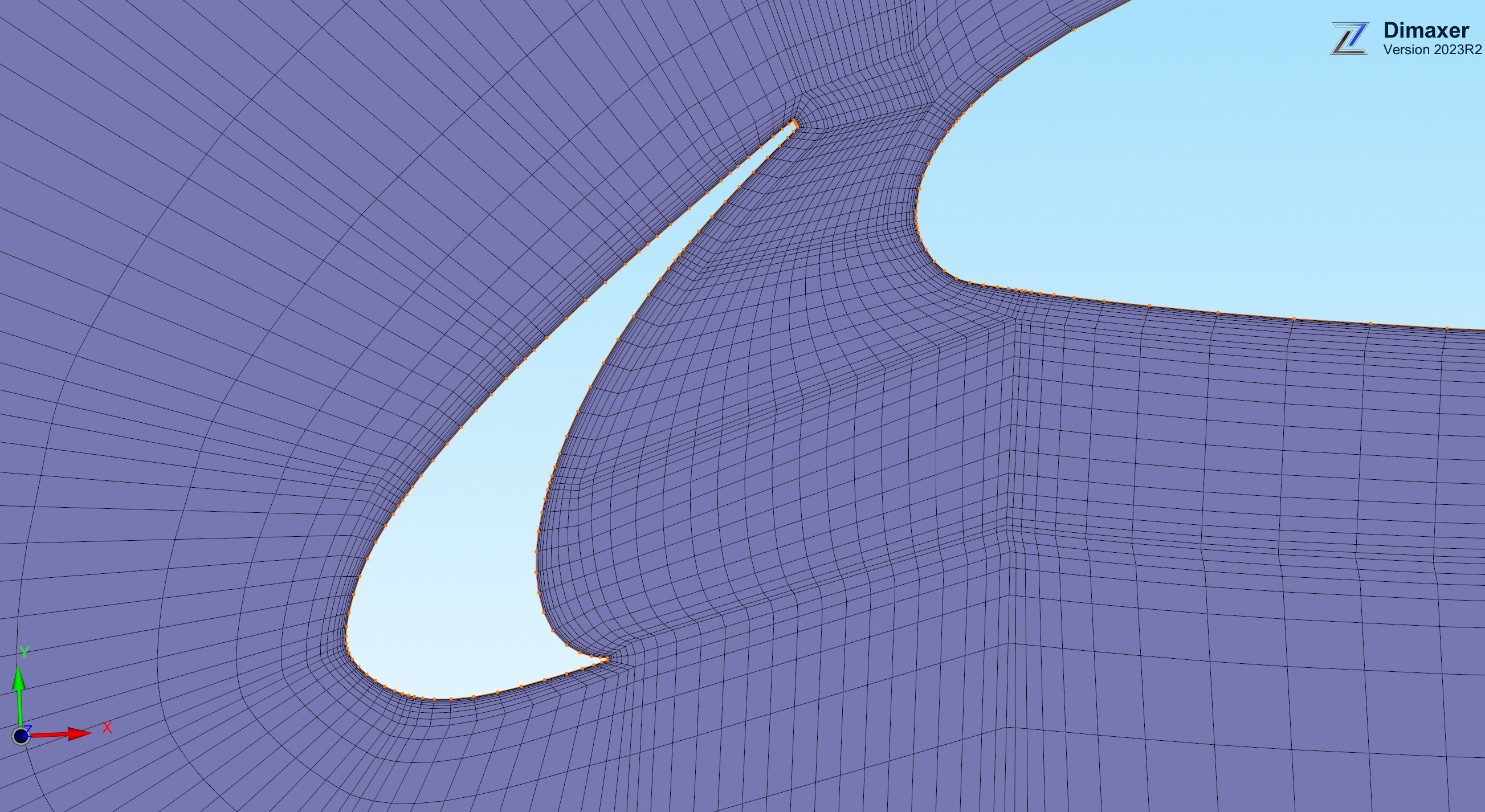}}
    \captionsetup{justification=centering} 
    \caption{\label{fig:meshes} Global view of the computational domain and detail views near the slat region for differnt meshes.}
\end{figure}

Figure~\ref{fig:meshes} shows a global view of the Medium mesh and detail views of the slat region of the Fine, Medium and Coarse mesh. The computational domain extends $30c_s$ in the forward as well as the vertical directions, and $36c_s$ in the rear direction. Due to the JAXA modified 30P30N has a sharp slat cusp, a cut is made to reduce mesh quantity and improve mesh quality. A translational periodic boundary condition is applied in the spanwise direction (z direction) with non-reflective boundary condition applied in the other two directions.
    
\subsection{\label{subsec:SPD}Simulation procedure Details}
A RANS solution is obtained at first as the initial field of each K3 WRLES simulation to speed up the convergence process. The stable K3 flow field then serves as the initial condition for the K4 WRLES simulation. As mentioned in Sec.~\ref{subsec:STE-KEP-FR} before, CERK is employed to explicitly advance the time step on local elements, with CFL equals to 0.6 and 0.3 for the K3 and K4 simulations. Each case is calculated until the flow field is stable, followed by an additional 10 flow pass times (FPTs, based on $c_s$). Data processing for the flow field is carried out only on the last 10 FPTs, with an acoustic data sampling rate of 100 kHz (except for the 0.5Re case, which has a sampling rate of 200 kHz).

Dimaxer (solver version 4610) conducts all simulations on a GPU node, using no more than two Nvidia Ada Lovelace GPUs per case. The basic 30P30N cases only require 23.86\textasciitilde143.26 GPU hours to time march 10 FPTs. For postprocessing, the computational mesh is divided into a significantly finer mesh, with each element partitioned into $K^3$ elements to visualize the high-order flow field.

\section{\label{sec:RaD}Results and Discussions}
\subsection{Influence of the FW-H Sampling Surfaces}
\begin{figure}[hbt!]
    \centering
    \includegraphics[width=.5\textwidth]{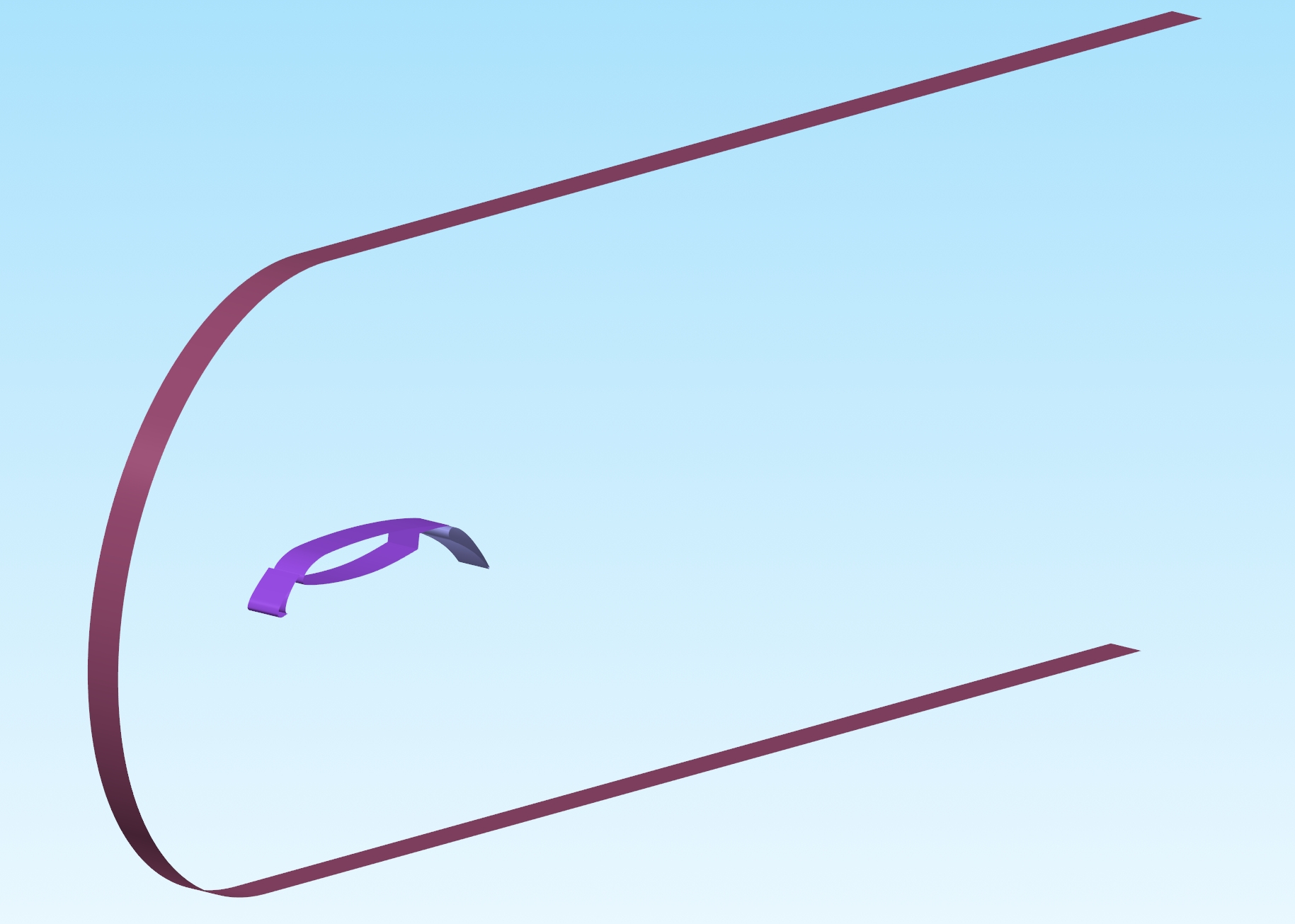}
    \caption{\label{fig:FWH_surfaces} Illustration of the FW-H sampling surface categories.}
\end{figure}

Two categories of surfaces are used to study the influence of the FW-H sampling surface. One is the solid surface category as shown in Fig.~\ref{fig:FWH_surfaces} with purple series color, the other is the porous surface category as shown in Fig.~\ref{fig:FWH_surfaces} with wine red color.

The solid surface category consists of four surfaces: slat, main wing, flap, and the airfoils above, all of which are directly extracted on the Fine mesh. The porous surface category includes two surfaces, both positioned $c_s$ in front of the airfoil and extending $5c_s$ in the wake direction, but they differ in density. The coarser surface is extracted directly on the Fine mesh, while the finer surface is refined to align with the K3 solution points.

\begin{figure}[hbt!]
    \centering
        \captionsetup{justification=raggedright, singlelinecheck=false}
        \subcaptionbox{\label{fig:Kevlar_solid}Solid surface category}{
        \includegraphics[width = 0.49\textwidth]{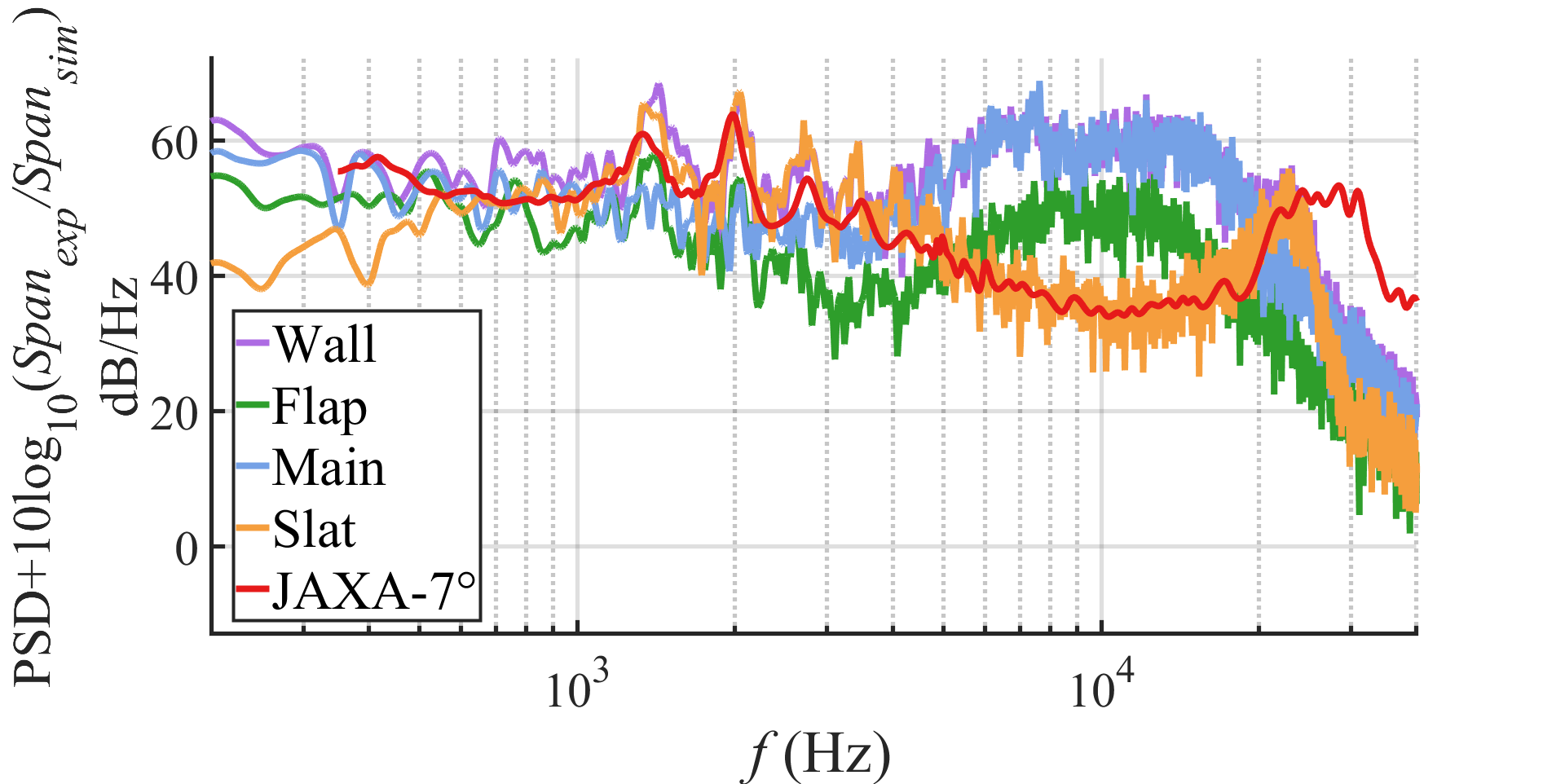}}
        \subcaptionbox{\label{fig:Kevlar_porous}Porous surface category}{\includegraphics[width = 0.49\textwidth]{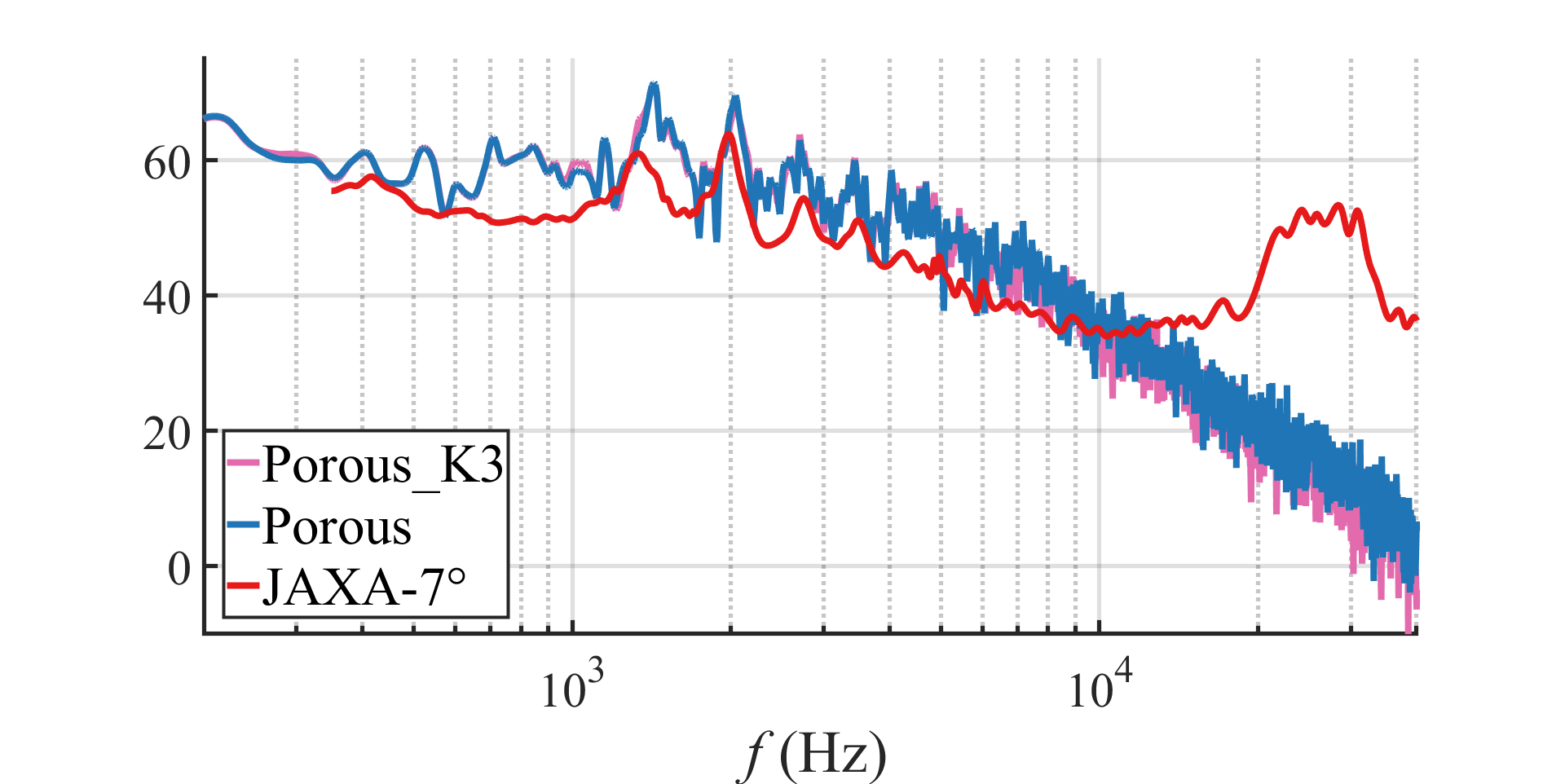}} 
    \caption{\label{fig:Kevlar} Acoustic signal results at the center of the JAXA Kevlar wall experiment phased-microphone array \cite{murayama2018experimental} through different FW-H sampling surfaces at $5.5^\circ$ AoA and 0.17 Ma.}
\end{figure}

\begin{figure}[hbt!]
    \centering
        \captionsetup{justification=raggedright, singlelinecheck=false}
        \subcaptionbox{\label{fig:Q_transition}Iso-surface of $Q=5e7$ colored by velocity magnitude}{
        \includegraphics[width = 0.475\textwidth]{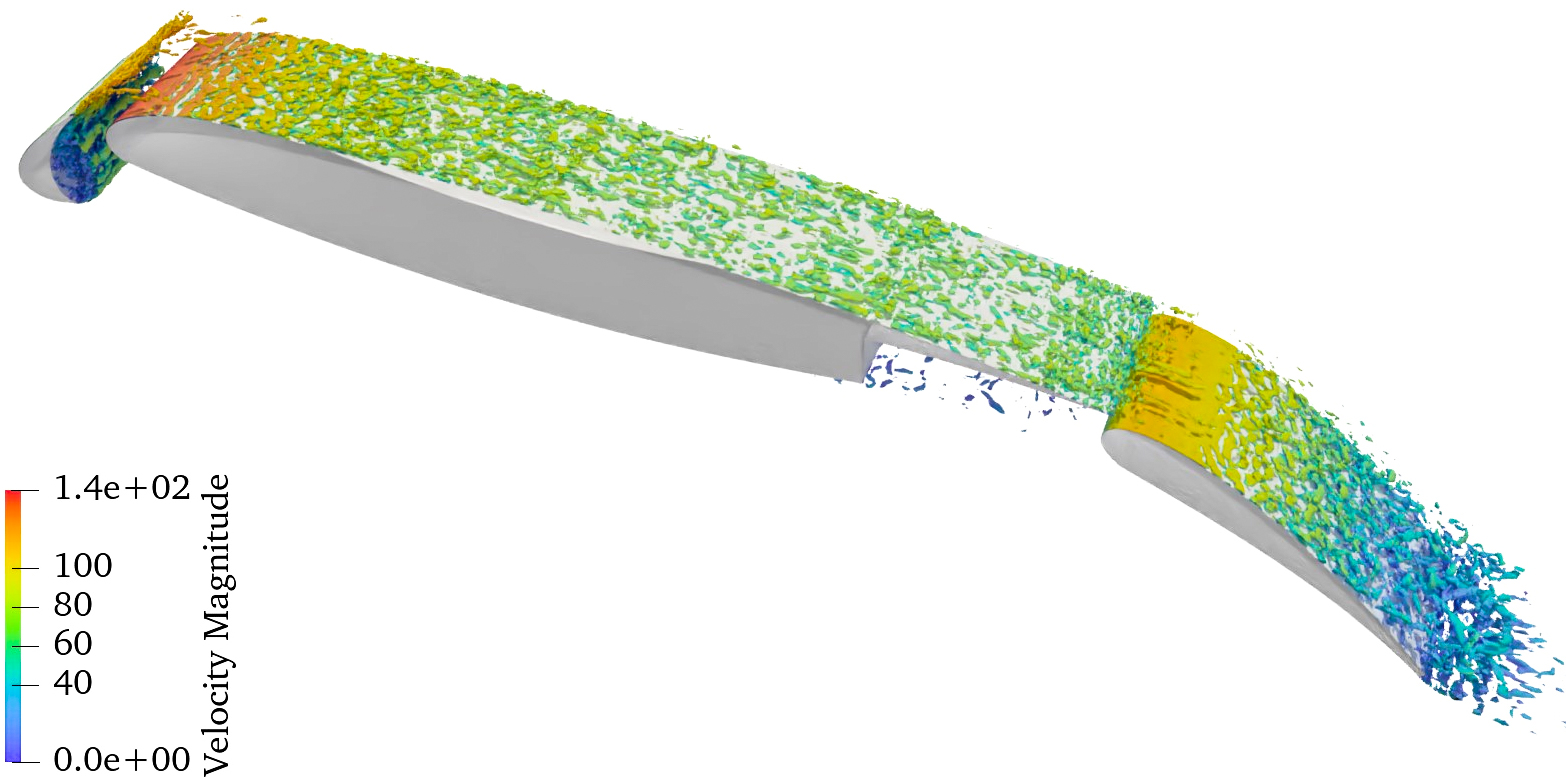}}
        \quad
        \subcaptionbox{\label{fig:Over_4kHz}Overall wall pressure level over 4 kHz}{\includegraphics[width = 0.49\textwidth]{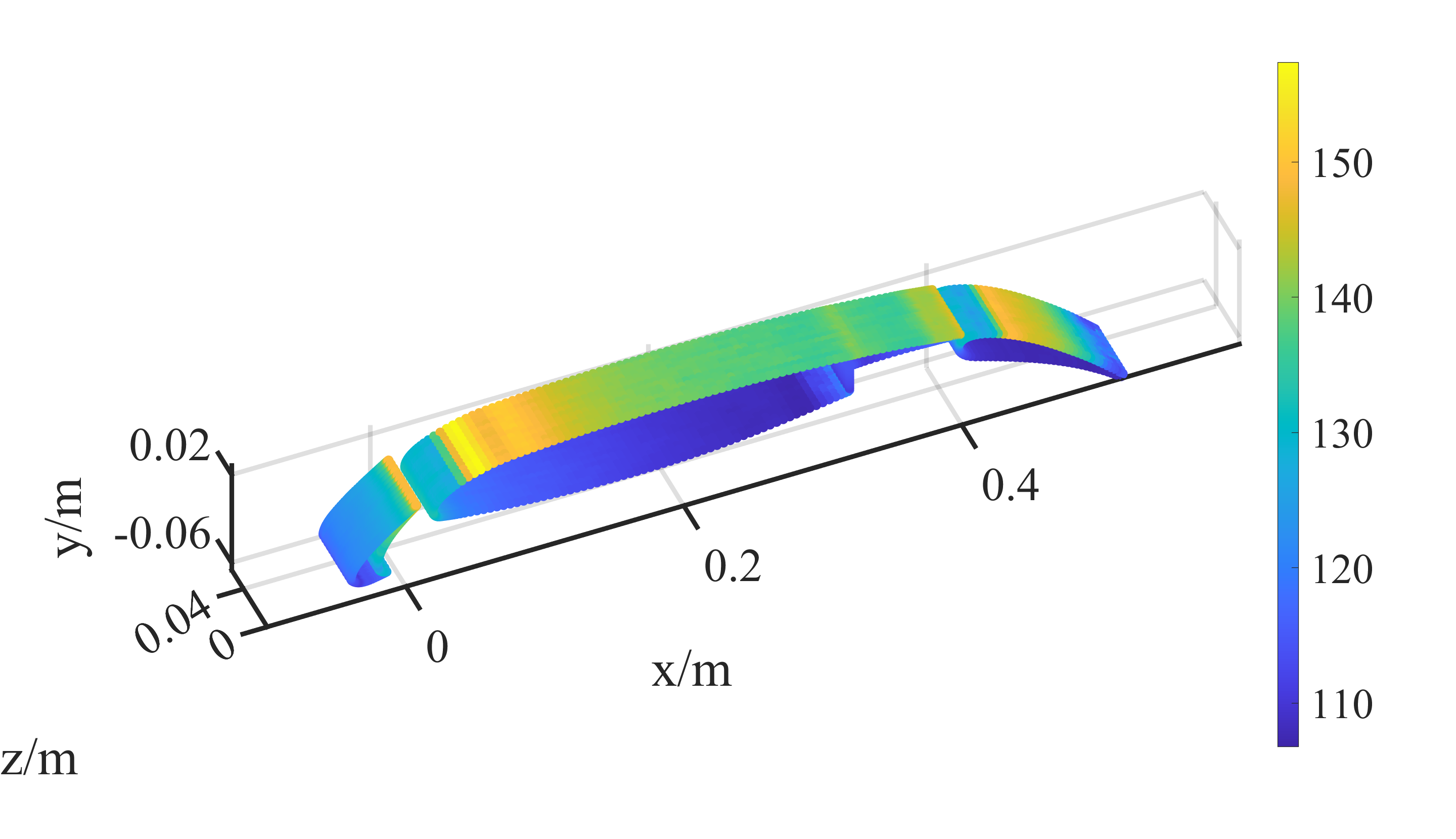}} 
    \captionsetup{justification=centering}
    \caption{\label{fig:Transition} Boundary-layer transition of the Fine mesh K3 simulation at $5.5^\circ$ AoA and 0.17 Ma.}
\end{figure}

\begin{figure}[hbt!]
    \centering
    \includegraphics[width=.49\textwidth]{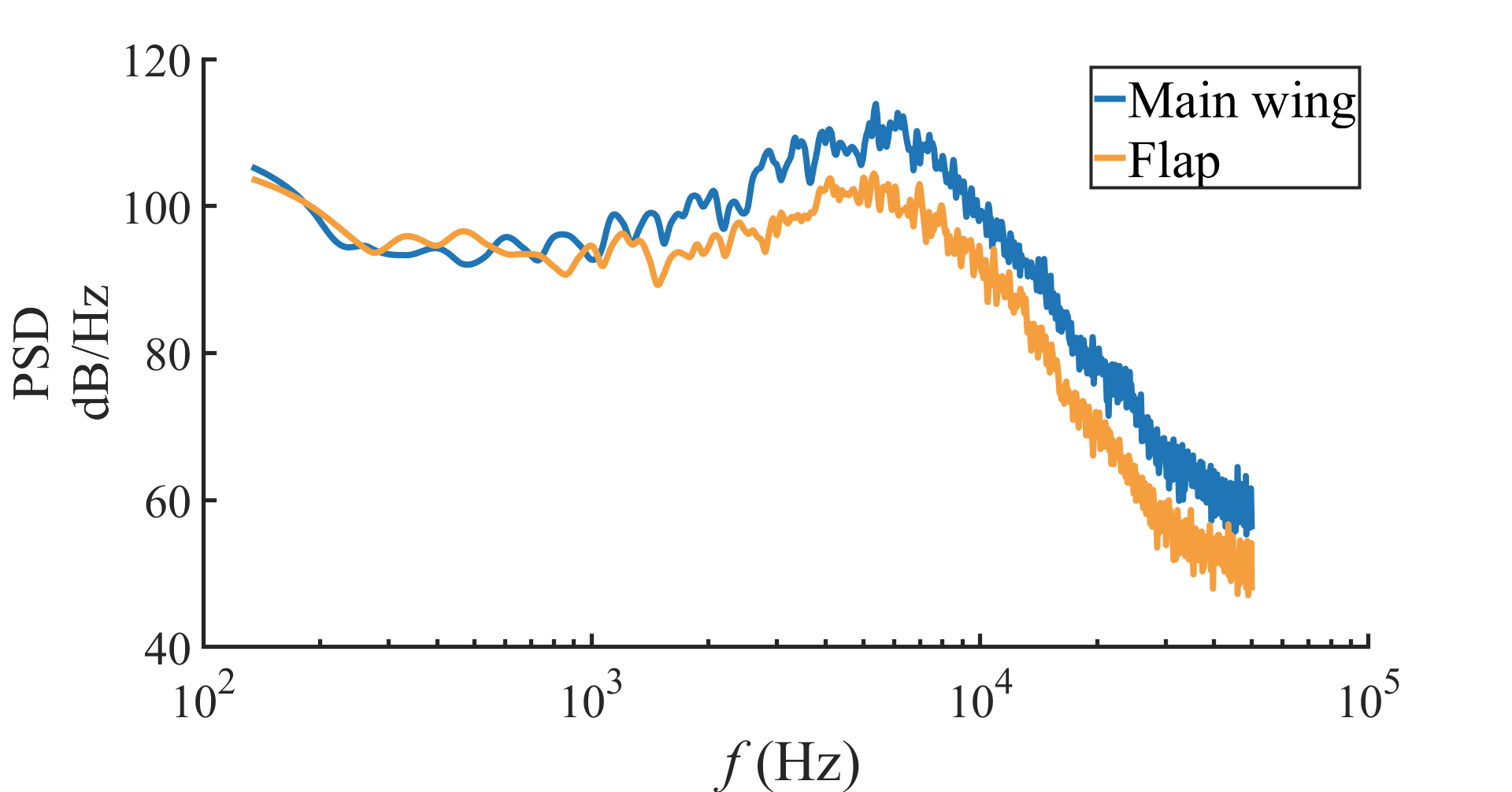}
    \caption{\label{fig:PSD_transition}Wall pressure spectrum on the position of the transition occurrence on the upper surface of the main wing and flap midplane in the Fine mesh K3 simulation at $5.5^\circ$ AoA and 0.17 Ma.}
\end{figure}

\begin{figure}[hbt!]
    \centering
    \includegraphics[width=.49\textwidth]{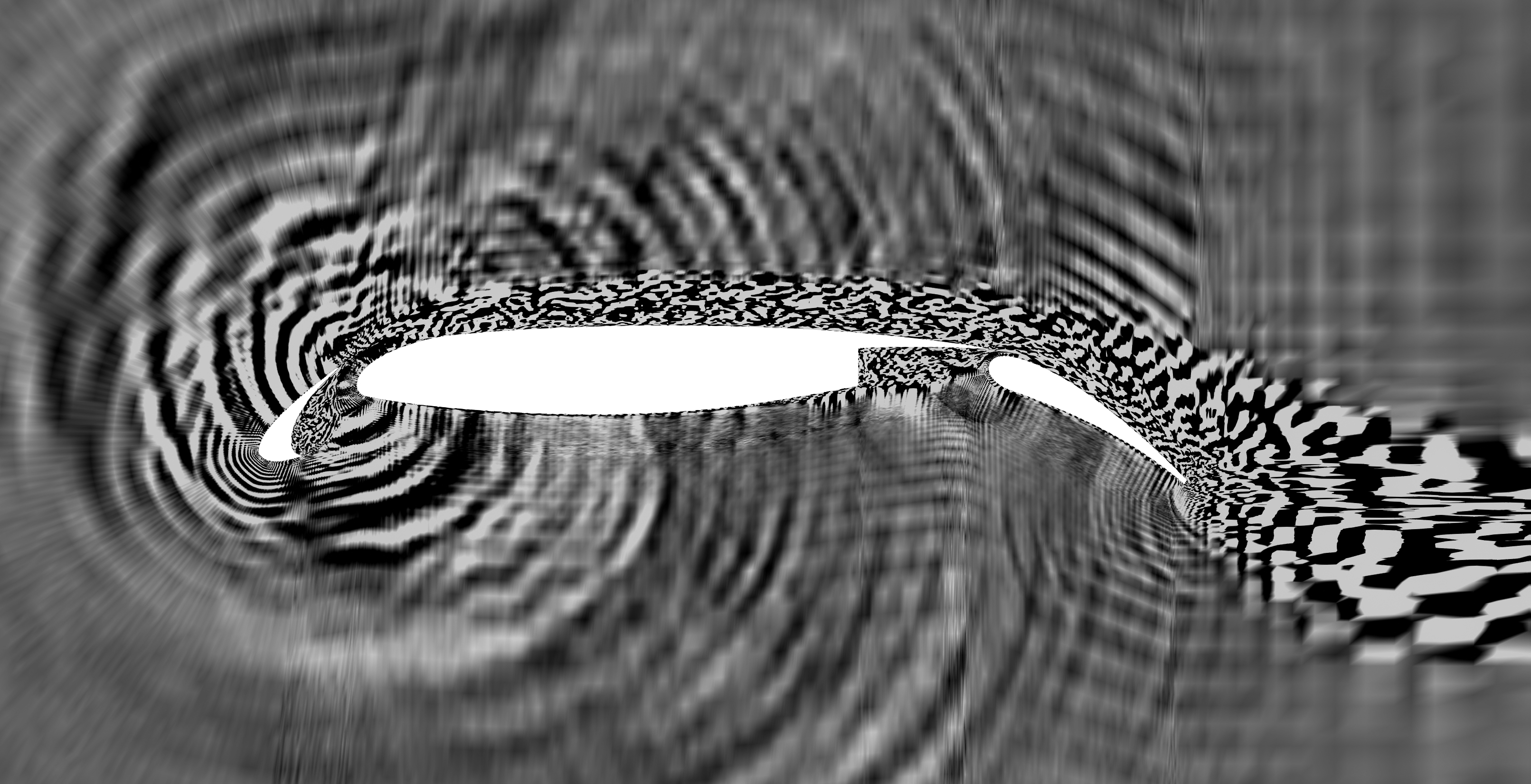}
    \caption{\label{fig:Div_Fine_K3}Instantaneous contour of $Div(\rho \boldsymbol{u})$ in the midspan plane at $5.5^\circ$ AoA and 0.17 Ma (-3<$Div(\rho \boldsymbol{u})$<3).}
\end{figure}

Figure~\ref{fig:Kevlar} presents the far-field noise results from various FW-H surfaces using the Fine mesh and K3 simulation to solve the flow field at $5.5^\circ$ AoA and 0.17 Ma. It can be seen in Fig.~\ref{fig:Kevlar_solid} that really good agreement has achieve with the JAXA Kevlar wall experiment \cite{murayama2018experimental} when the slat surface is used as the sampling surface, even the amplitude of the hump is match (but drops off quickly due to sampling frequency limitation). In contrast, the other three solid surfaces exhibit amplitude increases above 4 kHz. Figures~\ref{fig:Transition} and \ref{fig:PSD_transition} indicate that the pressure fluctuations caused by transition are the reason for these amplitude increases. Since the high-order WRLES simulation effectively captures the transition occurs on the upper surfaces of the main wing and flap, and the FW-H acoustic analogy propagates these pressure fluctuations into the far-field, which is inaccurate. This is because the pressure fluctuations from the transition do not emit sound waves, as presented in Fig.~\ref{fig:Div_Fine_K3}. Furthermore, the dilatation field indicates that the primary source of radiated noise originates from the slat TE and slat cove region. A comparison of Fig.~\ref{fig:Div_Fine_K3} with Fig.~7 from \citet{terracol2020wall} of another high-lift airfoil WRLES simulation with 2.6 billion cells reveals very similar sound wave propagation patterns.

Regarding the porous surface category, the results from the two surfaces are basically the same, except a minor difference can be observed in the high-frequency range, where the finer porous surface exhibits a slightly lower amplitude. This indicates that refining the sampling surface to match the solution points is not unnecessary.  Both porous surfaces show high-frequency decay in Fig.~\ref{fig:Kevlar_porous} due to the limited spatial grid resolution. Moreover, the results from the porous surfaces show a higher amplitude in the mid-low frequency range compared to the JAXA Kevlar wall experiment.

\begin{figure}[hbt!]
    \centering
    \includegraphics[width=.49\textwidth]{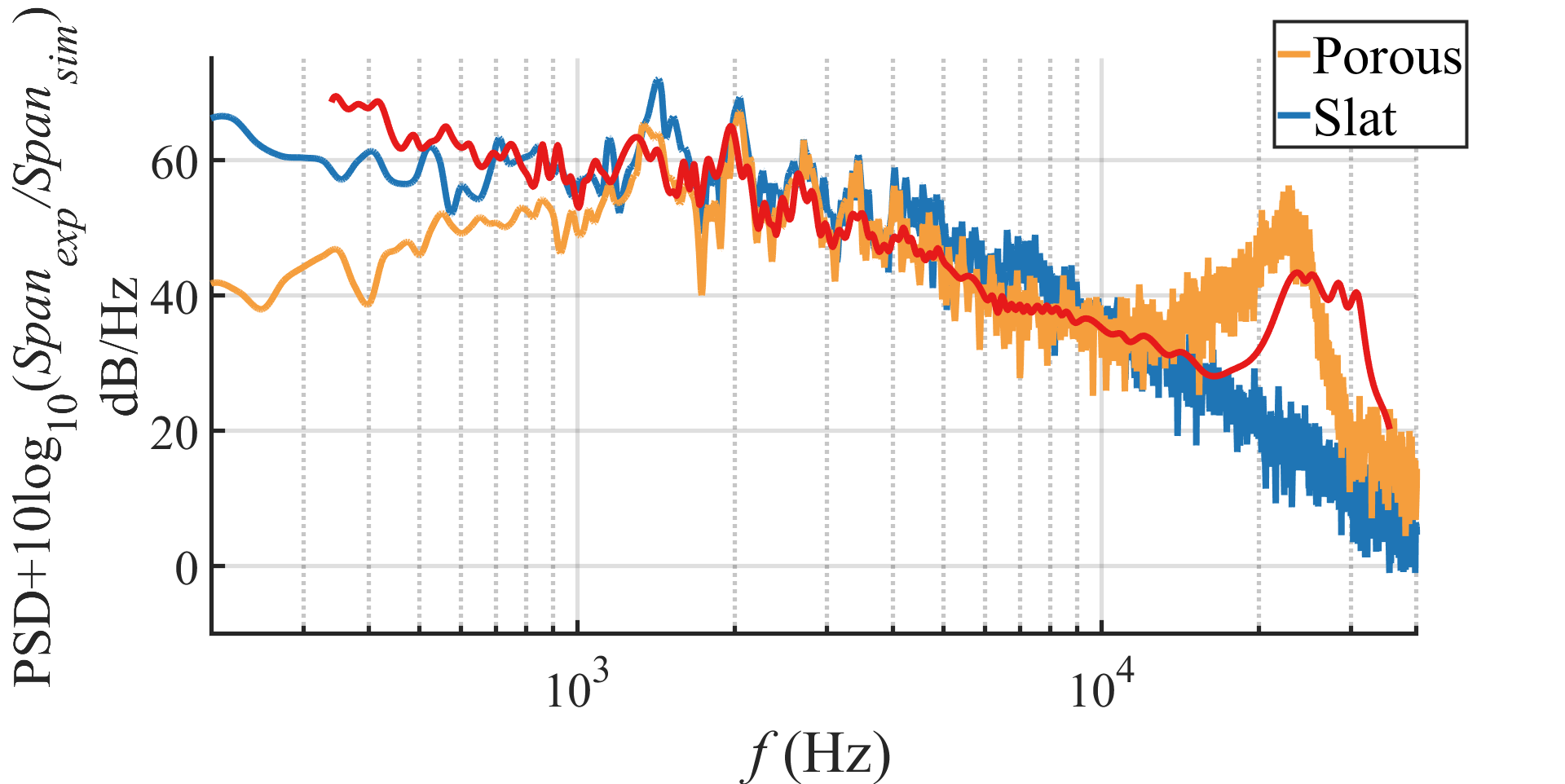}
    \caption{\label{fig:FWH_surfaces-Hard} Acoustic signal results at the center of the JAXA hard wall experiment phased-microphone array \cite{murayama2014experimental} through coaser porous and slat surface at $5.5^\circ$ AoA and 0.17 Ma.}
\end{figure}

When comparing the porous surface result to the JAXA hard experiment \cite{murayama2014experimental}, a much better agreement is achieved as presented in Fig.~\ref{fig:FWH_surfaces-Hard}. In the frequency ranges below 1 kHz, the porous surface result aligns more closely with the JAXA hard wall experiment \cite{murayama2014experimental} than the slat surface. However, in the frequency range above 1 kHz, where tonal noises and hump are present, the slat surface performs better. Additionally, the tonal noise amplitudes from both surfaces are more consistent with the hard wall experiment compared to the Kevlar wall experiment, which may be attributed to the Kevlar wall attenuated the high-frequency tonal noise amplitudes.

In general, using the slat surface as the FW-H sampling surface can prevent the pressure fluctuations caused by the main wing and flap upper sueface transition from polluting far-field noise results. Meanwhile, it shows better agreement with the results of both JAXA wind tunnel experiments in the tonal noise frequency range and demonstrates excellent consistency with the Kevlar wall experiment across the entire noise spectrum. Consequently, all subsequent far-field noise results will utilize the slat surface as the FW-H sampling surface.

\subsection{\label{sec:MOIS}Mesh and Order Independence Study}
Three meshes with four different DoFs are employed to verify the convergence of numerical solution at $5.5^\circ$ AoA and 0.17 Ma. which is the flow field condition discussed in BANC-III \cite{choudhari2015assessment}. Additionally, the spanwise length for these meshes are set to $s_z=50.8mm$, as recommended by the workshop.

As listed in Table~\ref{tab:GIVandO} the maximum DoF used is 30 million for the Fine mesh with K3 simulation, followed by the Medium mesh with K3 simulation and Coarse mesh with K4 simulation, both of which have approximately 15 million DoF. The aerodynamic force results show that the Medium\_K3 and Coarse\_K4 cases, exhibit errors of approximately 2\% and 3\% in comparison to the Fine\_K3 results for $C_L$ and $C_D$, respectively.
\begin{table}[hbt!]
    \caption{\label{tab:GIVandO} Computational cases and aerodynamic results for the mesh and order independence study}
    \centering
    \begin{tabular}{lccccccc}
    \hline
    Cases      &N      & DoF/million& GPU Hours for 10 FPTs& $C_L$& $Ratio_{\Delta C_L}$& $C_D$  & $Ratio_{\Delta C_D}$\\\hline
    Fine\_K3   &471,562& 30.2       & 83.62                & 2.865& $\diagdown$         & 4.65e-2& $\diagdown$ \\
    Medium\_K3 &235,750& 15.1       & 36.26                & 2.812& -1.85\%             & 4.77e-2& 2.74\%      \\
    Coarse\_K3 &117,929& 7.5        & 21.11                & 2.760& -3.68\%             & 4.86e-2& 4.57\%      \\
    Coarse\_K4 &117,929& 14.7       & 143.26               & 2.833& -1.11\%             & 4.50e-2& -3.11\%     \\
    \hline
\end{tabular}
\end{table}

\begin{figure}[hbt!]
    \centering
    \captionsetup{justification=raggedright, singlelinecheck=false}
    \subcaptionbox{\label{fig:Cp_overview_mesh}All three elements}{
    \includegraphics[width = 0.48\textwidth]{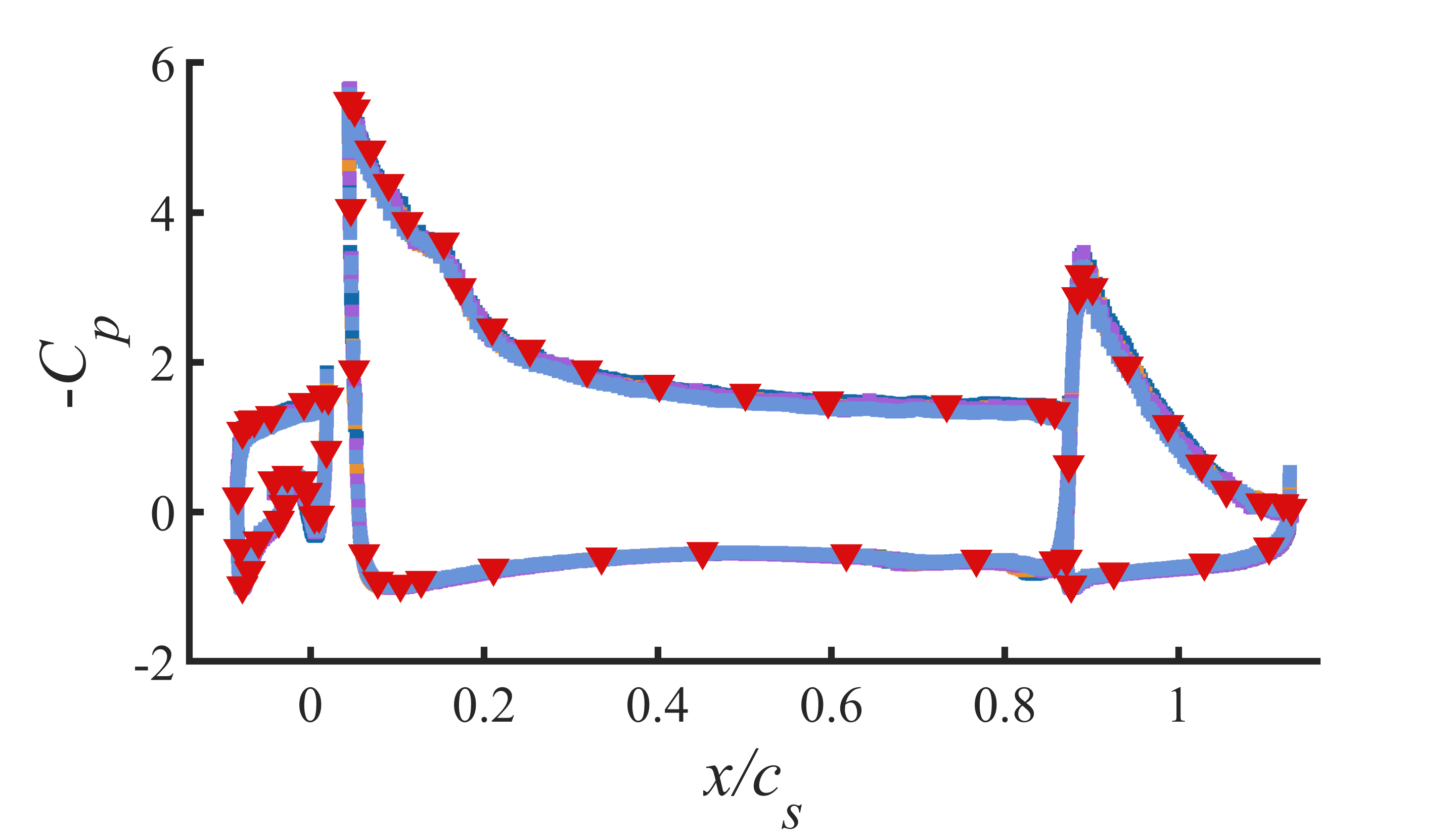}}
    \quad
    \subcaptionbox{\label{fig:Cp_slat_mesh}Around slat}{\includegraphics[width = 0.48\textwidth]{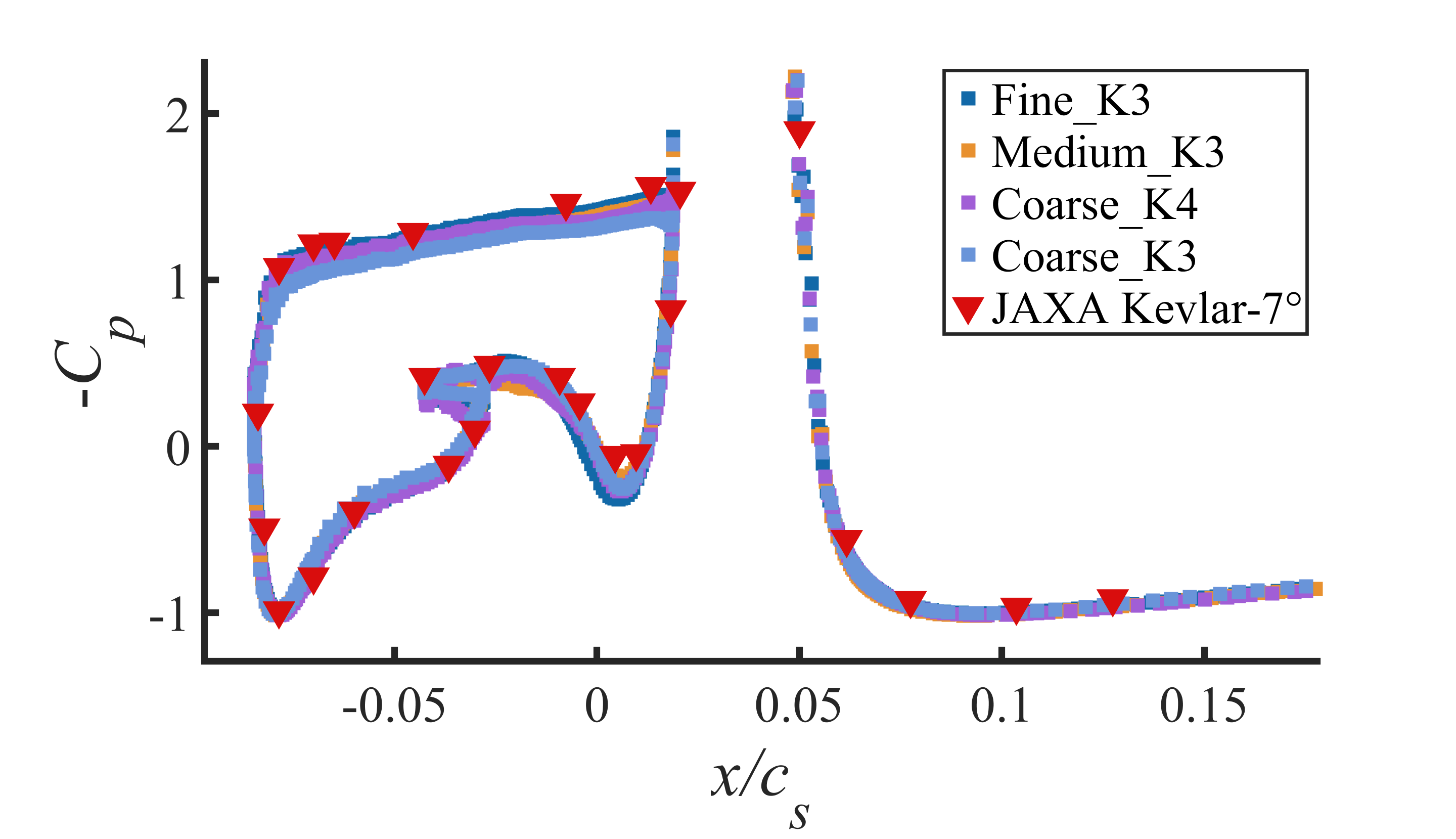}}
    \caption{\label{fig:Cp_mesh} Time-averaged $C_p$ distribution on the airfoil midplane compared with the JAXA Kevlar wall experiment \cite{murayama2018experimental} at $5.5^\circ$ AoA and 0.17 Ma.}
\end{figure}

Compared the pressure coefficient ($C_p$) distribution on the airfoil midplane with the JAXA Kevlar wall experiment \cite{murayama2018experimental} in Fig.~\ref{fig:Cp_mesh}. It can be seen that all results collapse well together with the experimental data, except for the Coarse\_K3 case, which exhibits a slightly lower negative pressure on all upper surfaces of the airfoil elements. This discrepancy accounts for the lowest $C_L$ for the Coarse\_K3 case. Additionally, Fig.~\ref{fig:Cp_mesh} demonstrates that by simply increasing the order of the simulation can greatly enhance the simulation resolution. However, the price is about 4 times the computational expense compared to the K3 simulation with a similar DoF, as detailed in Table~\ref{tab:GIVandO}. This increase in cost is due to the CFL being twice as small, as mentioned in Sec.~\ref{subsec:SPD}, along with the higher computational complexity at each solution point.

\begin{figure}[hbt!]
    \centering
    \includegraphics[width=.49\textwidth]{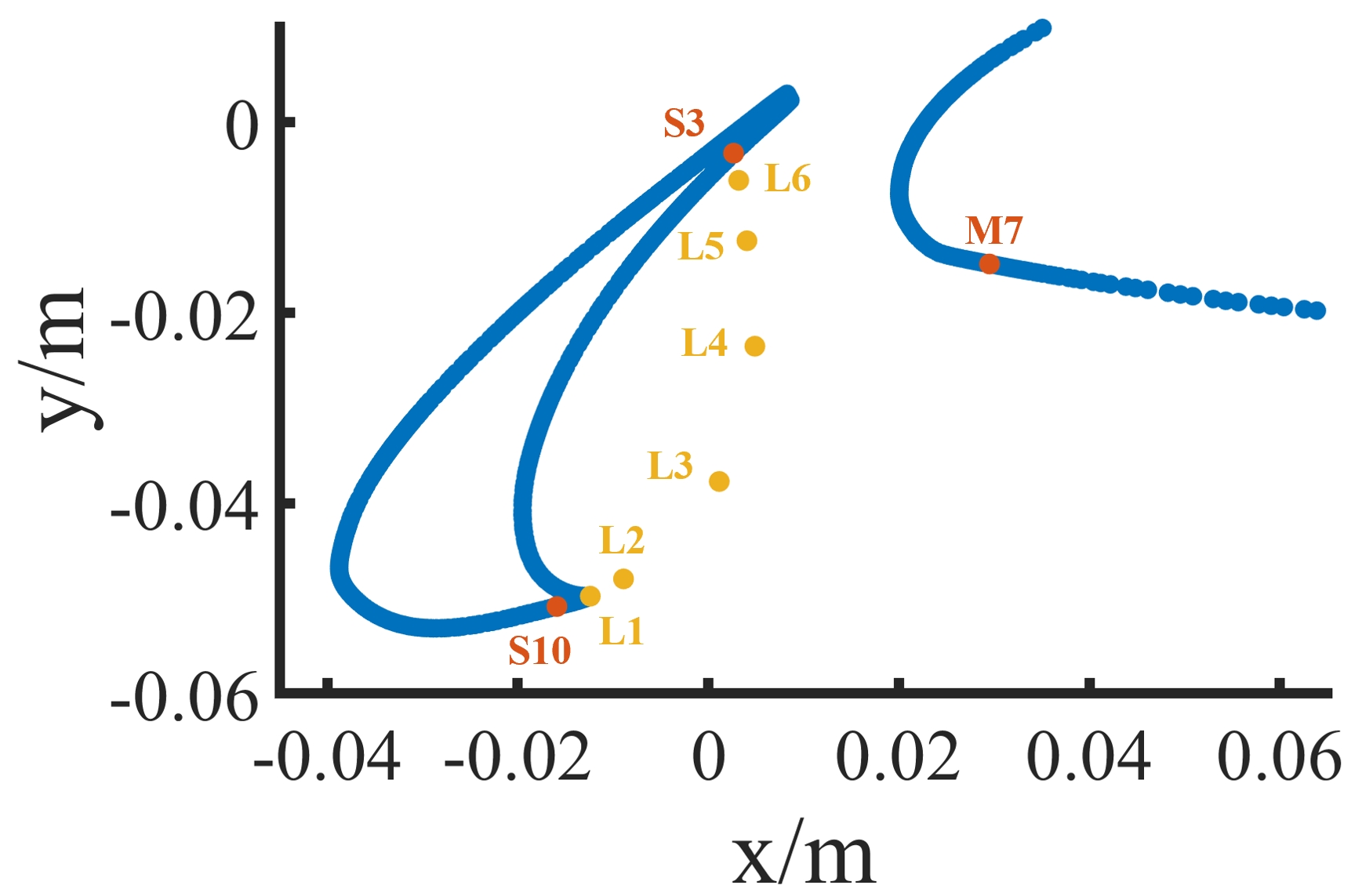}
    \caption{\label{fig:probe} Location of numerical probes in midplane, and S3, S10, M7 are defined in the JAXA experiment \cite{murayama2018experimental}.}
\end{figure}

\begin{figure}[hbt!]
    \centering
    \includegraphics[width=.49\textwidth]{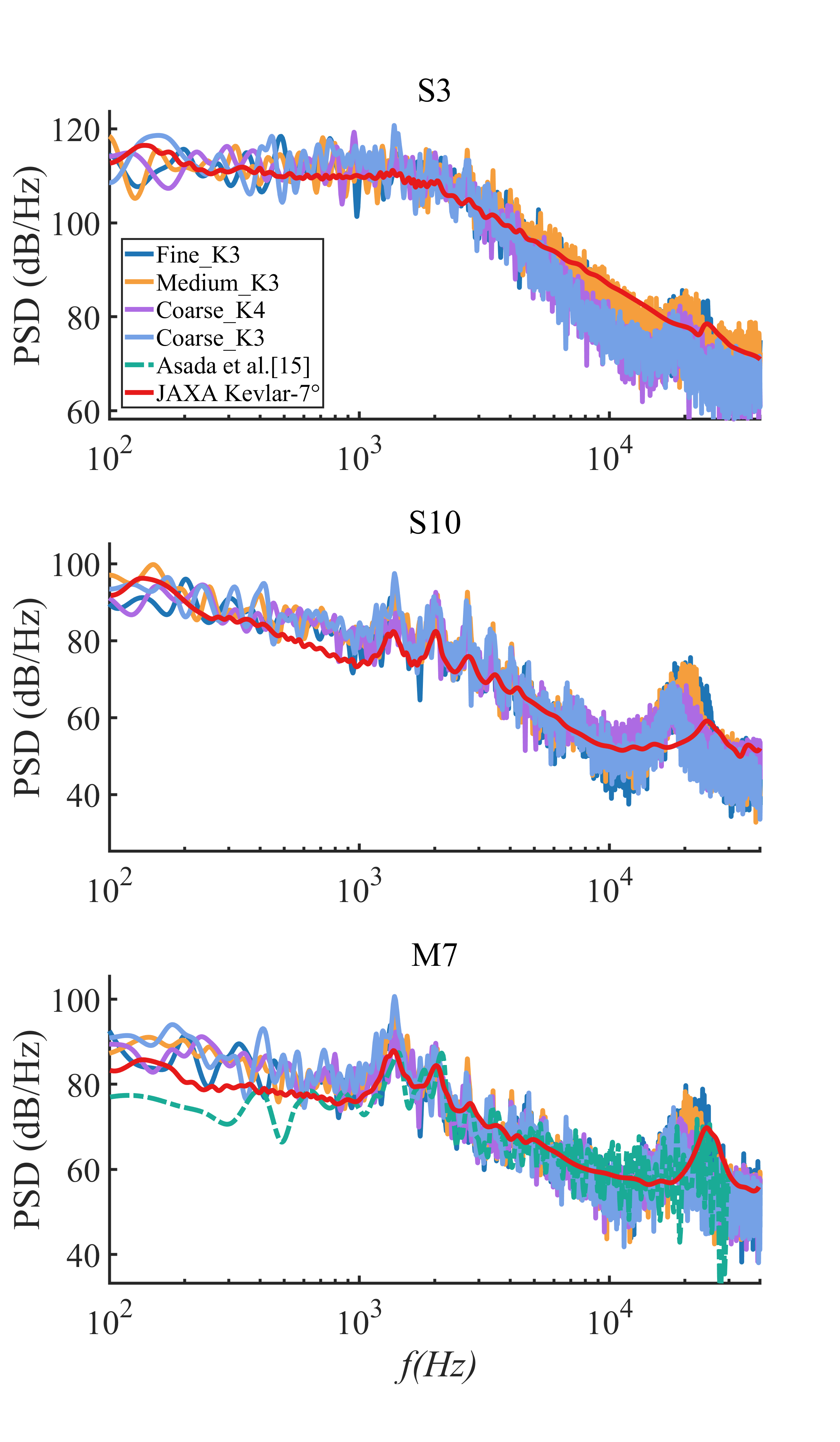}
    \caption{\label{fig:nf_mesh} Wall pressure spectra compare to the JAXA Kevlar wall experiment \cite{murayama2018experimental} and the WMLES solution of \citet{asada2024aeroacoustics} at $5.5^\circ$ AoA and 0.17 Ma.}
\end{figure}

Figure~\ref{fig:probe} depicted the numerical probes utilized in this study. The near-field wall pressure spectra at location S3, S10,and M7 are compared to experimental data in Fig.~\ref{fig:nf_mesh}. At probe S3, where the shear layer from the slat cusp impacts the slat surface at $5.5^\circ$ AoA, a broadband spectrum is obtained. All cases demonstrate good consistency with the experimental result in the low-mid frequency range, but only the Fine\_K3 and Medium\_K3 spectral keep up with experiment till 10 kHz, even the K4 simulation fails to improve the result of the Coarse mesh. At probe S10, in terms of the tonal noise, all cases show good consistency with the experiment, except for a relatively stronger amplitude in the higher frequency tonal noise. Furthermore, as the mesh is refined, the frequency of the hump shifts closer to the experimental results. At probe M7, result from  \citet{asada2024aeroacoustics} using 3.5 billion cells with WMLES is included, showing that the high-order results more accurately capture the two strongest tonal noise frequencies while requiring significantly less computational resources. Additionally, compared to the finer results, the Coarse\_K3 shows a higher amplitude for the tonal noise around 1.3 kHz.

\begin{figure}[hbt!]
    \centering
    \includegraphics[width=.49\textwidth]{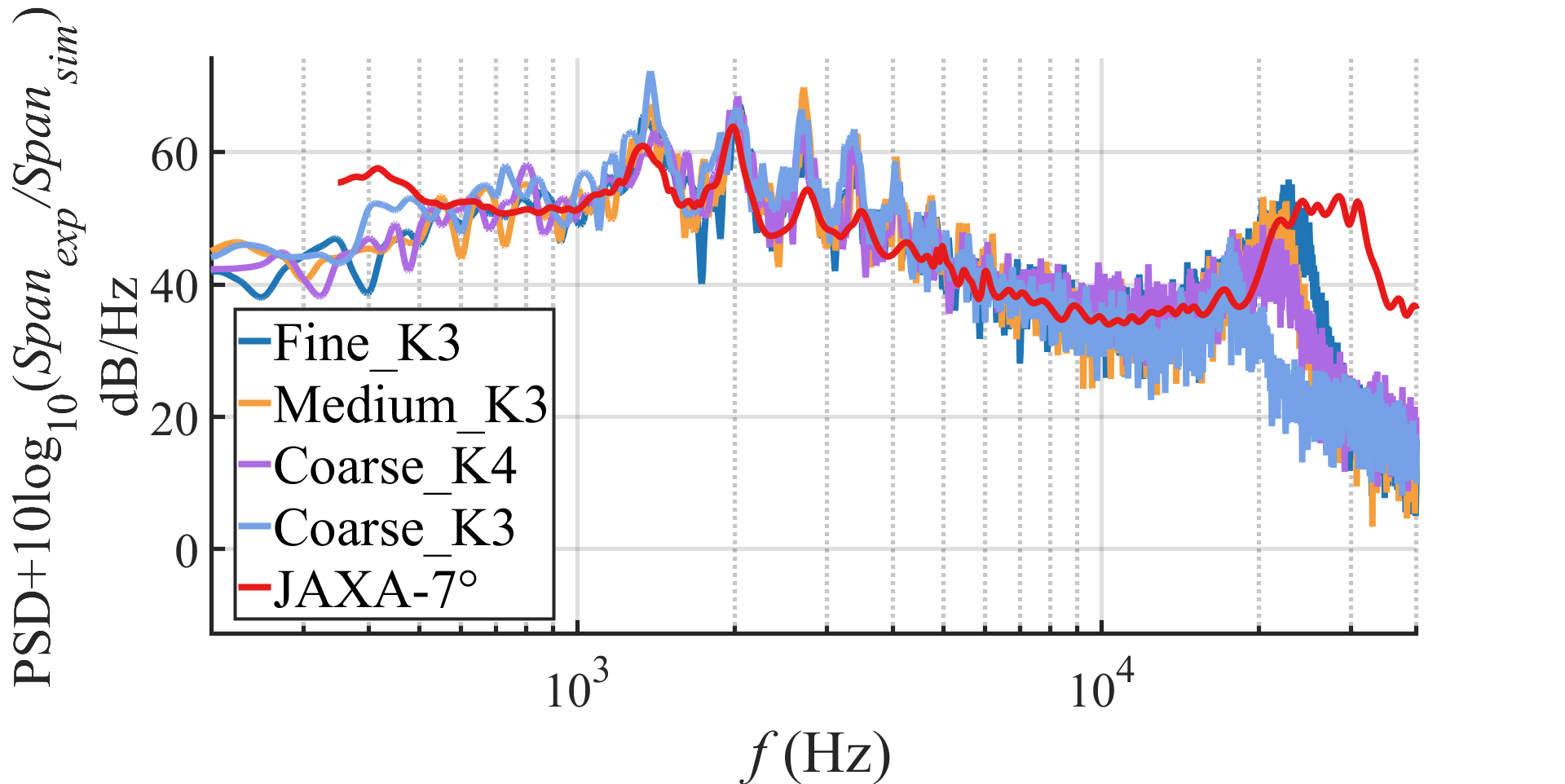}
    \caption{\label{fig:Mesh_ff} Far-field noise spectra compared to the JAXA Kevlar wall experiment \cite{murayama2018experimental} at $5.5^\circ$ AoA and 0.17 Ma.}
\end{figure}

The far-file noise results are plotted in Fig.~\ref{fig:Mesh_ff}. All simulation spectra align closely with the JAXA Kevlar wall experiment for frequency below 10 kHz, other than the Coarse\_K3 result, which shows a relatively higher peak at the tonal noise around 1.3 kHz.  Again, as the mesh is refined, the frequency of the hump shifts closer to the experimental results, and the amplitude also shifts in that direction in contrast to the probe S10 near-field spectra.

\begin{figure}[hbt!]
    \centering
    \includegraphics[width=.48\textwidth]{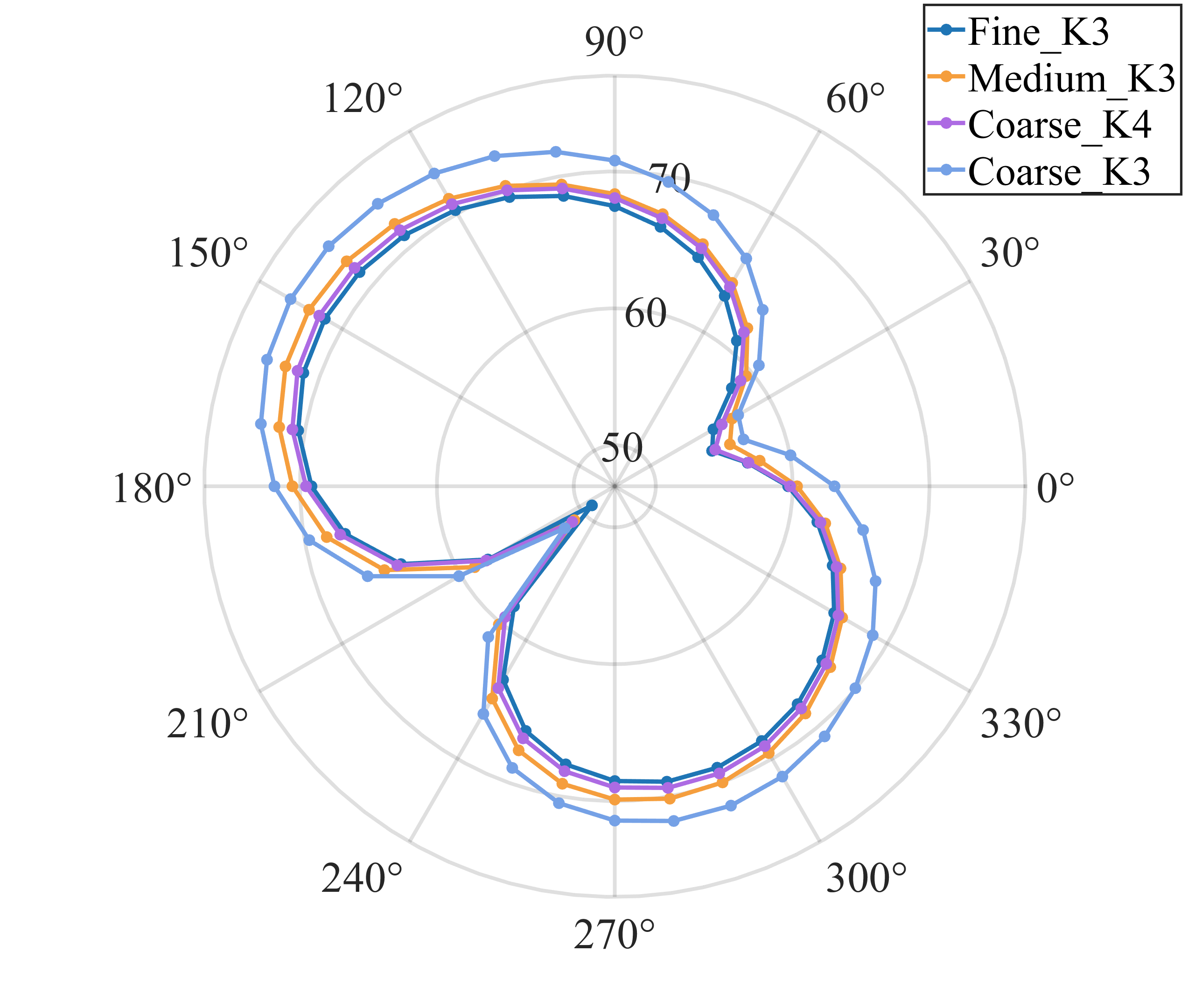}
    \caption{\label{fig:Mesh_OASPL} Far-field noise directivity patterns under $5.5^\circ$ AoA and 0.17 Ma at a distance of $10c_s$ in dB.}
\end{figure}

At the center of the 30P30N airfoil, taking the inflow direction is $0^\circ$ and the direction below the airfoil is $270^\circ$. Receiving points are arranged every $10^\circ$ around a circumference with a radius of $10c_s$, resulting in a total of 36 receiving points. The far-field overall sound pressure level (OASPL) integrated within 10 kHz (to remove the contribution from the vortex shedding at the slat TE) is plotted in Fig.~\ref{fig:Mesh_OASPL}. A typical dipole pattern is acquired for all cases, with the dipole axis aligned with the slat chord. It can be observed that, apart from the Coarse\_K3 result, all other results overlap well together.

\begin{figure}[hbt!]
    \centering
        \captionsetup{justification=raggedright, singlelinecheck=false}
        \subcaptionbox{\label{fig:Q_Fine_K3}Fine\_K3}{
        \includegraphics[width = 0.45\textwidth]{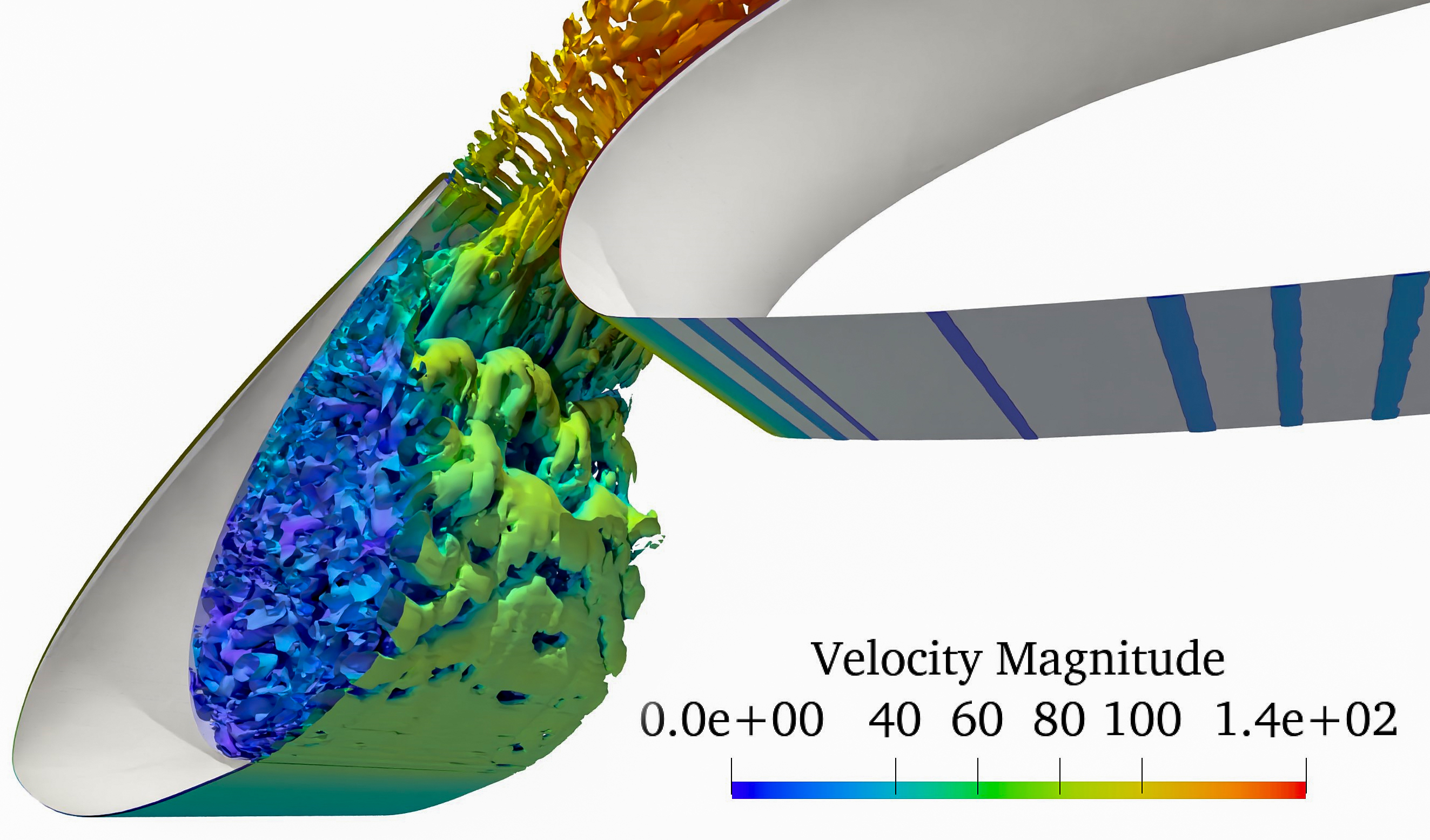}}
        \quad
        \subcaptionbox{\label{fig:Q_Medium_K3}Medium\_K3}{\includegraphics[width = 0.45\textwidth]{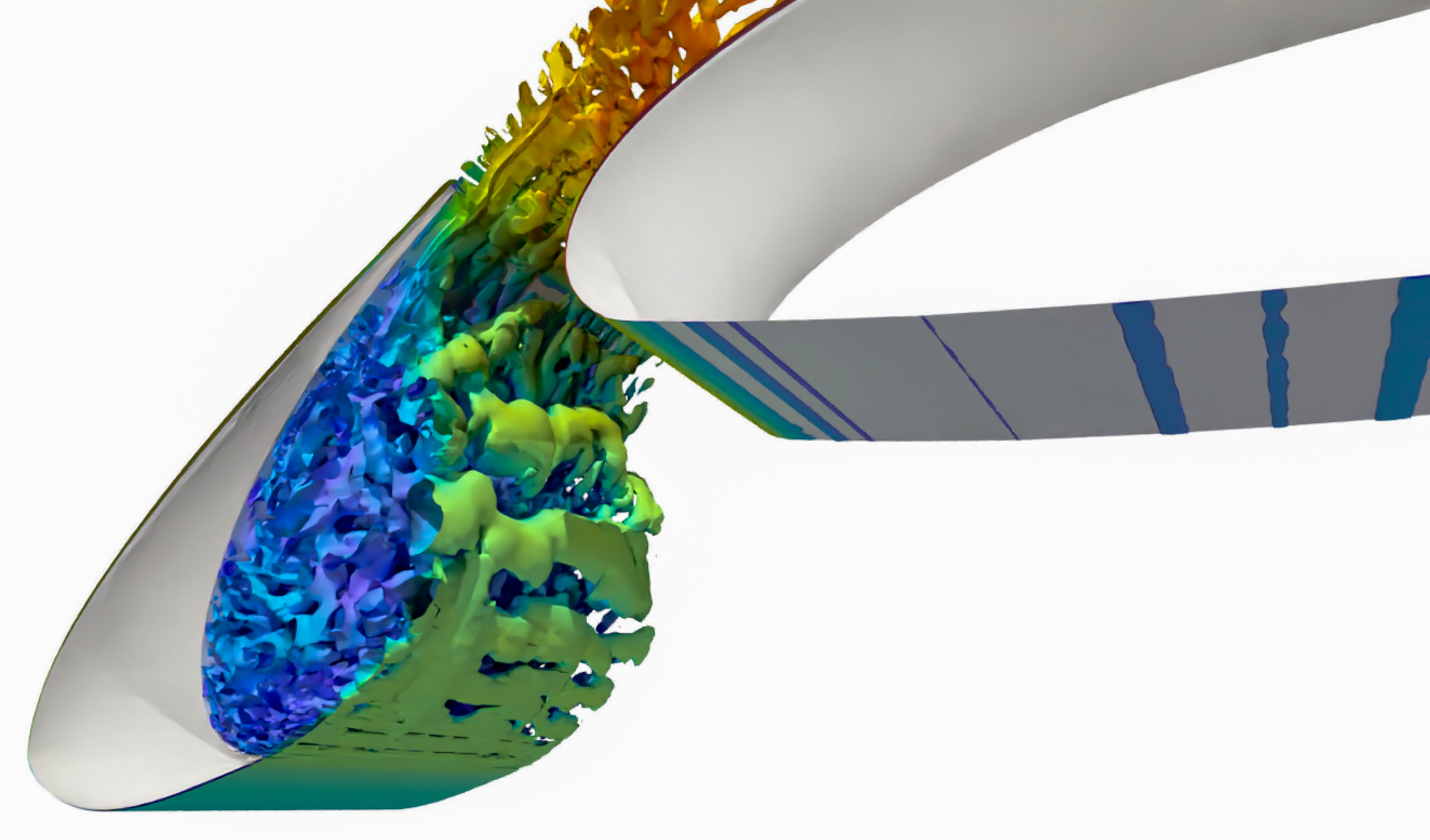}}
        \quad
        \subcaptionbox{\label{fig:Q_Coarse_K4}Coarse\_K4}{\includegraphics[width = 0.45\textwidth]{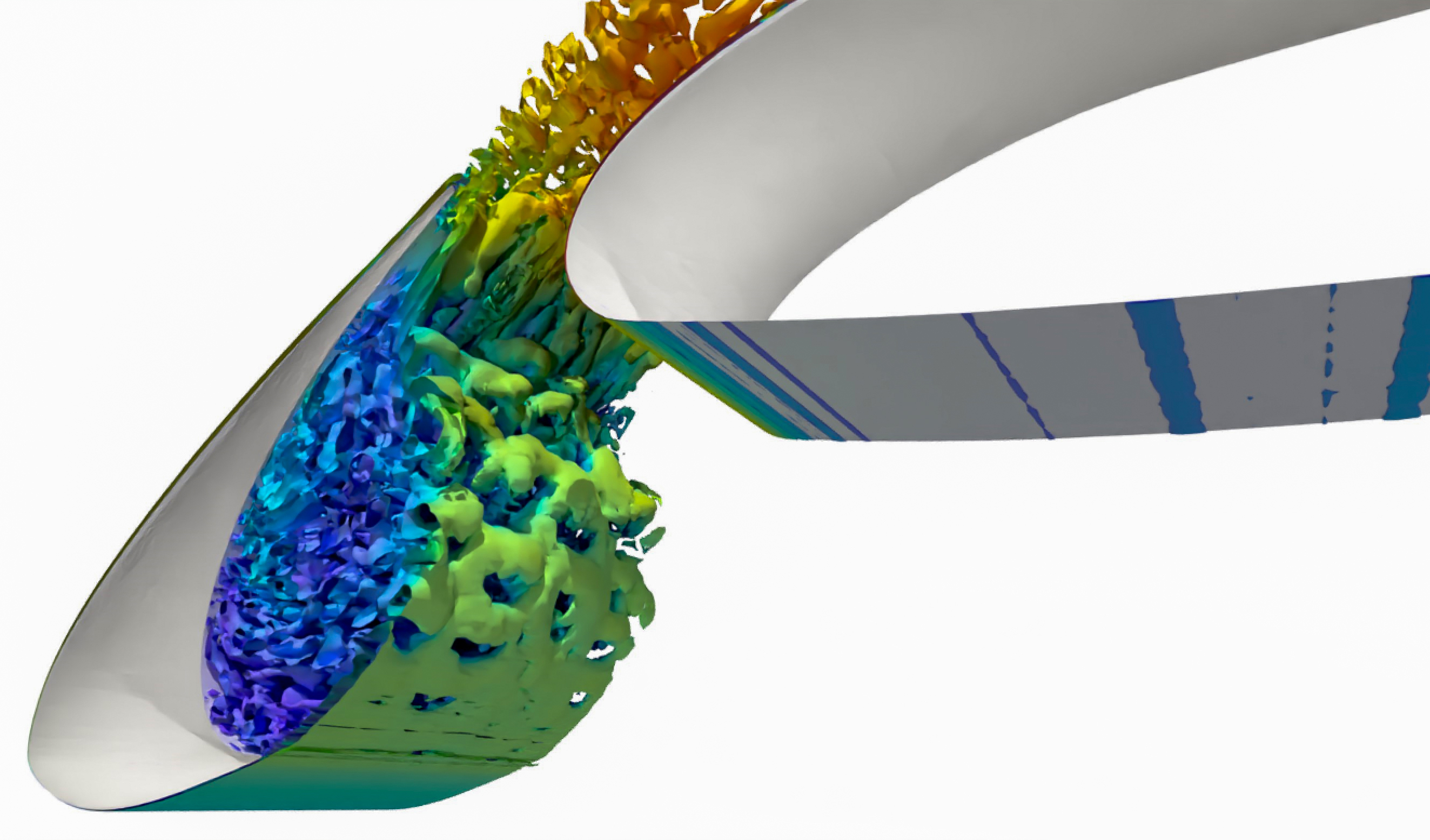}} 
        \quad
        \subcaptionbox{\label{fig:Q_Coarse_K3}Coarse\_K3}{\includegraphics[width = 0.45\textwidth]{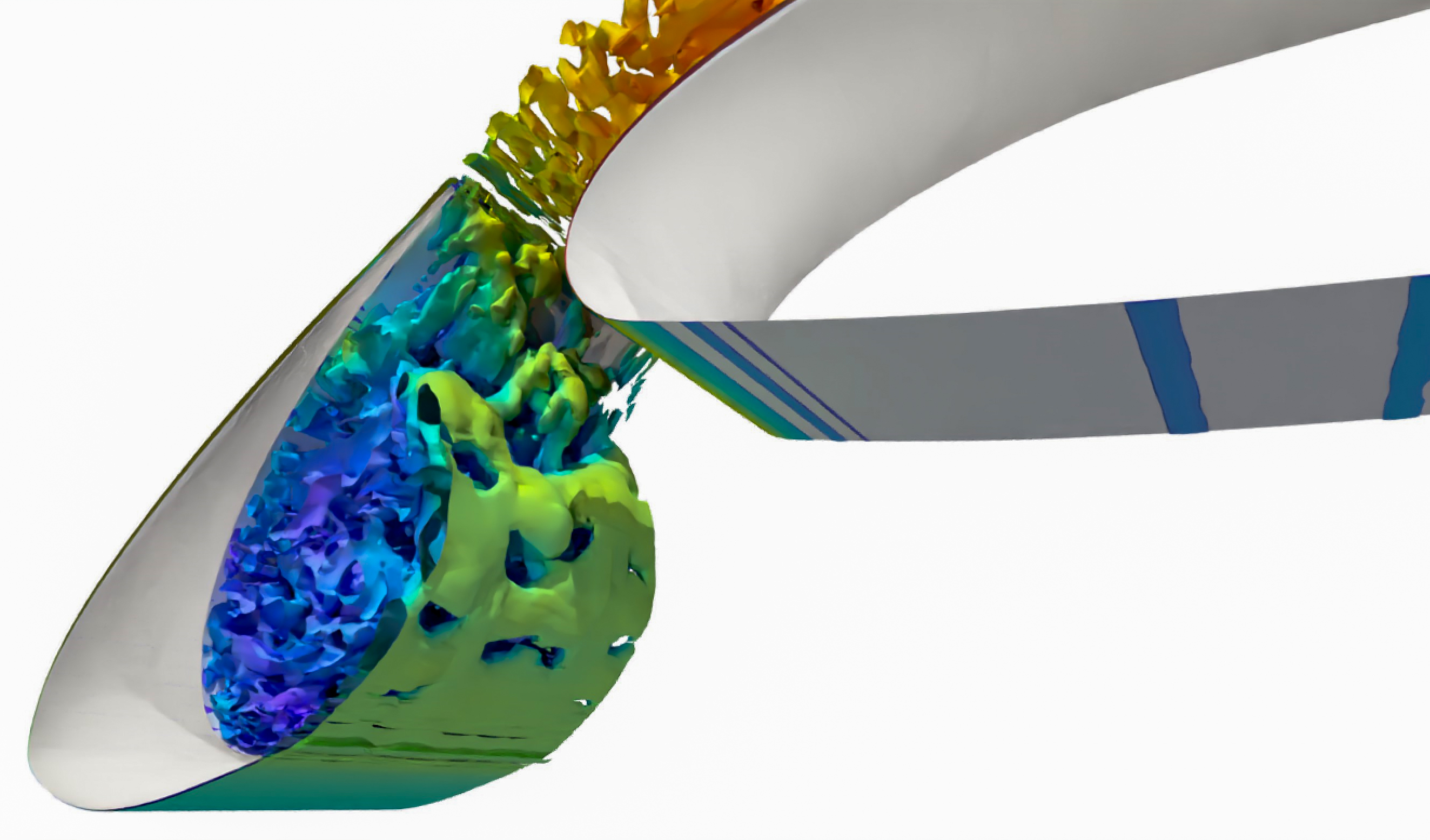}} 
    \caption{\label{fig:Q_Mesh} Instantaneous iso-surface of $Q=2e6$ in the slat cove region colored by velocity magnitude at $5.5^\circ$ AoA and 0.17 Ma.}
\end{figure}

Instantaneous iso-surface of the $Q$ criterion in the slat cove area is shown in Fig.~\ref{fig:Q_Mesh}. It is evident that a shear layer is generated at the slat cusp and developed along the slat cove vortex until the impingement point. A small portion of the large vortices in the shear layer break down into smaller structures and convect inside the slat cove, while large packets of vortices are elongated by the high-speed gap flow and burst toward the slat TE. The smaller the DoF is, the coarser these structures and vortices are resolved, resulting in larger vortices in the shear layer of the Coares\_K3 flow field. This leads to a stronger impact strength, which in turn causes a higher amplitude at the tonal noise around 1.3 kHz, as demonstrated in Figs~.\ref{fig:nf_mesh} and \ref{fig:Mesh_ff}.

Overall, the Medium\_K3 case yields aerodynamic forces, aeroacoustic spectra, and flow field that are in good consistent with the Fine\_K3 case and experimental data. Meanwhile, it costs less than half of the computing resources. Therefore, the Medium mesh and its associated meshes will be employed for K3 simulations in subsequent research.

\subsection{Spanwise Independence Study}
Using the 2D plane grid from the Medium mesh, two associated meshes with different spanwise lengths but identical spanwise resolution have been created, as listed in Table~\ref{tab:meshes}. This setup aims to investigate the influence of spanwise length on the flow field and slat noise. The simulation condition is also $5.5^\circ$ AoA and 0.17 Ma.

Table~\ref{tab:SIV} presents a comparison of the aerodynamic force results with the Fine\_K3 case. All cases have an error within 2\% compared to the Fine\_K3 result for $C_L$. While for $C_D$, the Medium\_25.6\% mesh has the lowest error at 0.11\% compared to Fine\_K3, followed by the Medium\_4.44\% mesh with an error of 1.64\%, and lastly, the Medium\_11.1\% mesh (which corresponds to the Medium\_K3 case discussed in Sec.~\ref{sec:MOIS}) with an error of 2.75\%.
\begin{table}[hbt!]
    \caption{\label{tab:SIV} Computational cases and aerodynamic results of spanwise length study}
    \centering
    \begin{tabular}{lccccccc}
    \hline
    Cases          & DoF/million& GPU Hours for 10 FPTs& $C_L$& $Ratio_{\Delta C_L}$& $C_D$  & $Ratio_{\Delta C_D}$\\\hline
    Medium\_4.44\% & 6.0   & 19.36                & 2.810& -1.85\%             & 4.72e-2& 1.64\%      \\
    Medium\_11.1\% & 15.1  & 36.26                & 2.812& -1.92\%             & 4.77e-2& 2.74\%      \\
    Medium\_25.6\% & 34.7  & 91.98                & 2.809& -1.97\%             & 4.86e-2& 0.11\%      \\
    \hline
\end{tabular}
\end{table}

\begin{figure}[hbt!]
    \centering
        \captionsetup{justification=raggedright, singlelinecheck=false}
        \subcaptionbox{\label{fig:Q_Medium_444_K3}Medium\_4.44\%}{
        \includegraphics[width = 0.45\textwidth]{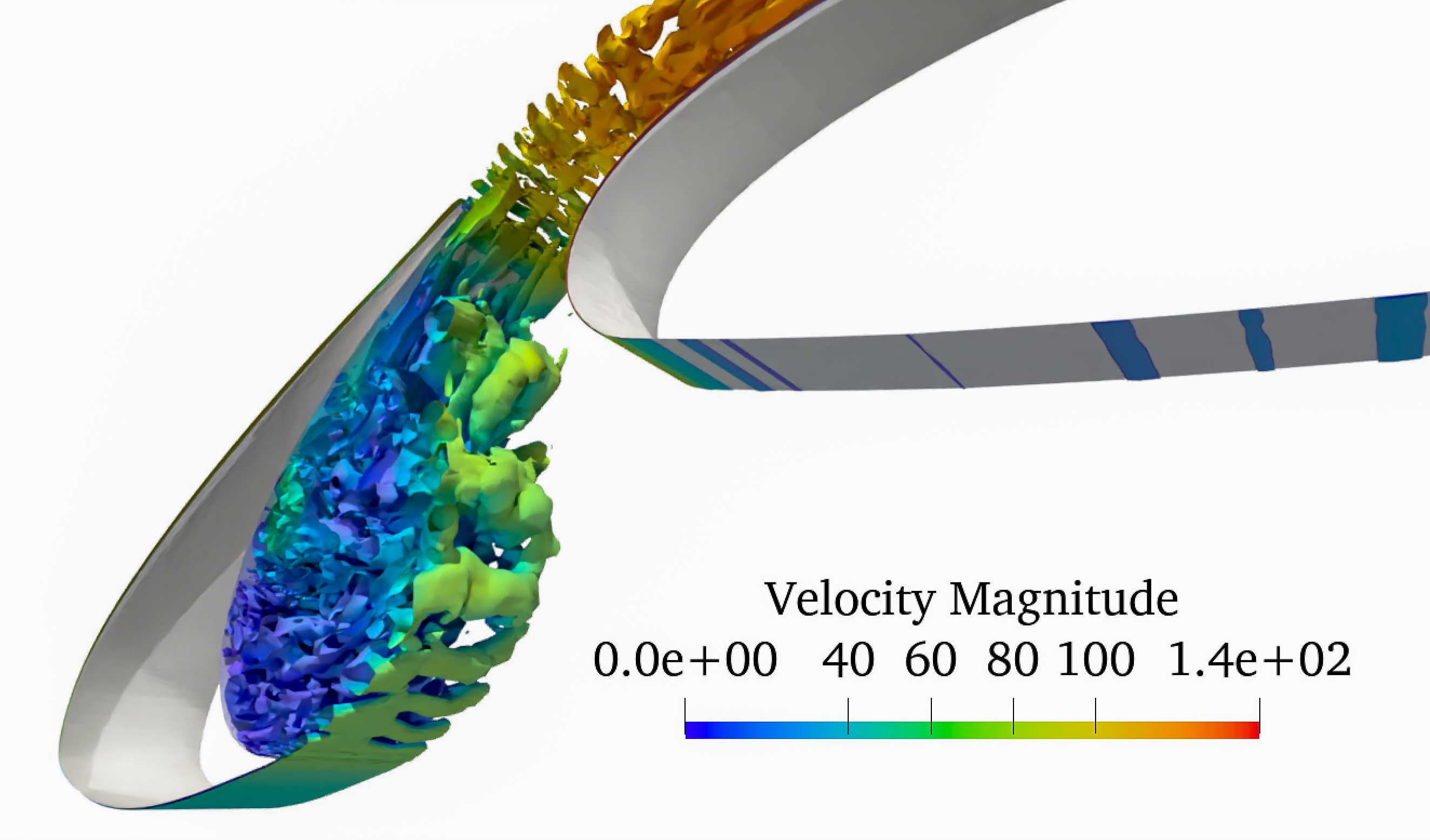}}
        \quad
        \subcaptionbox{\label{fig:Q_Medium_256_K3}Medium\_25.6\%}{\includegraphics[width = 0.45\textwidth]{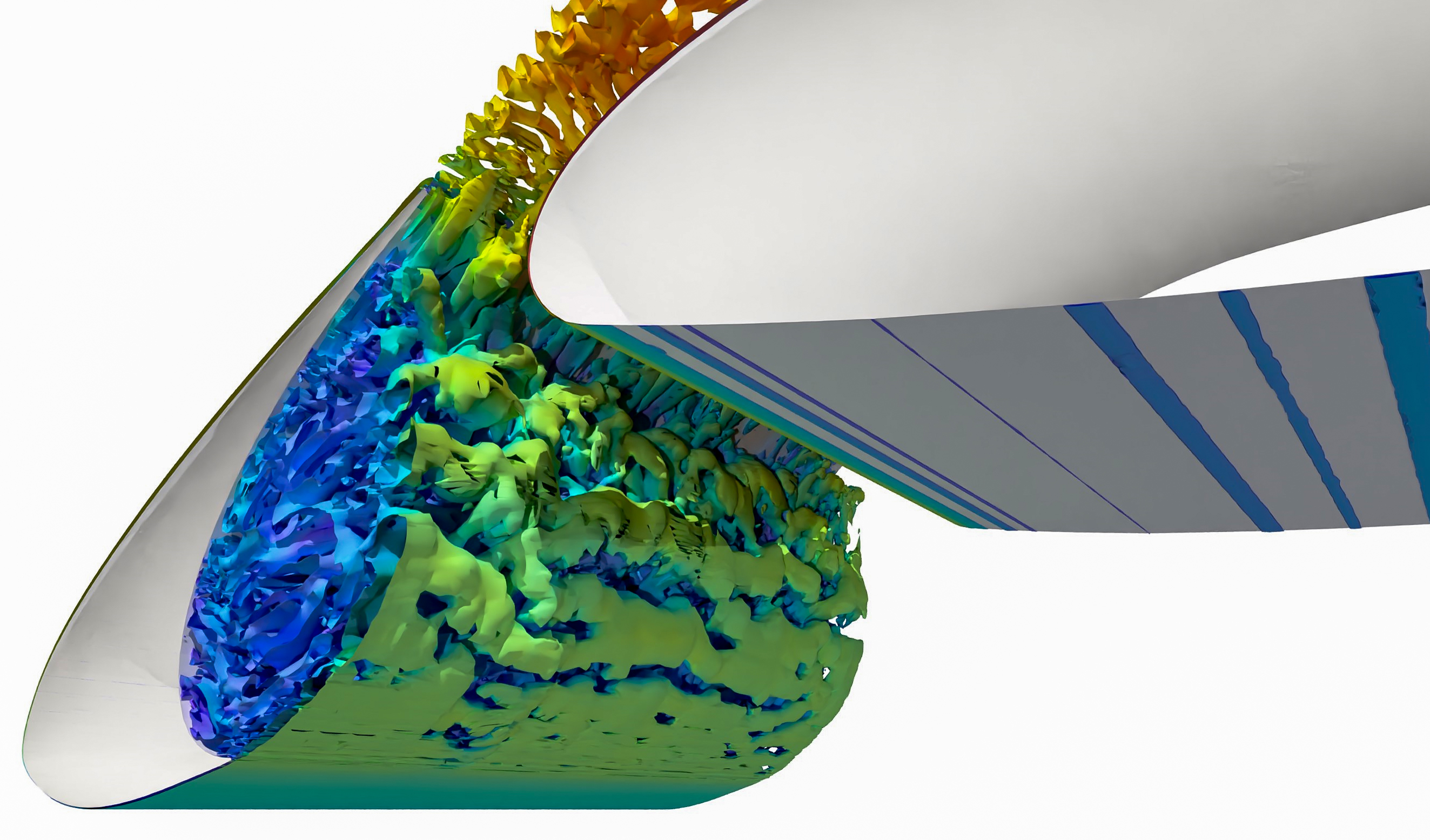}} 
    \caption{\label{fig:Q_Span} Instantaneous iso-surface of $Q=2e6$ in the slat cove colored by velocity magnitude for different spanwise cases at $5.5^\circ$ AoA and 0.17 Ma.}
\end{figure}

Figure~\ref{fig:Q_Span} displays the instantaneous iso-surface of the $Q$ criterion in the slat cove area for the Medium\_4.44\% and Medium\_25.6\% cases. It can be seen that even with a spanwise length of $4.44\%c_s$, the flow field features are well resolved.

\begin{figure}[hbt!]
    \centering
    \includegraphics[width=.9\textwidth]{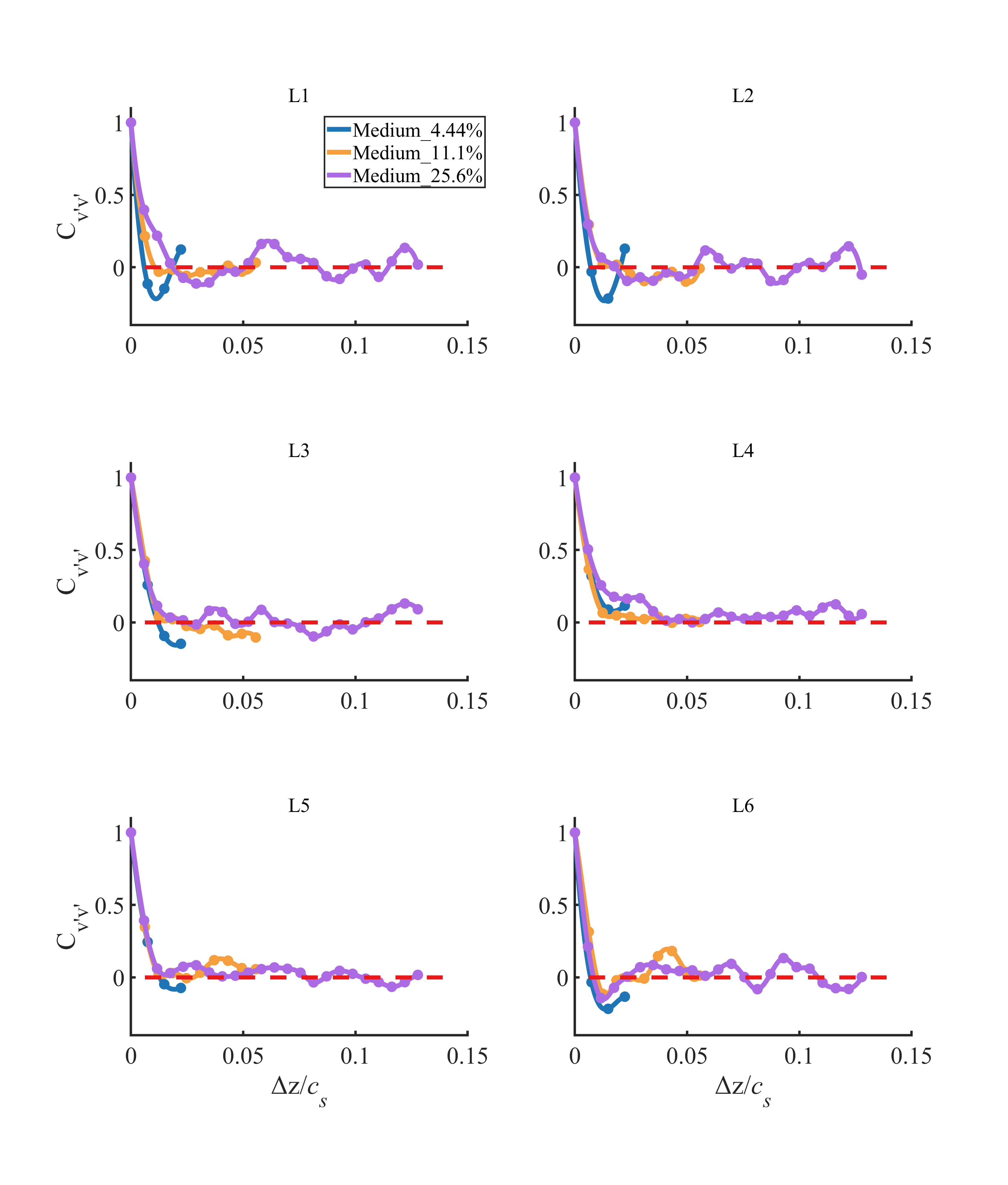}
    \caption{\label{fig:span_vv} Two-point spanwise correlation coefficient of vertical velocity fluctuation along the slat shear layer.}
\end{figure}

To further explore the effect of the spanwise length, the two-point spanwise correlation coefficient for vertical velocity fluctuations along the slat shear layer are plotted in Fig.~\ref{fig:span_vv}. For simplicity, other velocity fluctuation components have the same trend are not shown. It can be observed that there is a rapid decay of correlation across all cases. However, the Medium\_4.44\% case achieves a relatively high negative correlation coefficient at probe L1, L2, and L5, which might imply the influence of the periodic boundary condition in the spanwise direction. In contrast, the Medium\_11.1\% case follow the trend of the Medium\_25.6\% case, reaching a fully spanwise decorrelation. It appears that a spanwise length of $0.1c_s$ ($2/3c_{slat}$) is sufficient for achieving spanwise decorrelation, which is similar with the suggestion of $0.8c_{slat}$ from \citet{terracol2016investigation}.

\begin{figure}[hbt!]
    \centering
    \includegraphics[width=.49\textwidth]{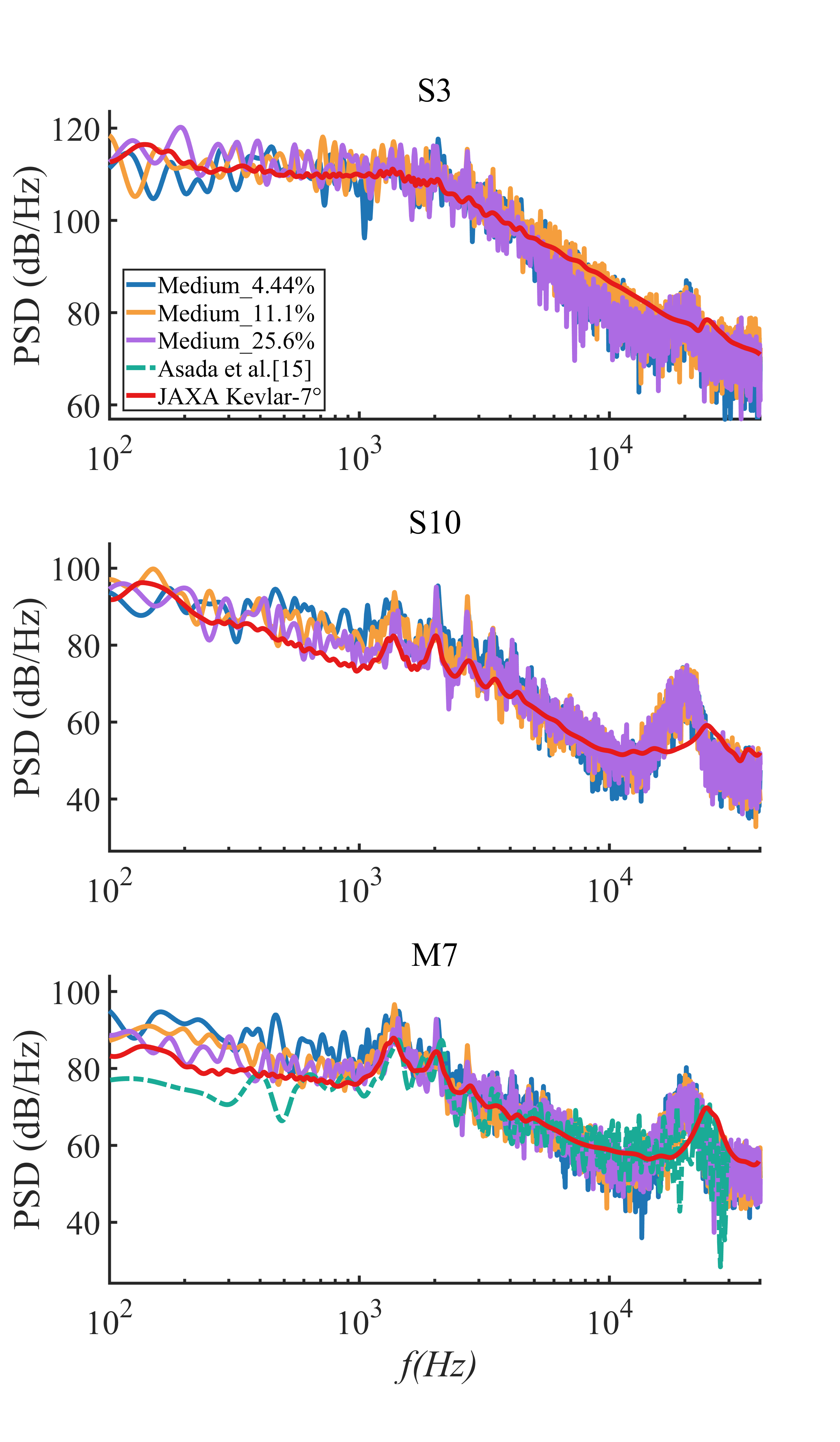}
    \caption{\label{fig:nf_span} Wall pressure spectrum of different spanwise length meshes compare to the JAXA Kevlar wall experiment \cite{murayama2018experimental} and the WMLES solution of \citet{asada2024aeroacoustics} at $5.5^\circ$ AoA and 0.17 Ma.}
\end{figure}

\begin{figure}[hbt!]
    \centering
    \includegraphics[width=.49\textwidth]{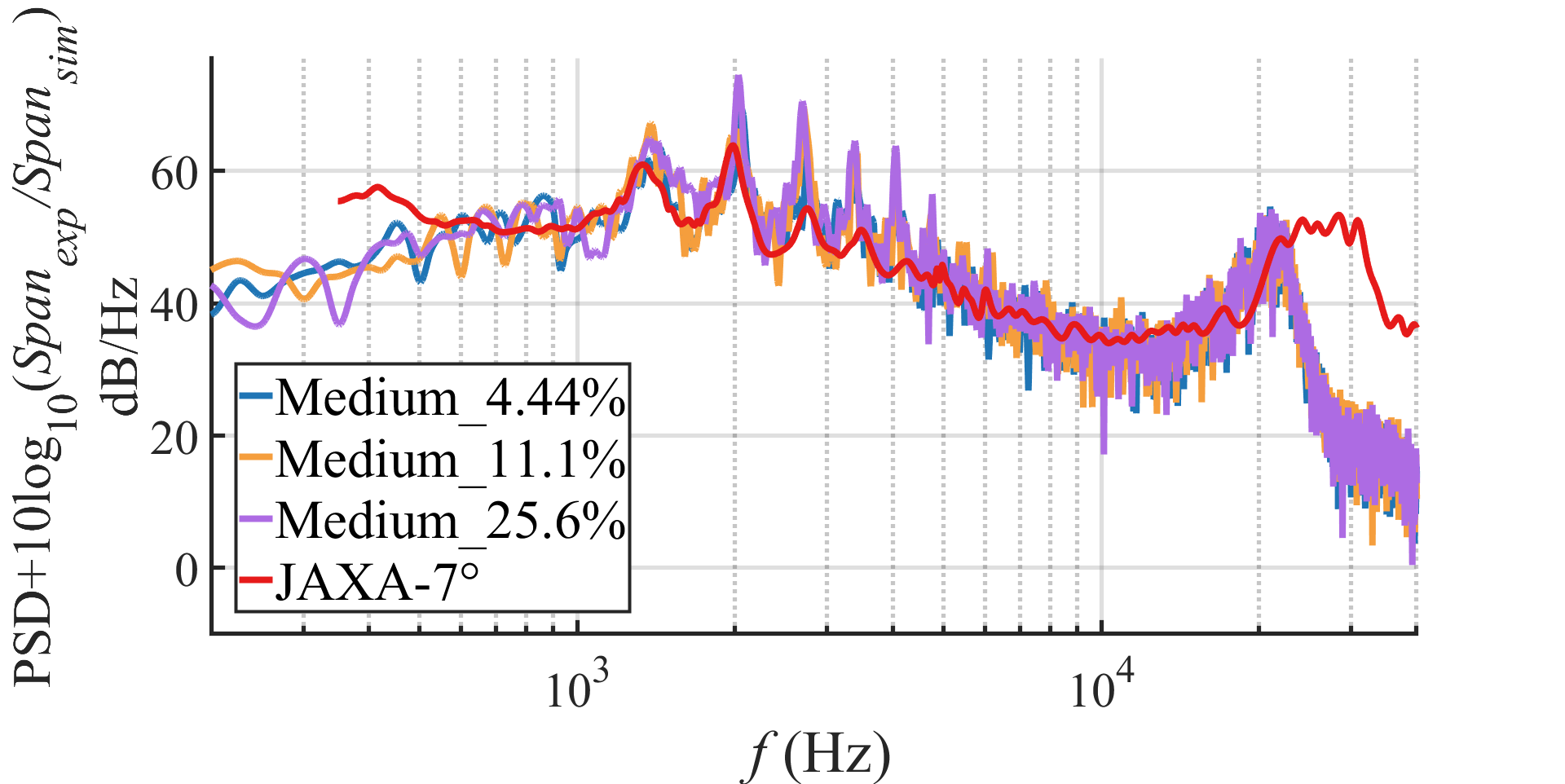}
    \caption{\label{fig:Span_ff} Far-field noise spectrum of different spanwise length meshes compare to the JAXA Kevlar wall experiment \cite{murayama2018experimental} at $5.5^\circ$ AoA and 0.17 Ma.}
\end{figure}

Near-field and far-field noise spectrum of different spanwise lengths are shown in Figs.~\ref{fig:nf_span} and \ref{fig:Span_ff}, respectively. At probe S3, it can be seen that the spanwise length does not affect the low-mid frequency range, and the case with the longest spanwise length exhibits the lowest amplitude in the mid-high frequency range. For probes S10 and M7, all cases effectively capture the tonal noise frequencies. In terms of the low frequency broadband noise, the shortest spanwise case has a slightly higher amplitude compared to the other two. In terms of far-field noise, all three cases display similar spectra, but the shortest spanwise length case yields the best tonal noise amplitude compared to the experimental result.

\begin{figure}[hbt!]
    \centering
        \captionsetup{justification=raggedright, singlelinecheck=false}
        \subcaptionbox{\label{fig:Span_OASPL_10}Spanwise correction of $10log{SD}$}{
        \includegraphics[width = 0.5\textwidth]{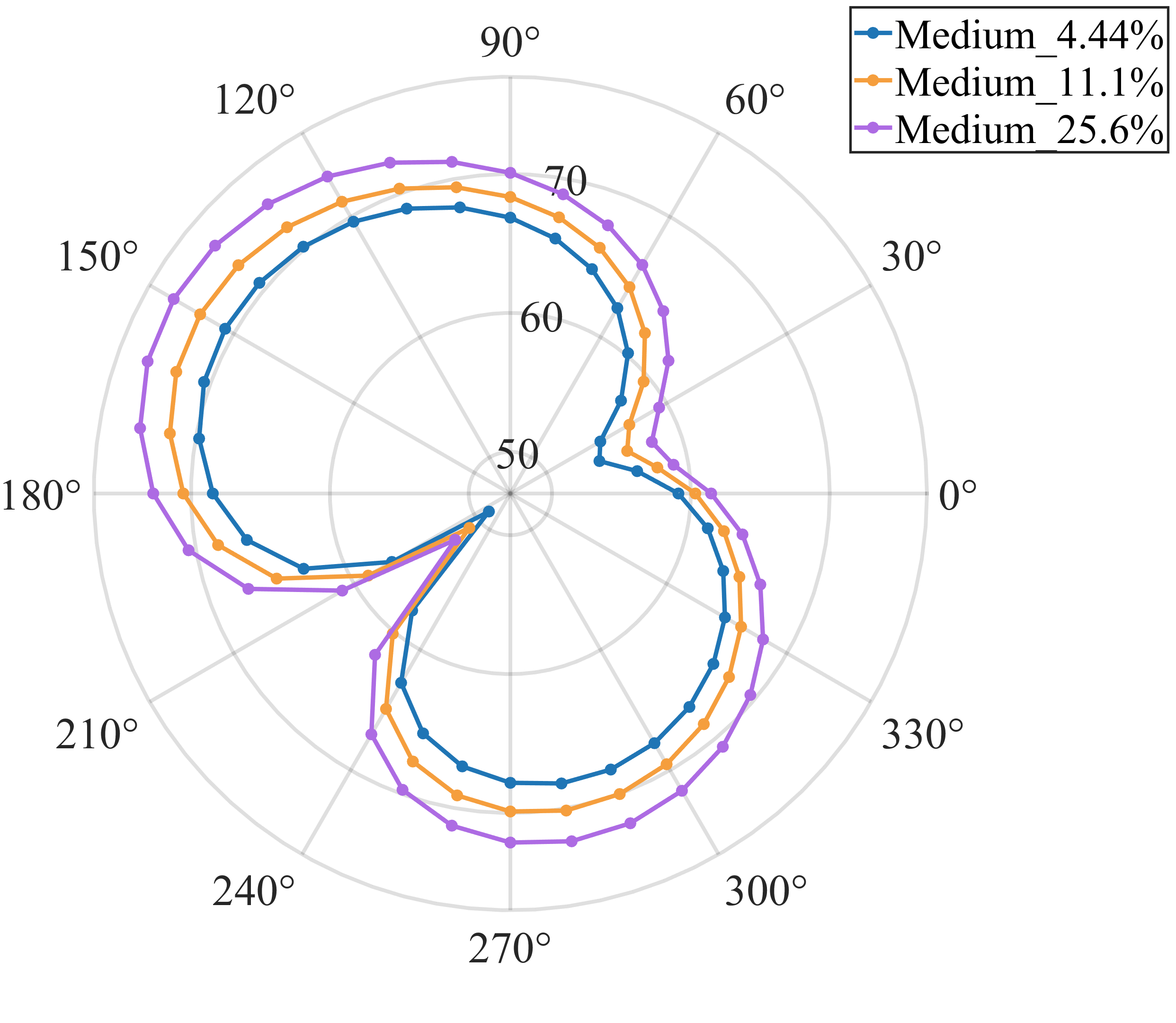}}
        \quad
        \subcaptionbox{\label{fig:Span_OASPL_15}Spanwise correction of $15log{SD}$}{\includegraphics[width = 0.41\textwidth]{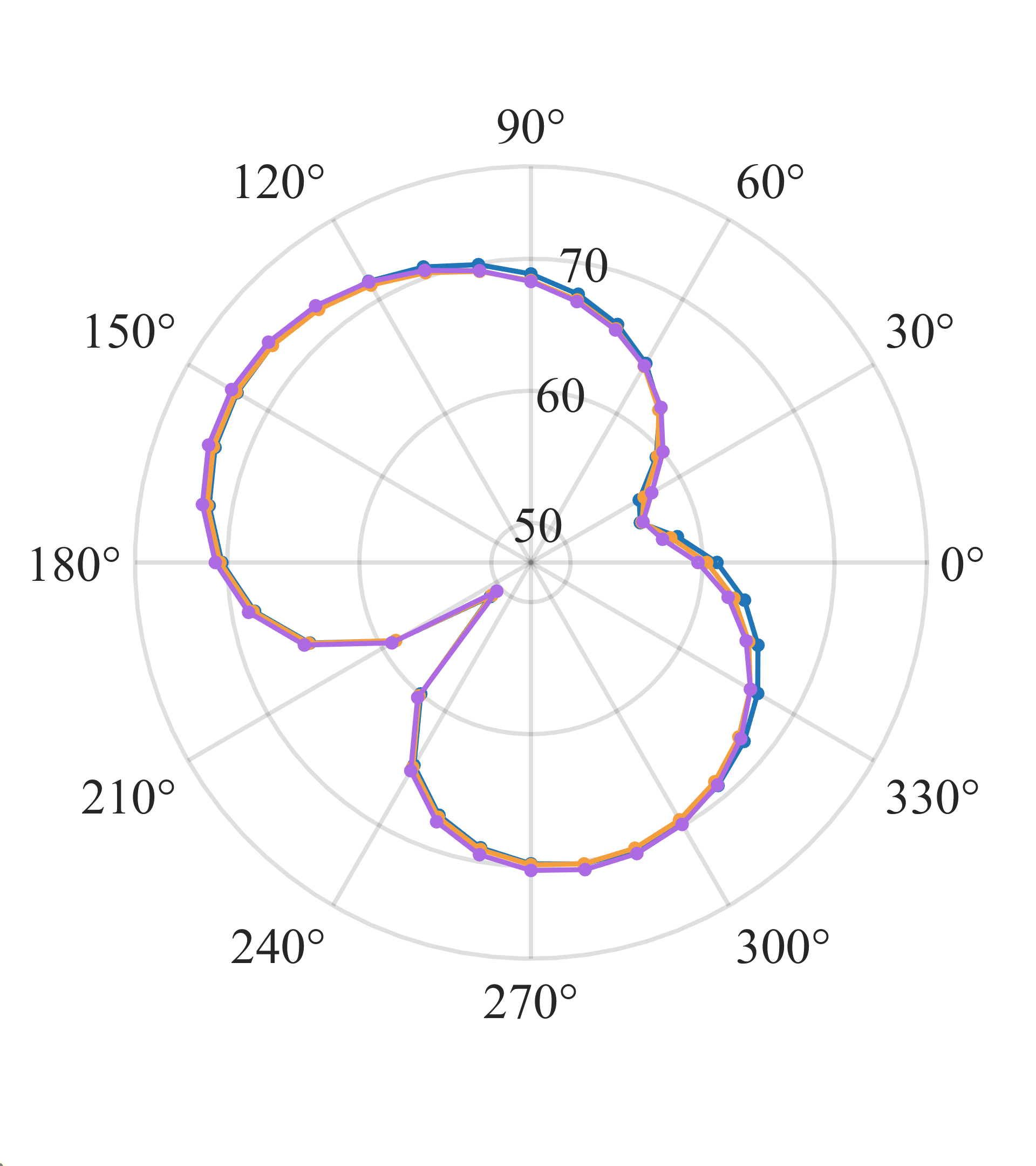}} 
    \caption{\label{fig:Span_OASPL} Far-field noise directivity patterns under $5.5^\circ$ AoA and 0.17 Ma at a distance of $10c_s$ away with different spanwise correction in dB.}
\end{figure}

Figure~\ref{fig:Span_OASPL} presents the far-field noise directivity patterns with two different spanwise corrections. According to \citet{manoha_correlating_2017}, a correction of $10log{SD}$ (SD refers to spanwise difference) is appropriate for fully decorrelated sources, while a $20log{SD}$ correction is suitable for fully correlated sources. As shown in Fig.~\ref{fig:Span_OASPL_15}, a coefficient of 15 is necessary to scale far-field noise together, suggesting that slat noise exhibits partial correlation behavior. Referring back to Fig.~\ref{fig:Span_ff}, it can be concluded that both the broadband noise and the hump are fully decorrelated, whereas the tonal noise shows some level of correlation, which accounts for the variations in tonal noise amplitude.

All things considered, a spanwise length of $11.1\%c_s$ is enough to achieve fully spanwise decorrelation, yielding nearly identical noise results as the $25.6\%c_s$ case and the experimental data, yet demanding relatively low computational resources. Thus, the Medium\_11.1\% case setup will be used for the subsequent simulations.

\subsection{\label{sec:Re}Reynolds Number Effect}
In previous studies, the effect of Reynolds number on aeroacoustic noise was mainly explored through pressurization wind tunnels \cite{mouton2024aeroacoustic}, cryogenic wind tunnels \cite{ahlefeldt2013aeroacoustic,clark2020requirements} or testing the same model with varying characteristic lengths in multiple wind tunnels \cite{verges2022benchmarking,ivanova2024experimental}. The first two approaches necessitate specially designed wind tunnels, while the latter involves multiple tests across various wind tunnels, all of which incur significant expenses. However, with Dimaxer, Reynolds number effect can be easily explored by simply scaling the simulation model while keeping other variables remain unchanged. The chord length of the basic JAXA modified 30P30N airfoil is scaled from 0.5 to 10 times, resulting in a Reynolds number range from $8.55\times 10^5$ to $1.71\times 10^7$, reaching the level of real aircraft. Additionally, as the size of the entire computational domain scales while maintaining a consistent $y_{mesh}^+$, the mesh sizes vary as detailed in Table~\ref{tab:meshes}.
\begin{table}[hbt!]
    \caption{\label{tab:Re} Computational resources required for different Reynolds numbers.}
    \centering
    \begin{tabular}{lccccccc}
    \hline
    Cases        & DoF/million& GPU Hours for 10 FPTs \\\hline
    Medium\_0.5Re& 15.8  & 30.79         \\
    Medium\_1Re  & 15.1  & 36.26         \\
    Medium\_5Re  & 17.6  & 127.89        \\
    Medium\_10Re & 19.1  & 256.22        \\
    Medium\_10Re*& 18.6  & 272.84        \\
    \hline
\end{tabular}
\end{table}

The corresponding FPT will be scaled linearly since the inflow velocity remains constant, and the computational resource costs will also scale linearly as shown in Table~\ref{tab:Re}. This is an important feature of the STE-KEP-FR scheme, i.e. as the Reynolds number increases, the computational cost only increases linearly rather than exponentially as other CFD solvers. Additionally, even in the scenario requiring the most computing resources, the total cost is only 272.84 GPU hours. For wind tunnel experiments, a test section of $5\times5m$ is necessary for this $c_s=4.572m$ airfoil, and this type of wind tunnel is typically applied for testing 3D aircraft models.

\begin{figure}[hbt!]
    \centering
    \includegraphics[width=.9\textwidth]{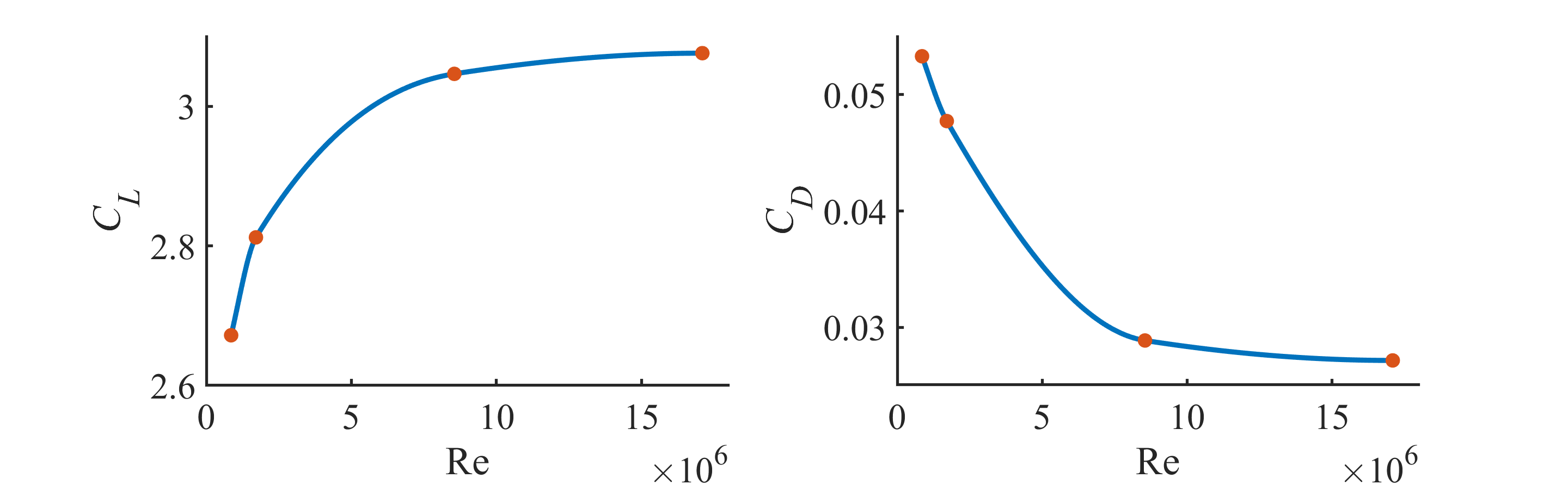}
    \caption{\label{fig:Re_clcd} Time-averaged force coefficients for different Reynolds numbers at $5.5^\circ$ AoA and 0.17 Ma.}
\end{figure}

The aerodynamic effect of the Reynolds number at $5.5^\circ$ AoA and 0.17 Ma is presented in Fig.~\ref{fig:Re_clcd}. As the Reynolds number increases, the $C_L$ also increased, which is consistent with the pressurized wind tunnel experiments on the CRM-HL configuration conducted by \citet{mouton2024aeroacoustic}. Meanwhile, the $C_D$ decreases as the Reynolds number increases.

\begin{figure}[hbt!]
    \centering
        \captionsetup{justification=raggedright, singlelinecheck=false}
        \subcaptionbox{\label{fig:Re_ff_St_10}St based on $c_{slat}$ with spanwise scale $10logSD$}{
        \includegraphics[width = 0.49\textwidth]{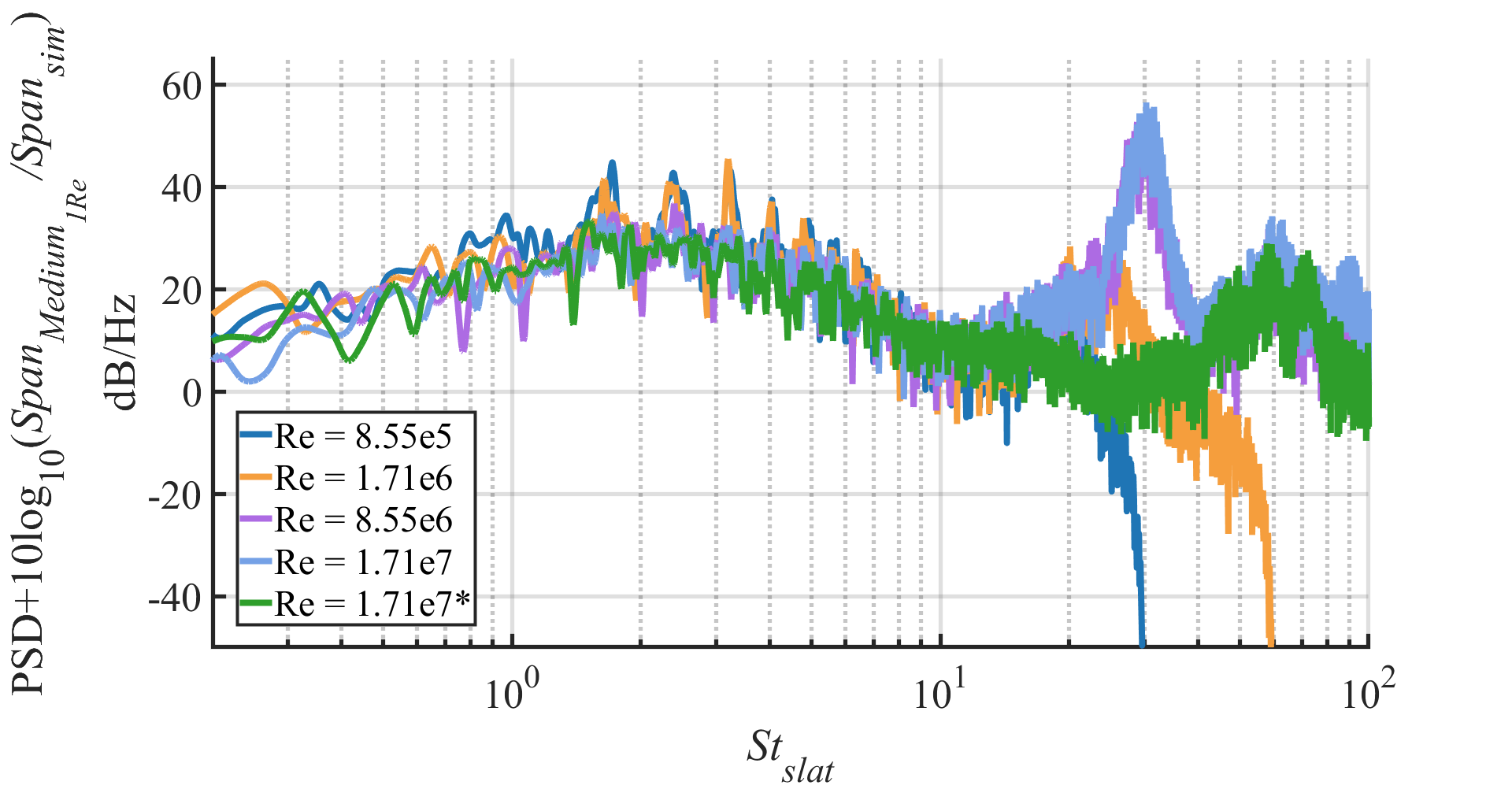}}
        \subcaptionbox{\label{fig:Re_ff_St_5.9}St based on $c_{slat}$ with spanwise scale $5.9logSD$}{\includegraphics[width = 0.49\textwidth]{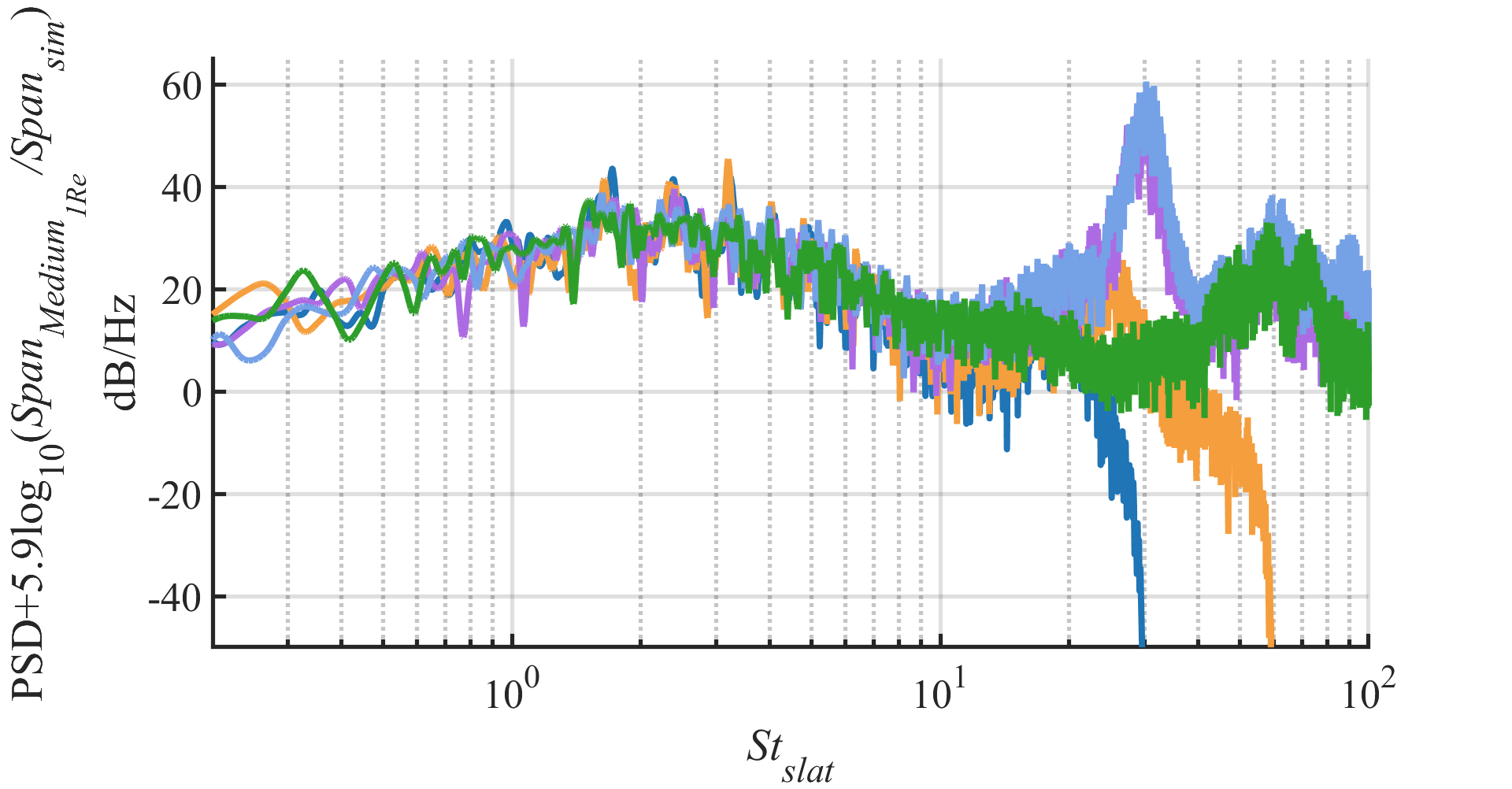}} 
        \subcaptionbox{\label{fig:Re_ff_St_4}St based on $c_{slat}$ with spanwise scale $4logSD$}{\includegraphics[width = 0.49\textwidth]{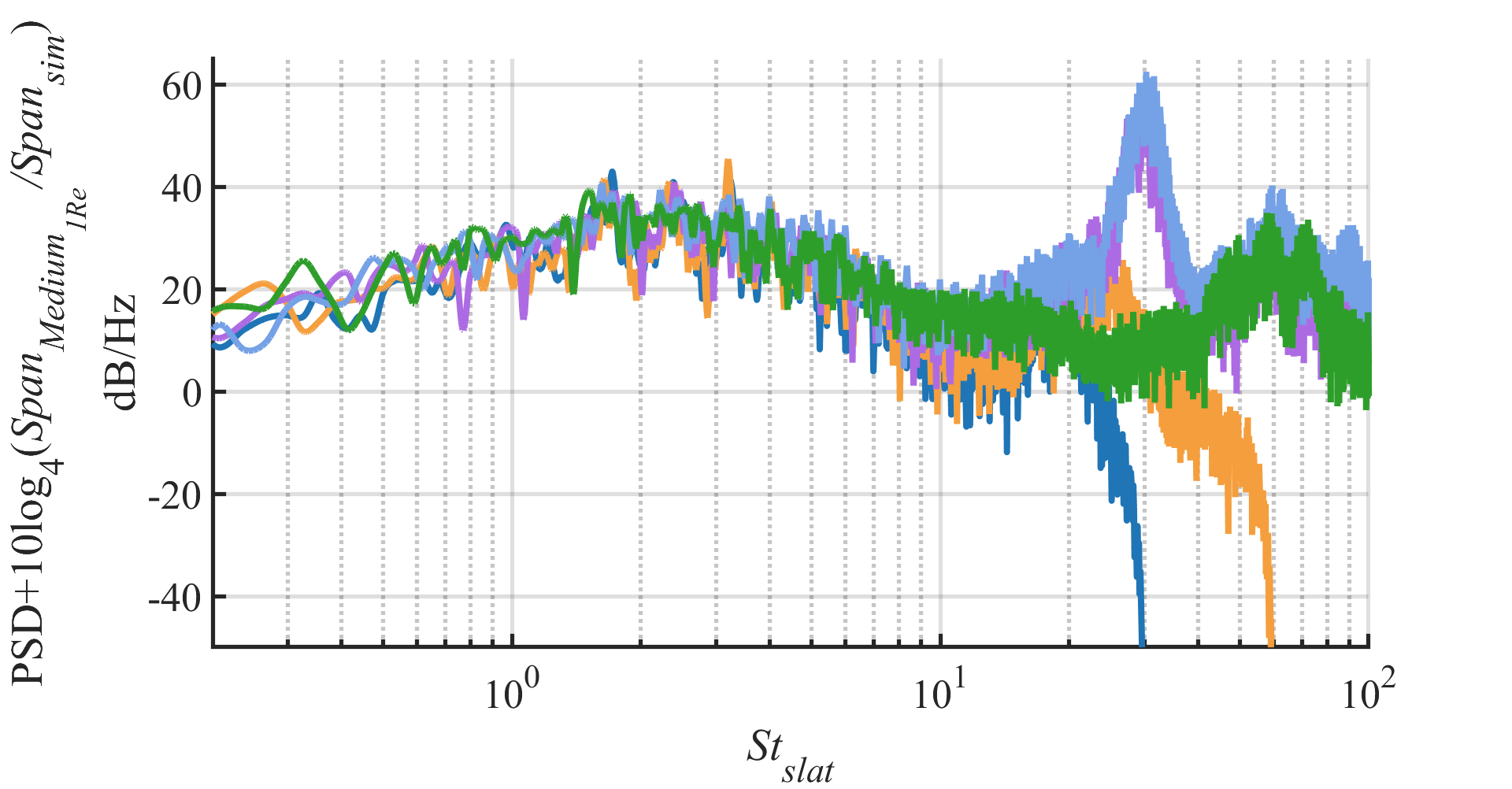}} 
        \subcaptionbox{\label{fig:Re_ff_St_st}St based on $s_t$ with spanwise scale $10logSD$}{\includegraphics[width = 0.49\textwidth]{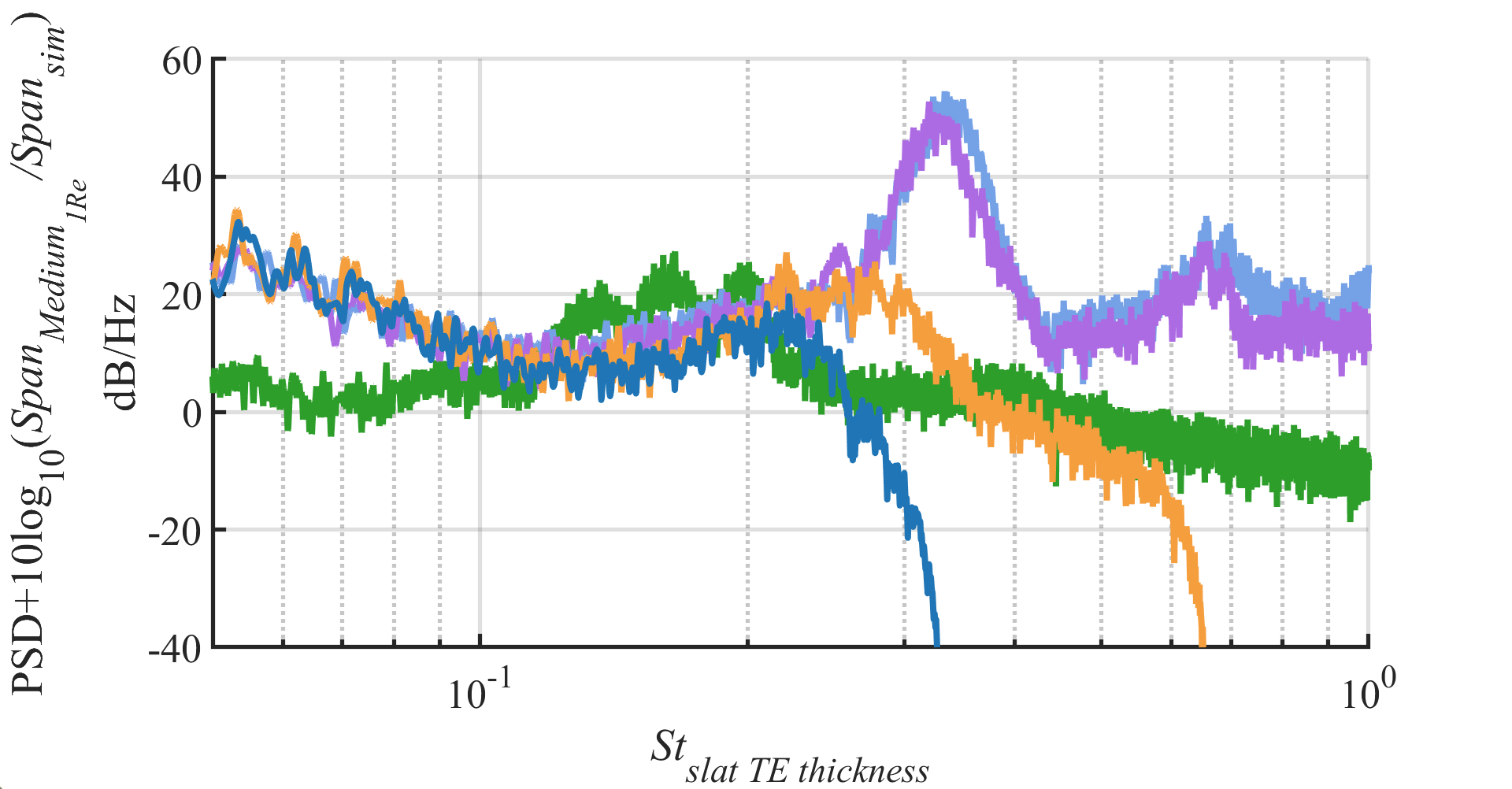}} 
    \caption{\label{fig:Re_ff}Acoustic signal spectra from the receiver $10c_s$ right below the airfoil with different Reynolds numbers at $5.5^\circ$ AoA and 0.17 Ma.}
\end{figure}

\begin{figure}[hbt!]
    \centering
    \includegraphics[width=.49\textwidth]{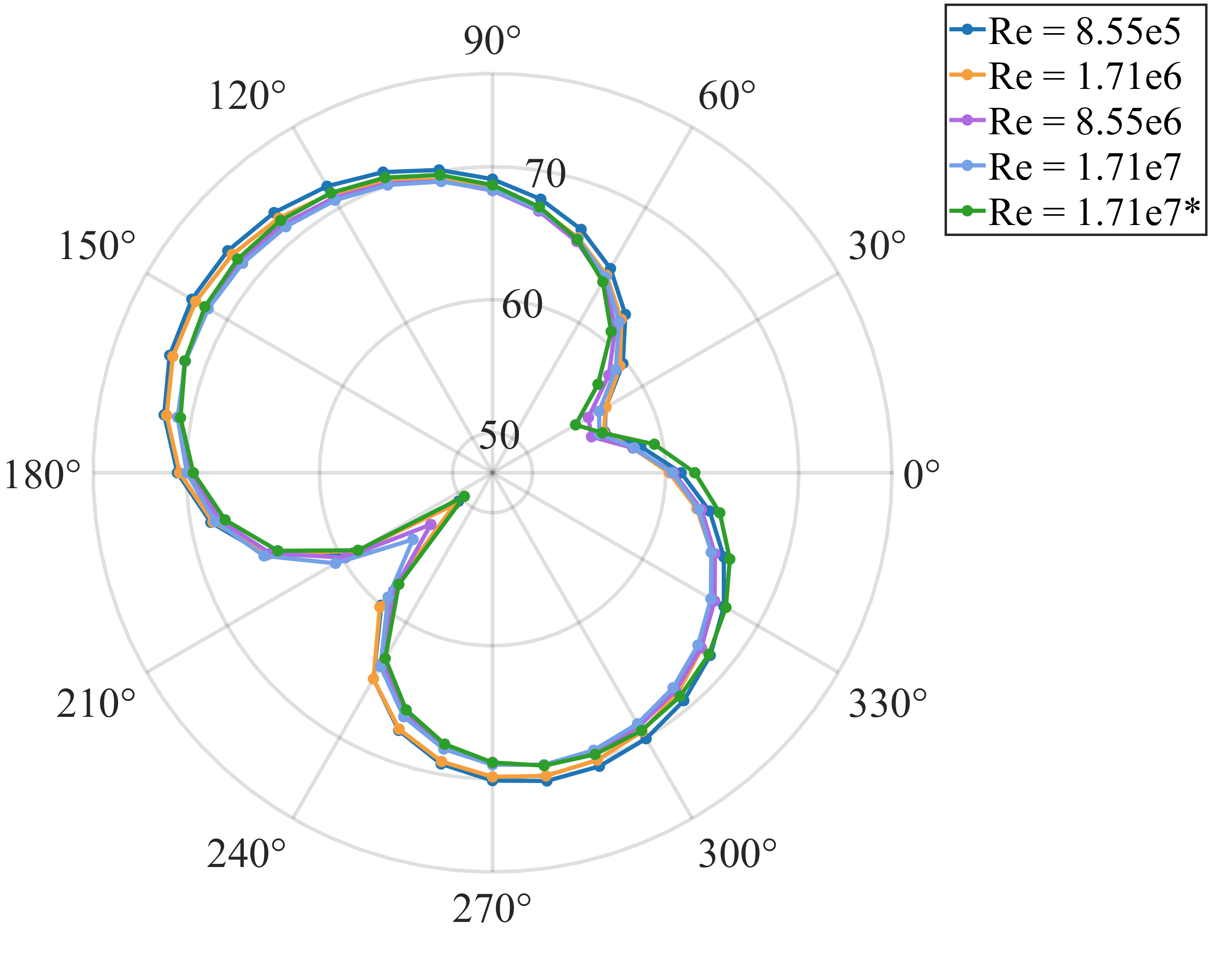}
    \caption{\label{fig:Re_OASPL} Far-field noise directivity patterns for different Reynolds number cases under $5.5^\circ$ AoA and 0.17 Ma at a distance of $10c_s$ away with a spanwise correction of $5.9log{SD}$ in dB.}
\end{figure}

Far-field noise spectra from receivers $10c_s$ right beneath the airfoil are presented in Fig.~\ref{fig:Re_ff}. It can be seen that, based on the Strouhal number of slat chord length, the tonal noise frequencies across various Reynolds numbers match together. Furthermore, Figs.~\ref{fig:Re_ff_St_10} and \ref{fig:Re_ff_St_4} respectively show that a spanwise correction of $10log{SD}$ is appropriate for the mid-high frequency broadband noise, while $4log{SD}$ is more suitable for the tonal noise and low frequency broadband noise. Figure~\ref{fig:Re_OASPL} presents the far-field noise OASPL integrated within $St=10$ with a spanwise correction of $5.9log{SD}$, which yields the smallest OASPL differences of 1.57dB consider all directions except at $220^\circ$. The corresponding far-field noise spectra are plotted in Fig.~\ref{fig:Re_ff_St_5.9}. In addition, the spanwise correction of $5.9log{SD}$ may suggest that as the Reynolds number increases, the relative intensity of the tonal noise decreases.

\begin{figure}[hbt!]
    \centering
        \captionsetup{justification=raggedright, singlelinecheck=false}
        \subcaptionbox{\label{fig:Cwz_10Re}$s_t/c_{slat}=1.11\%$}{
        \includegraphics[width = 0.41\textwidth]{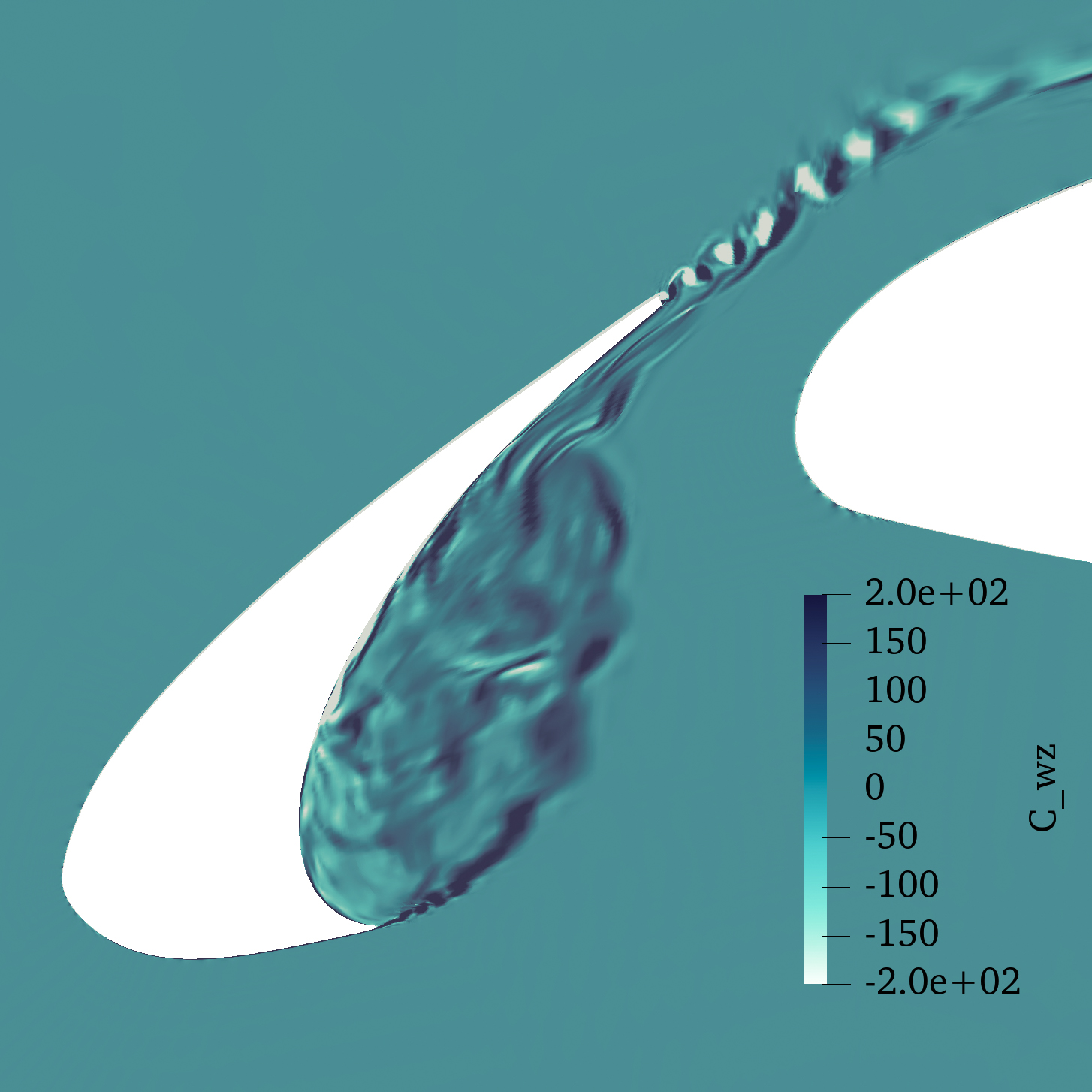}}
        \quad
        \subcaptionbox{\label{fig:Cwz_10Re_star}$s_t/c_{slat}=0.28\%$}{\includegraphics[width = 0.41\textwidth]{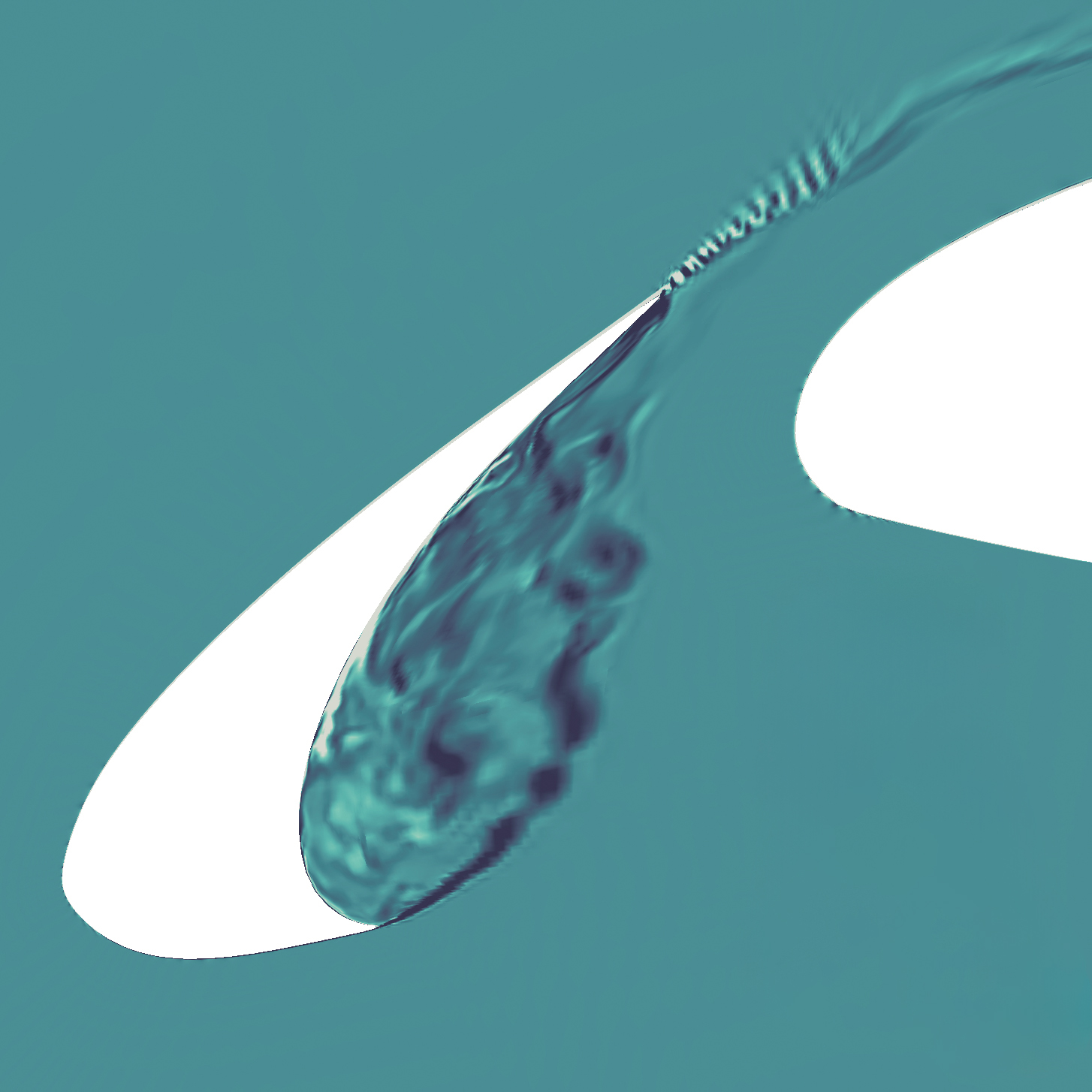}} 
    \caption{\label{fig:Re_Cwz} Contours of instantaneous normalized spanwise vorticity ($C_{\omega z} = \omega_zc_s/U_0$) on the airfoil midplane plane for different slat TE thickness at $5.5^\circ$ AoA, 0.17 Ma and $Re=1.71\times 10^7$.}
\end{figure}

\begin{figure}[hbt!]
    \centering
        \captionsetup{justification=raggedright, singlelinecheck=false}
        \subcaptionbox{\label{fig:Div_10Re}$s_t/c_{slat}=1.11\%$}{
        \includegraphics[width = 0.41\textwidth]{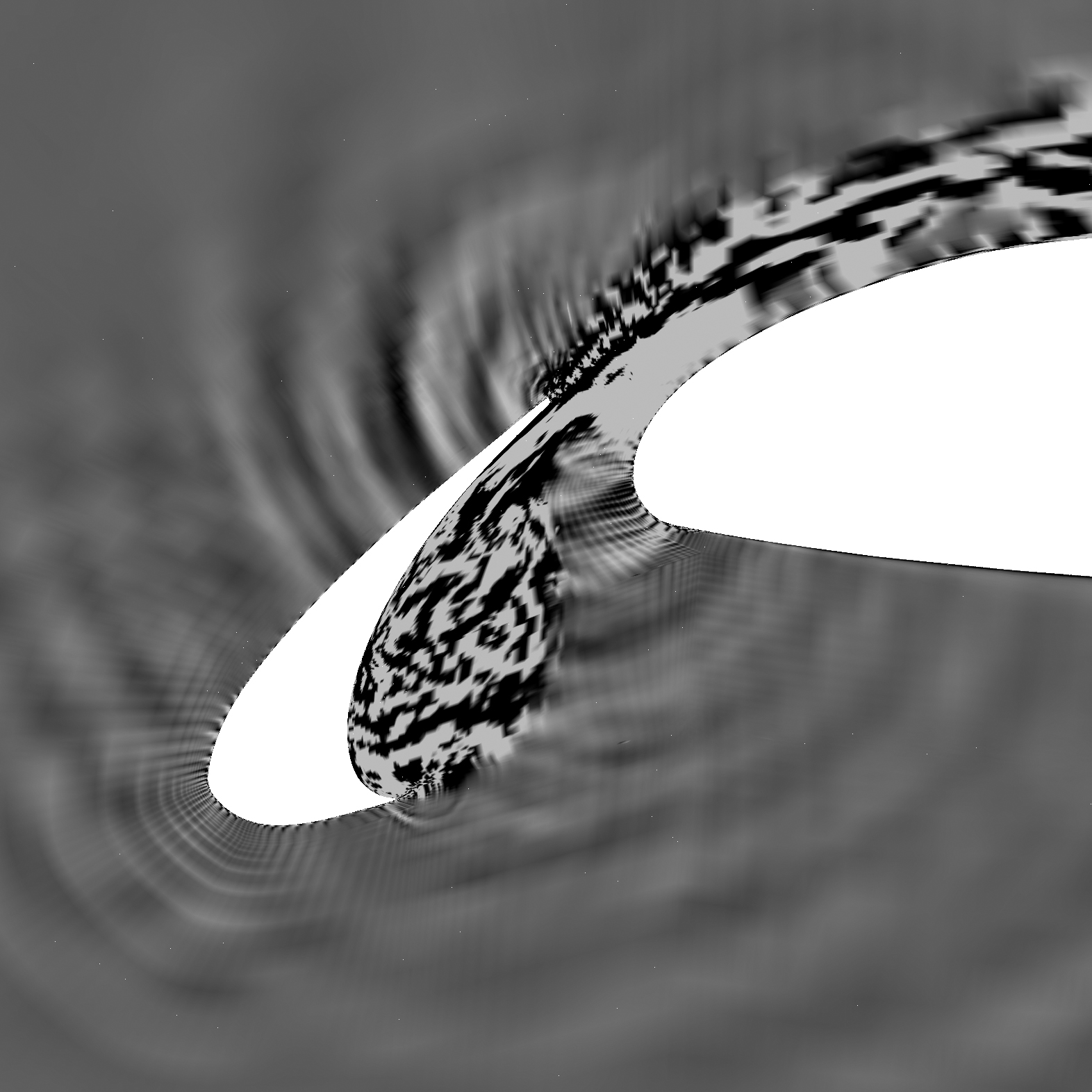}}
        \quad
        \subcaptionbox{\label{fig:Div_10Re_star}$s_t/c_{slat}=0.28\%$}{\includegraphics[width = 0.41\textwidth]{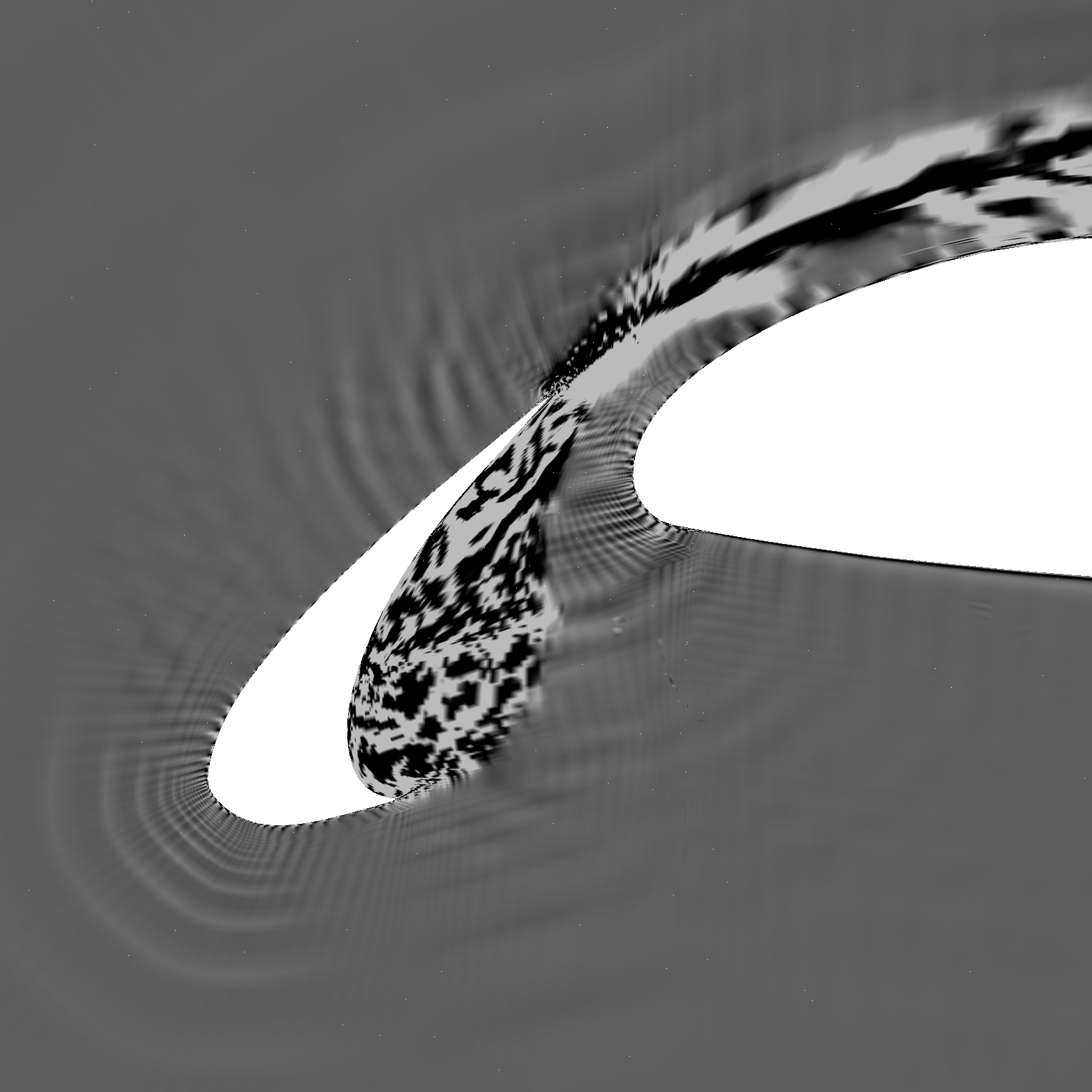}} 
    \caption{\label{fig:Re_Div} Instantaneous contour of $Div(\rho \boldsymbol{u})$ in the midspan plane around slat region for different slat TE thickness at $5.5^\circ$ AoA, 0.17 Ma and $Re=1.71\times 10^7$ (-10<$Div(\rho \boldsymbol{u})$<10).}
\end{figure}

Regarding the high-frequency vortex shedding hump, when the thickness of the slat TE is reduced to match that of actual aircraft, Fig.~\ref{fig:Re_ff_St_10} illustrates a significant reduction in the amplitude of the hump. The reason behind is shown in Figs.~\ref{fig:Re_Cwz} and \ref{fig:Re_Div}, i.e. a thinner slat TE leads to a much lower high-frequency vortex shedding intensity. Contrary to the mainstream views that the high-frequency vortex shedding hump is merely a byproduct of the thicker slat TE in reduced-scale models and does not apply to real aircraft \cite{dobrzynski2001slat,himeno2021spod}, Fig.~\ref{fig:Re_ff_St_10} suggests the hump remains a strong source of noise even when the slat TE thickness is adjusted to that of real aircraft. Furthermore, the St (based on $s_t$) associated with the hump increases with the Reynolds number, rising from 0.22 at $Re=8.55\times 10^5$ to 0.33 at $Re=1.71\times 10^7$, as illustrated in Fig.~\ref{fig:Re_ff_St_st}.

In general, by considering the Strouhal number, spanwise length, and distance from the receiver, the far-field noise spectrum of such high-lift airfoils under various Reynolds numbers can be scaled and matched together, at least within the range of $8.55\times 10^5$\textasciitilde$1.71\times 10^7$. This finding is extremely beneficial for employing reduced-scale high-lift airfoils in wind tunnel tests and CFD simulations, as simply scaling the measured or predicted noise spectrum can immediately yield the spectrum corresponding to the original model. Thus, for wind tunnel experiment:
\begin{enumerate}
\item smaller wind tunnels can be utilized to study slat noise;
\item closed section experiments can employ smaller scale models to avoid blocking effects;
\item open-jet wind tunnel tests can use smaller scale models to minimize mean flow deflection effects \cite{terracol2016investigation}.
\end{enumerate}
As for CFD, a significant reduction in computational expenses can be seen in Table~\ref{tab:Re}.

\subsection{\label{sec:AoA}AoA Effect}
\begin{figure}[hbt!]
    \centering
    \includegraphics[width=.9\textwidth]{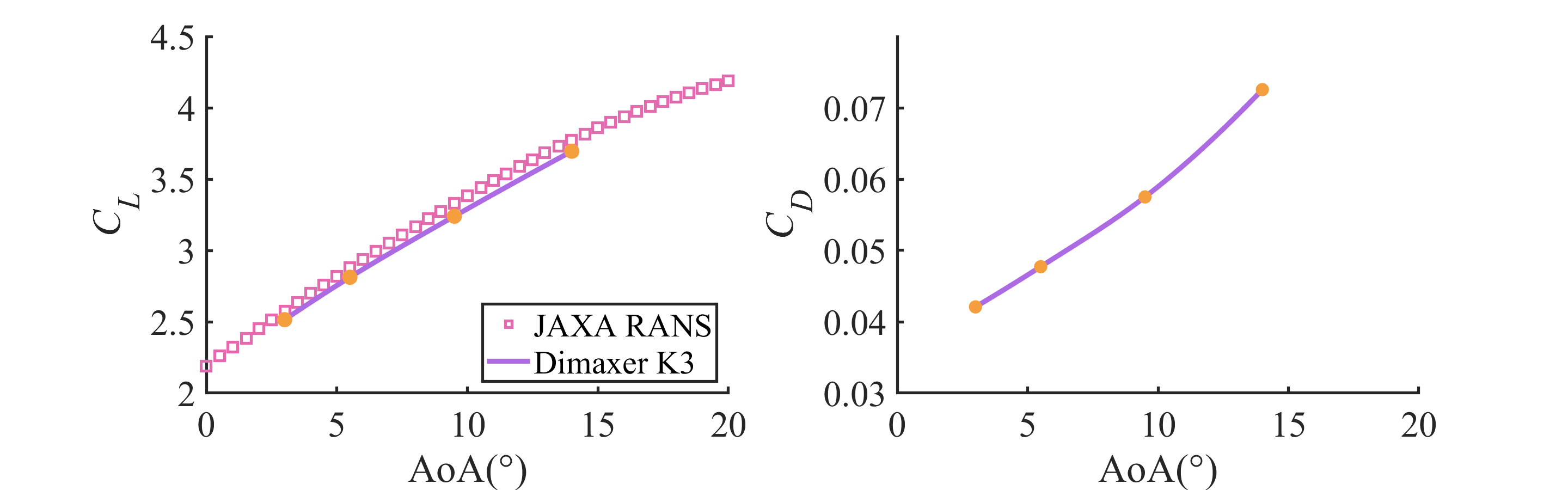}
    \caption{\label{fig:AoA_clcd} Time-averaged force coefficients for different AoAs at 0.17 Ma with $C_L$ compared to 2D RANS results from JAXA \cite{murayama2018experimental}.}
\end{figure}

Four AoAs are selected to investigate the AoA effect on slat noise with K3 simulations. The time-averaged aerodynamic force results are presented in Fig.~\ref{fig:AoA_clcd}. As the AoA increased, the $C_L$ increased linearly, while the $C_D$ exhibited a more nonlinear increment. Compared to the $C_L$ results obtained from 2D RANS using the Spalart-Allmaras turbulence model \cite{murayama2018experimental}, relatively good agreements have achieved with discrepancies of less than 2.7\% across all AoAs.
\begin{figure}[hbt!]
    \centering
        \captionsetup{justification=raggedright, singlelinecheck=false}
        \subcaptionbox{\label{fig:Cp-3}$AoA_{sim}=3^\circ, AoA_{exp}=4.5^\circ$}{
        \includegraphics[width = 0.483\textwidth]{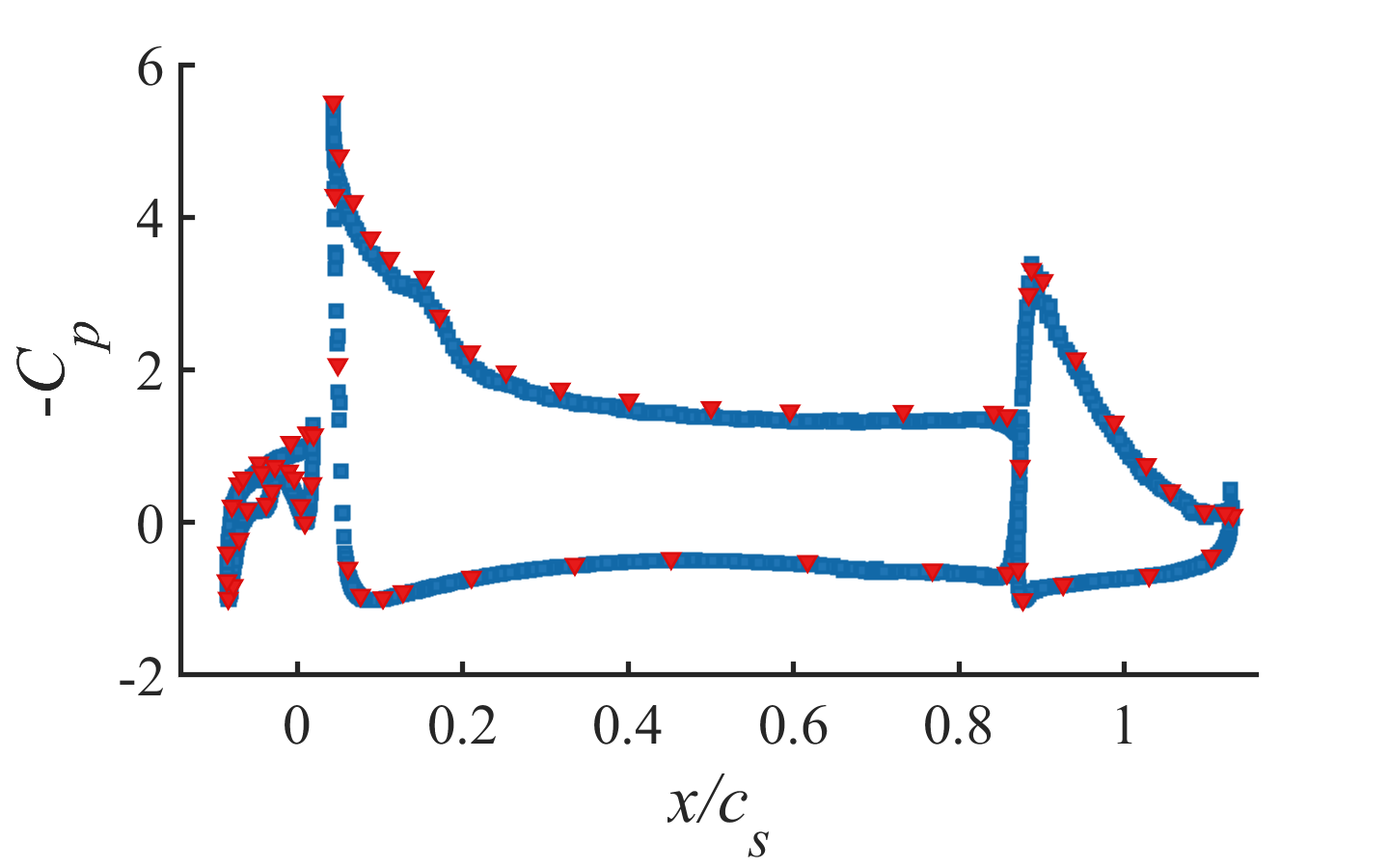}}
        \quad
        \subcaptionbox{\label{fig:Cp-5.5}$AoA_{sim}=5.5^\circ, AoA_{exp}=7^\circ$}{\includegraphics[width = 0.483\textwidth]{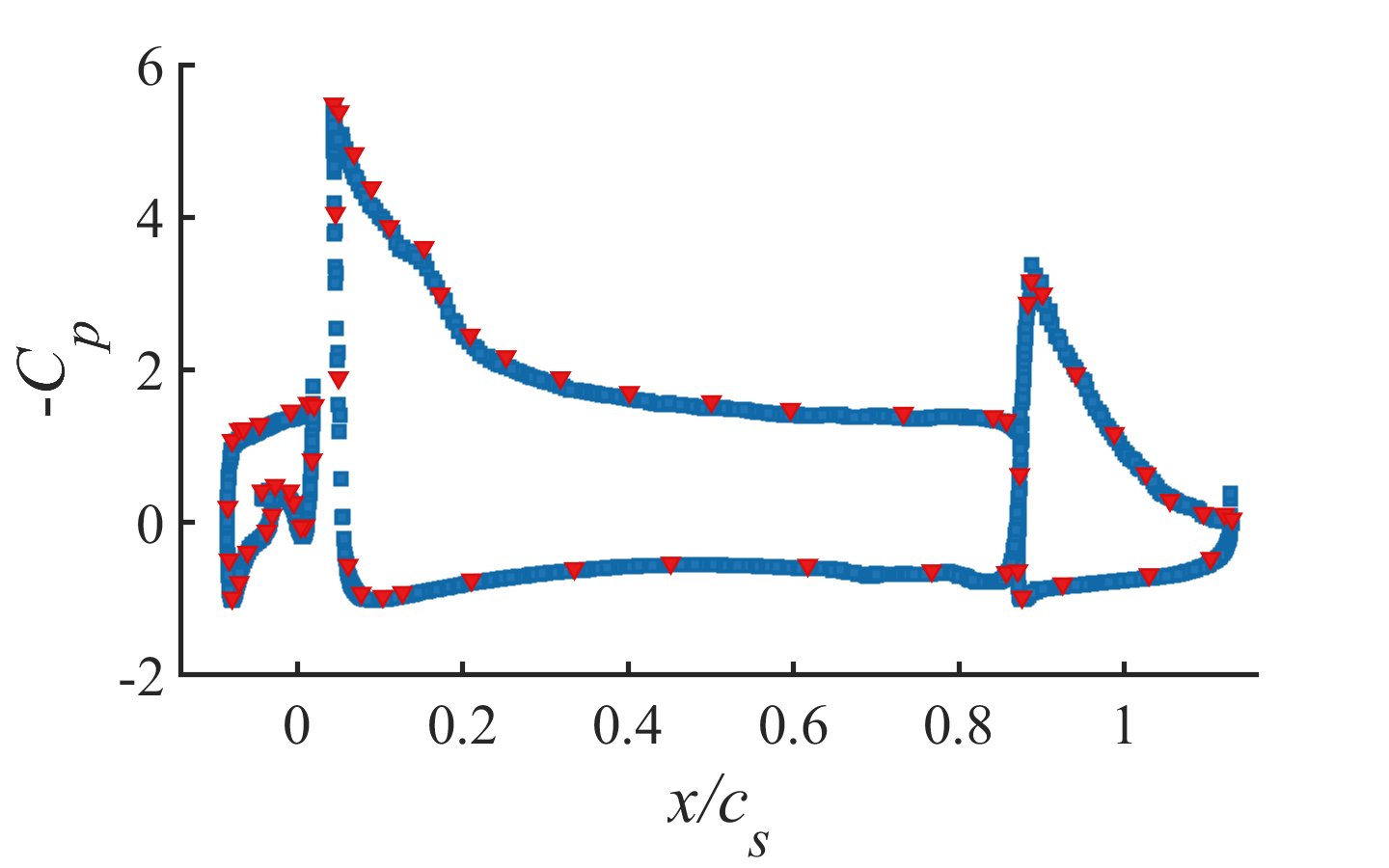}}
        \quad
        \subcaptionbox{\label{fig:Cp-9.5}$AoA_{sim}=9.5^\circ, AoA_{exp}=11^\circ$}{\includegraphics[width = 0.483\textwidth]{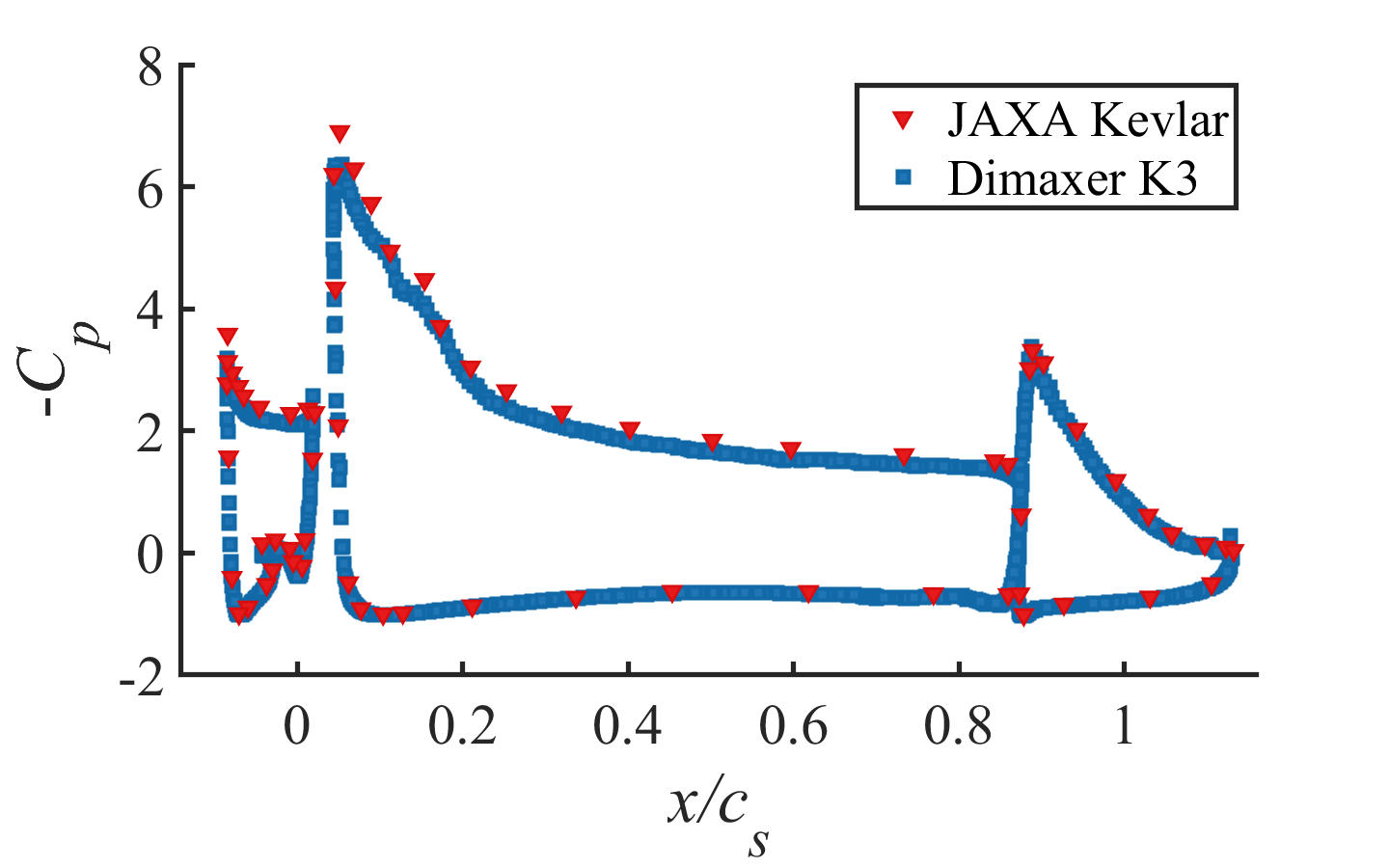}} 
        \quad
        \subcaptionbox{\label{fig:Cp-14}$AoA_{sim}=14^\circ$}{\includegraphics[width = 0.483\textwidth]{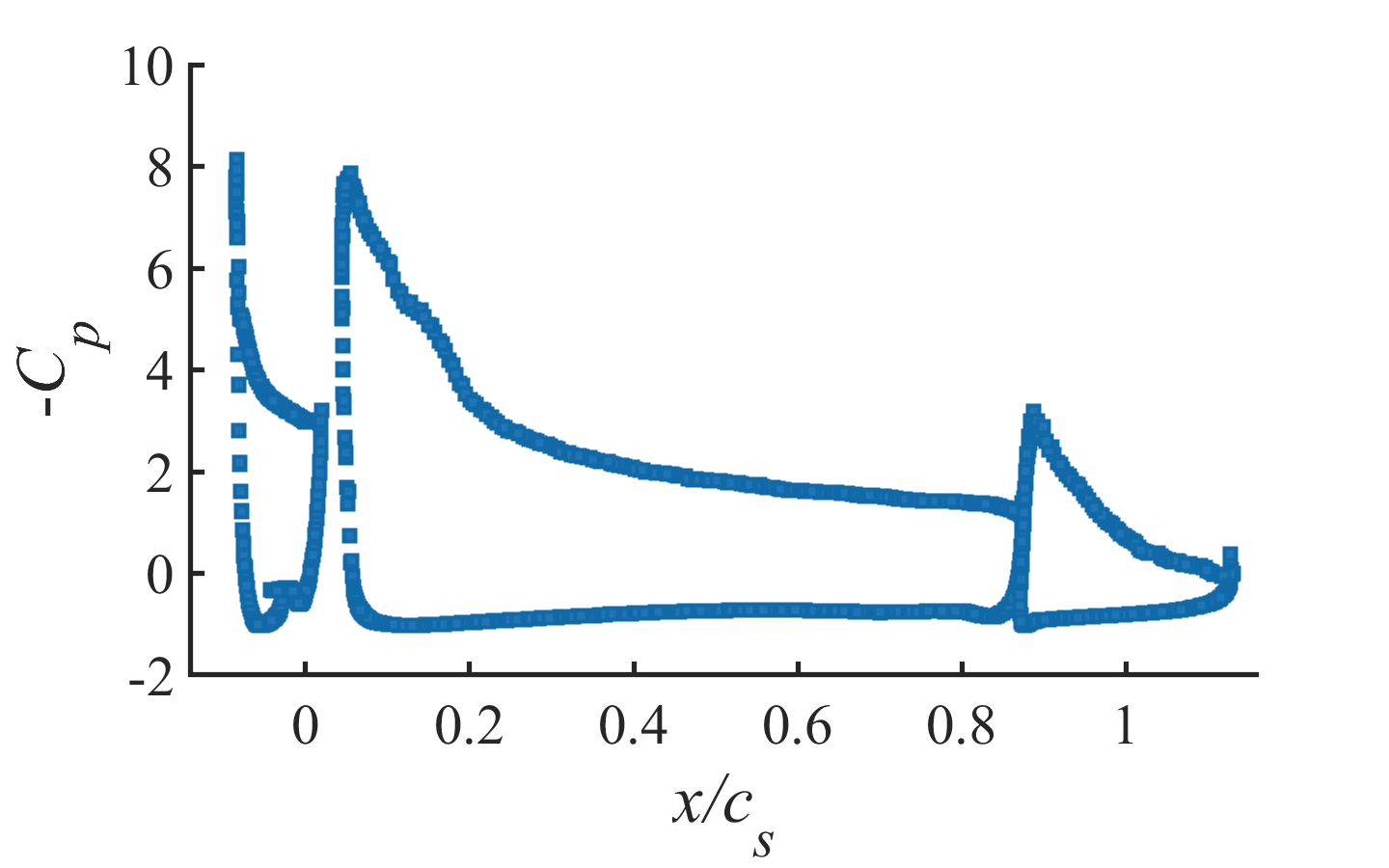}} 
    \caption{\label{fig:AoA_Cp} Time-averaged $C_p$ distribution on the airfoil midplane compared with the JAXA Kevlar wall experiment \cite{murayama2018experimental} for different AoAs at 0.17 Ma.}
\end{figure}

Time-averaged $C_p$ distribution on the airfoil midplane for different AoAs is presented in Fig.~\ref{fig:AoA_Cp}. Overall good agreements are achieved compared with the JAXA Kevlar wall experiment for the three lower AoAs with experimental $C_p$ distribution data. It is evident that, as AoA increases, the negative pressure on the upper surface of the slat increases, while the negative pressure on the upper surface of the main wing increases relatively slower. As for the negative pressure on the upper surface of the flap, it basically remained unchanged.

\begin{figure}[hbt!]
    \centering
        \captionsetup{justification=raggedright, singlelinecheck=false}
        \subcaptionbox{\label{fig:ff-3}$AoA_{sim}=3^\circ, AoA_{exp}=4.5^\circ$}{
        \includegraphics[width = 0.483\textwidth]{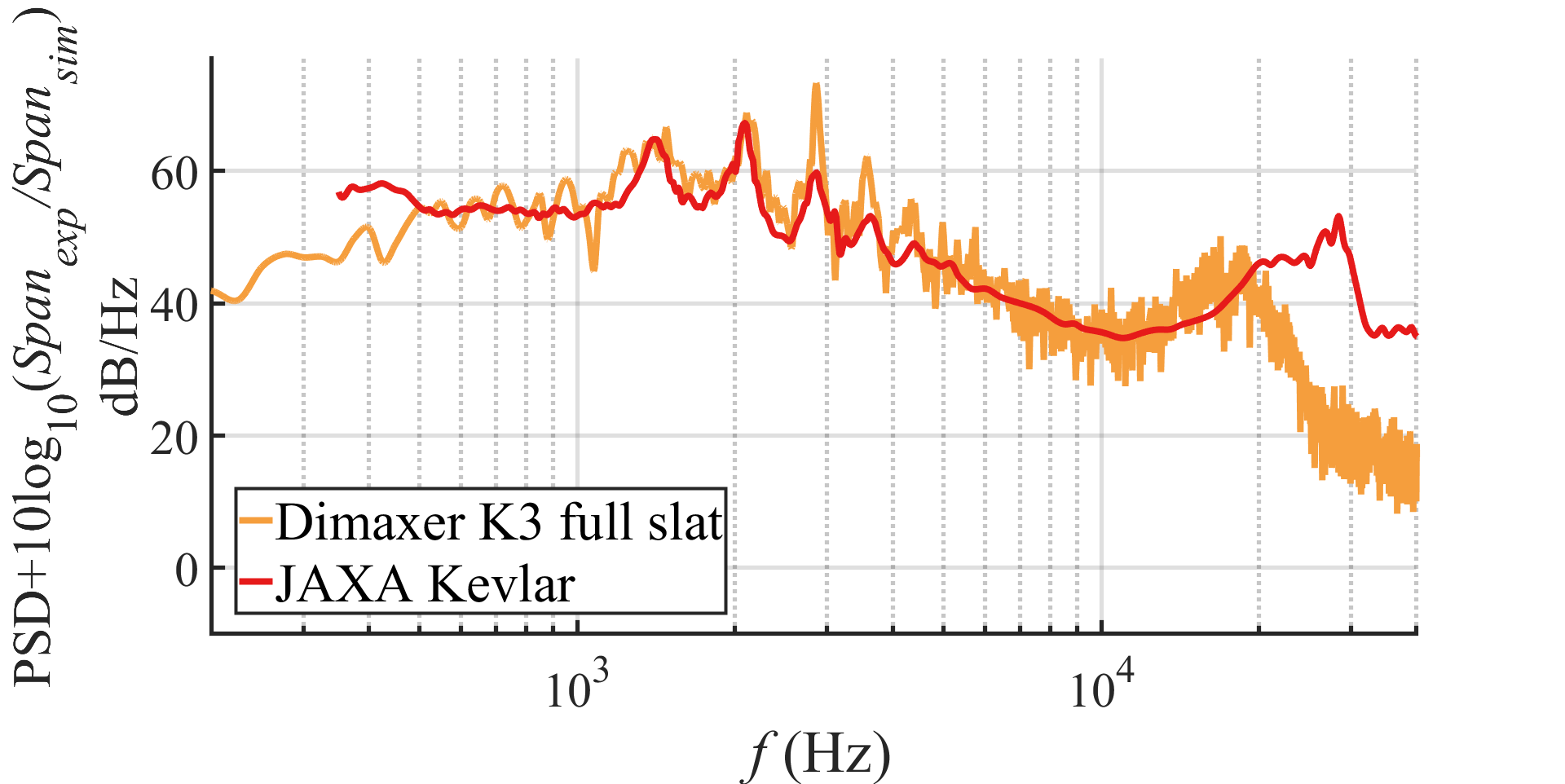}}
        \quad
        \subcaptionbox{\label{fig:ff-9.5}$AoA_{sim}=9.5^\circ, AoA_{exp}=11^\circ$}{\includegraphics[width = 0.483\textwidth]{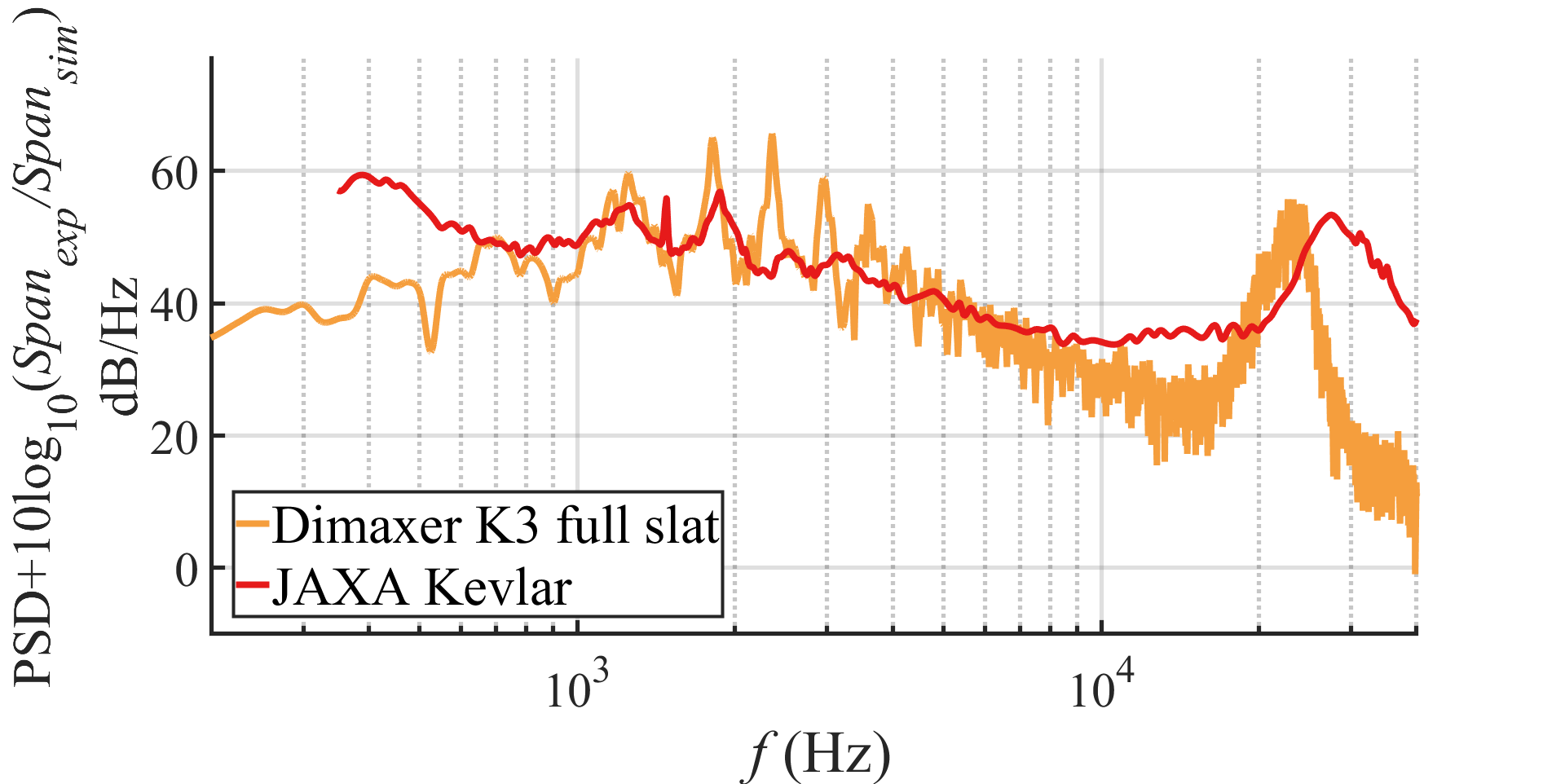}} 
        \subcaptionbox{\label{fig:ff-14}$AoA_{sim}=14^\circ,AoA_{exp}=16^\circ$}{\includegraphics[width = 0.483\textwidth]{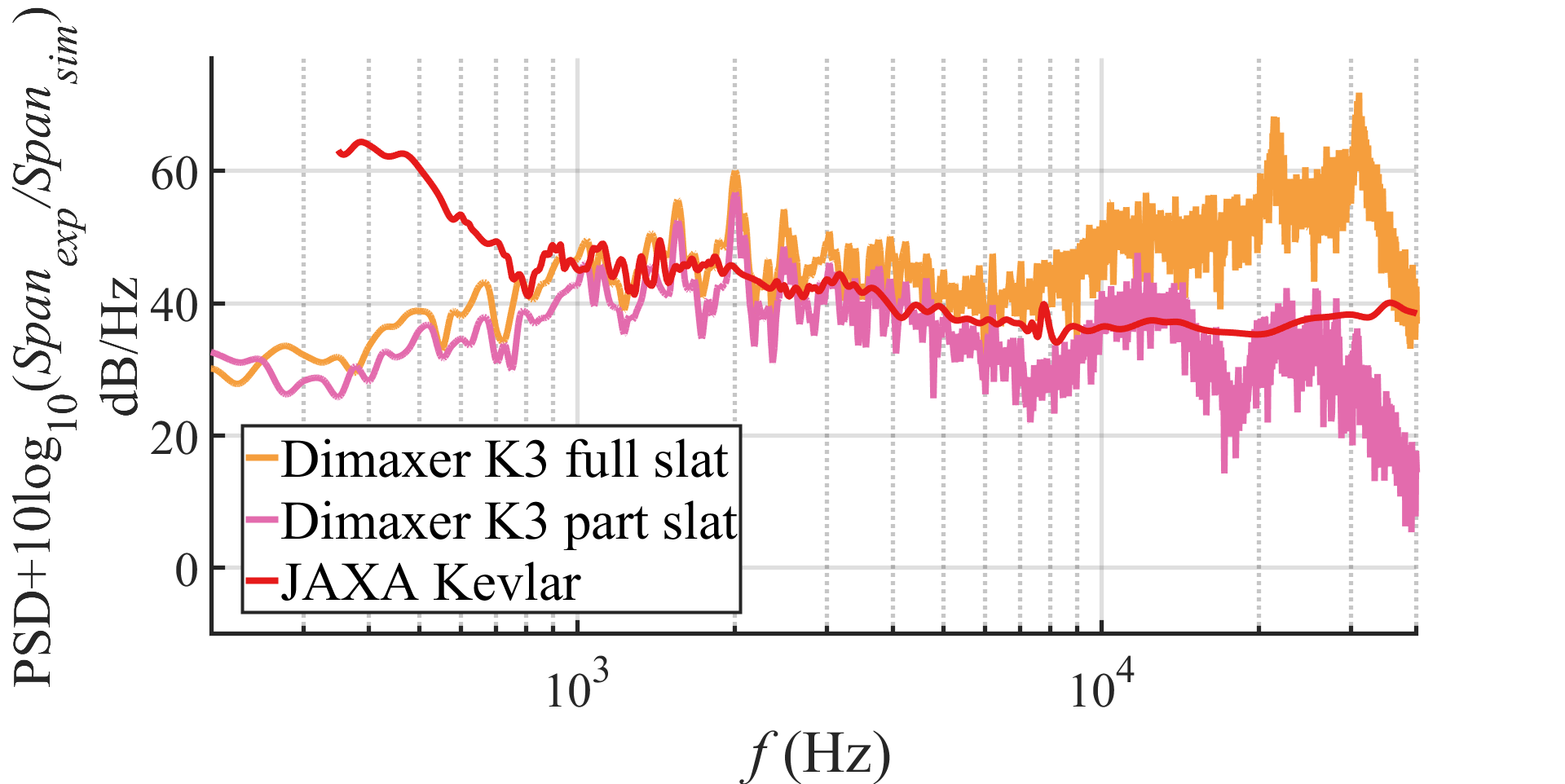}} 
        \quad
        \subcaptionbox{\label{fig:AoA_ff}K3 simulation results}{\includegraphics[width = 0.483\textwidth]{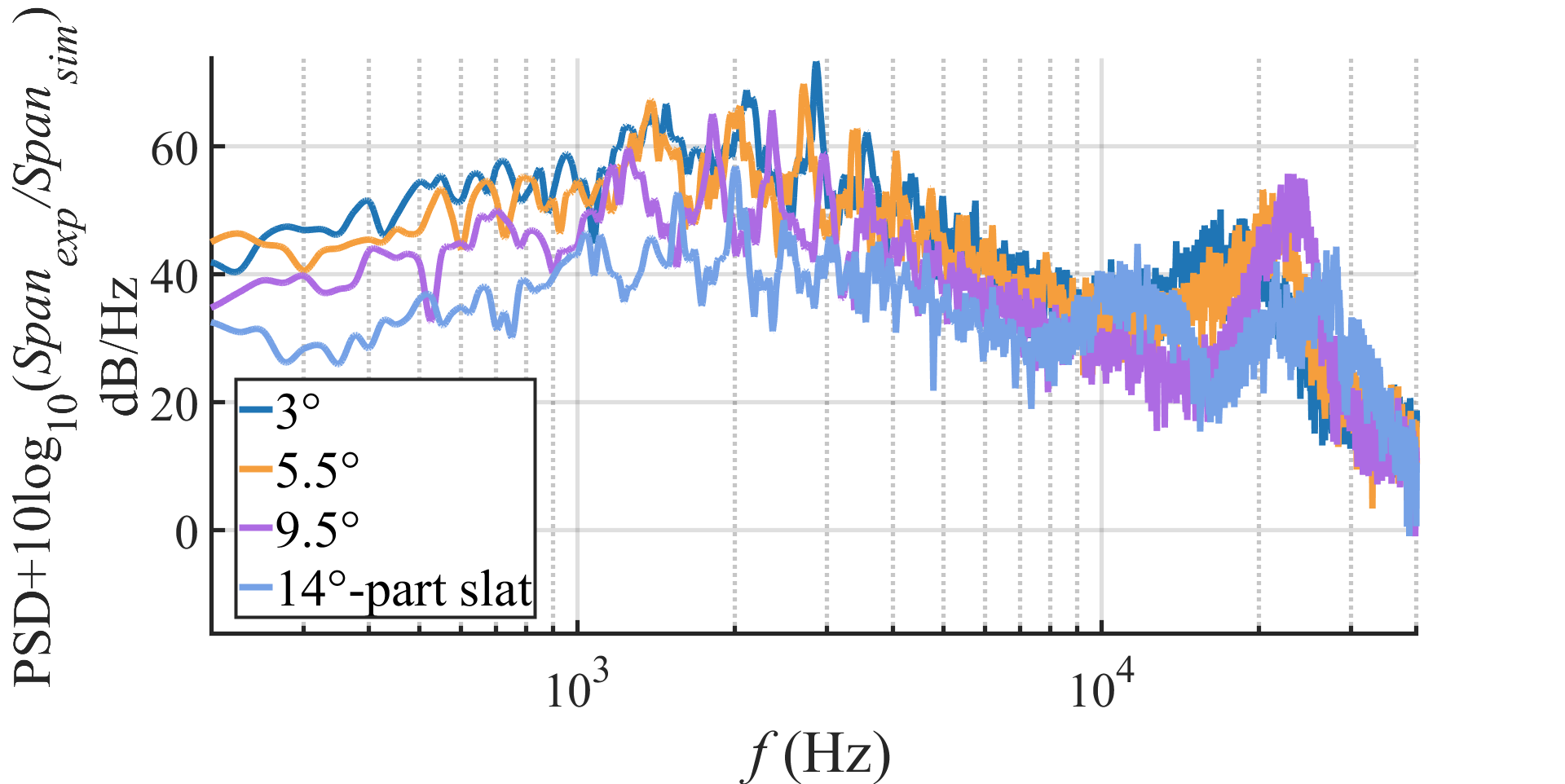}}
        \quad
    \caption{\label{fig:AoA_far} Far-field noise spectra compared to the JAXA Kevlar wall experiment \cite{murayama2018experimental} at 0.17 Ma under different AoAs (part slat removes the upper surface of the slat where transition occurs at $14^\circ$ AoA).}
\end{figure}

The comparison of the far-field noise spectrum with the JAXA Kevlar wall experiment \cite{murayama2018experimental} under $3^\circ, 9.5^\circ, 14^\circ$ is plotted  in Fig.~\ref{fig:AoA_far}. The spectrum of $5.5^\circ$ is already plotted in Fig.~\ref{fig:Mesh_ff}. It can be seen that the far-field noise spectrum of the two lowest AoA collapses well together with the experimental data between 0.5 kHz and 20 kHz. As the AoA increases, the agreement with the experimental result decreases. For $9.5^\circ$, the range of good consistency is narrowed down to between 0.7 kHz and 9 kHz, and for $14^\circ$, it further narrows to between 0.8 kHz and 6 kHz. The primary difference appears to be associated with low-frequency broadband noise, which may be influenced by the higher AoA blocking effects in the Kevlar wall experiment, as the amplitude of experimental broadband noise increases with rising AoA. In addition, the spectrum of $14^\circ$ from the slat surface exhibits a transition hump feature above 8 kHz, akin to that observed when using the main wing and flap as the FW-H sampling surface.

\begin{figure}[hbt!]
    \centering
        \captionsetup{justification=raggedright, singlelinecheck=false}
        \subcaptionbox{\label{fig:Q_3}$3^\circ$}{
        \includegraphics[width = 0.48\textwidth]{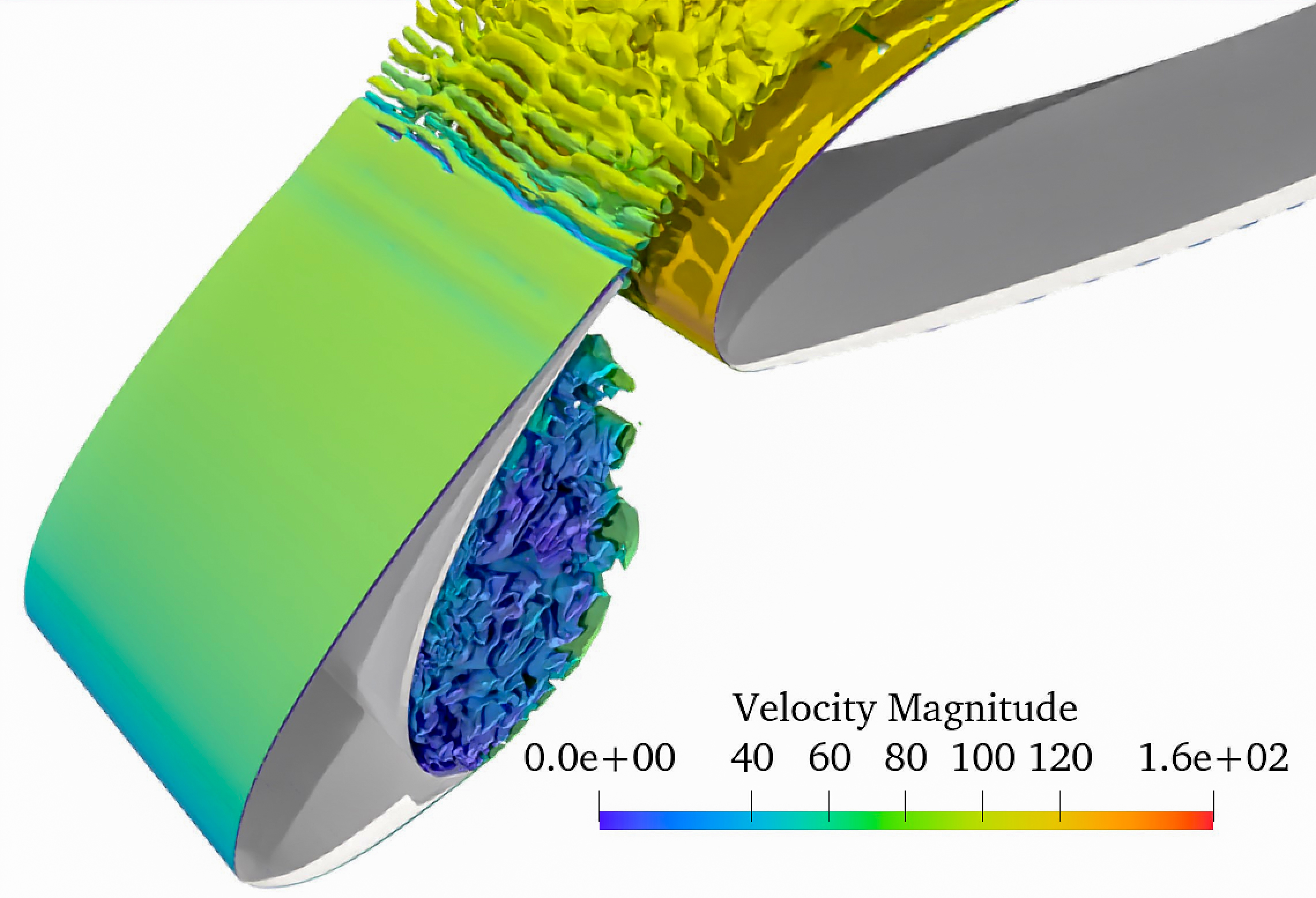}}
        \quad
        \subcaptionbox{\label{fig:Q_5.5}$5.5^\circ$}{\includegraphics[width = 0.48\textwidth]{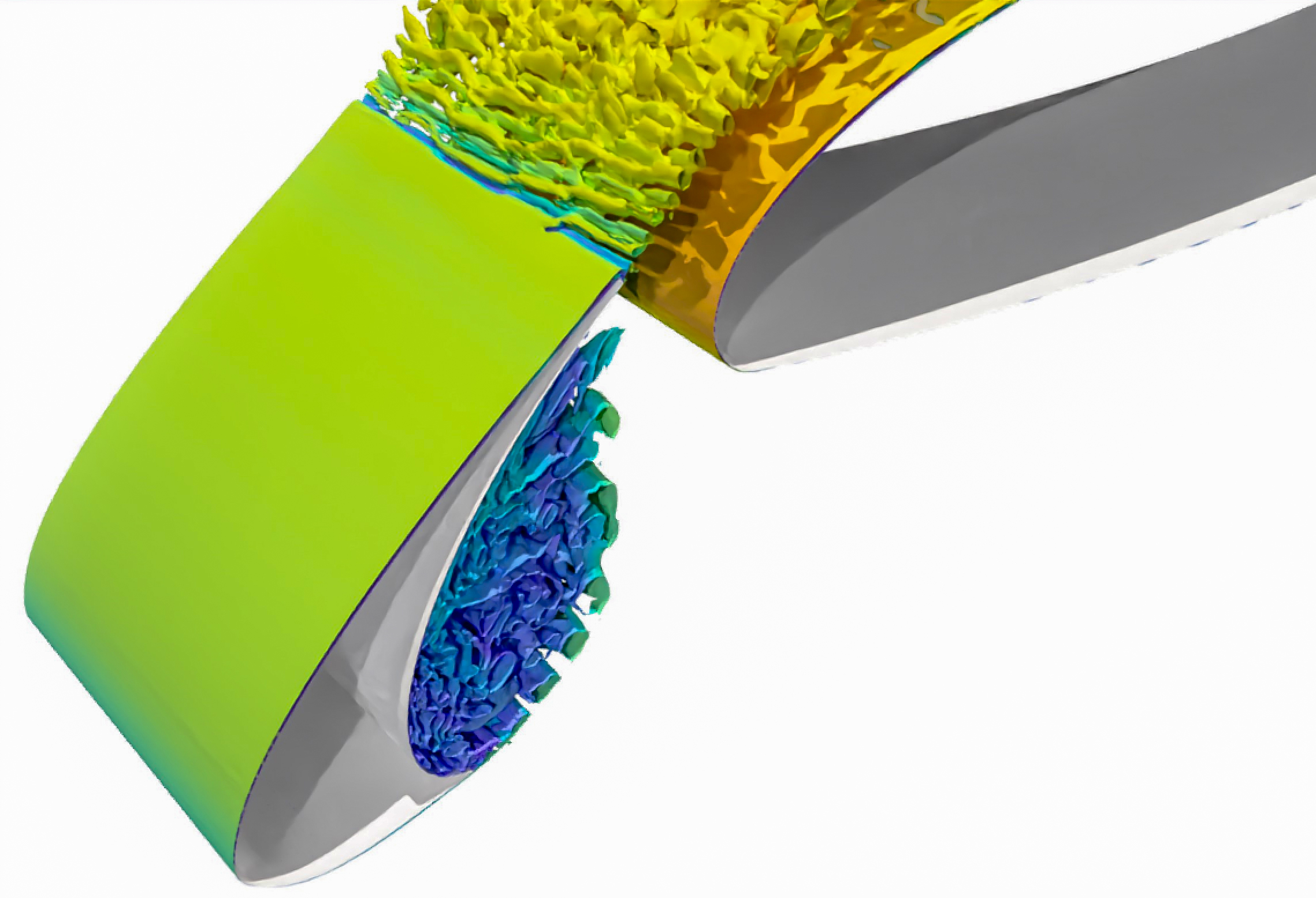}} 
        \subcaptionbox{\label{fig:Q_9.5}$9.5^\circ$}{\includegraphics[width = 0.48\textwidth]{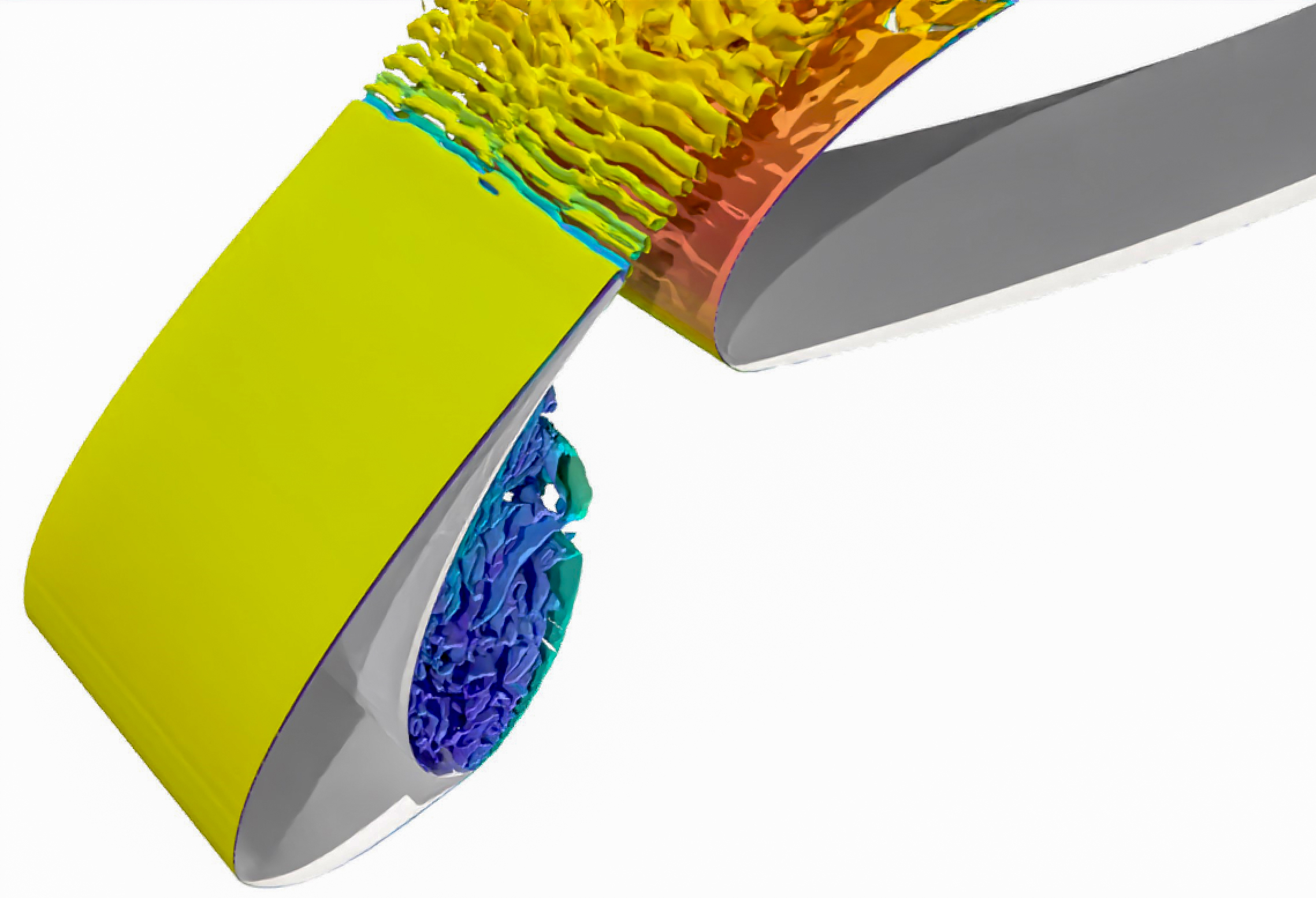}} 
        \quad
        \subcaptionbox{\label{fig:Q_14}$14^\circ$}{\includegraphics[width = 0.48\textwidth]{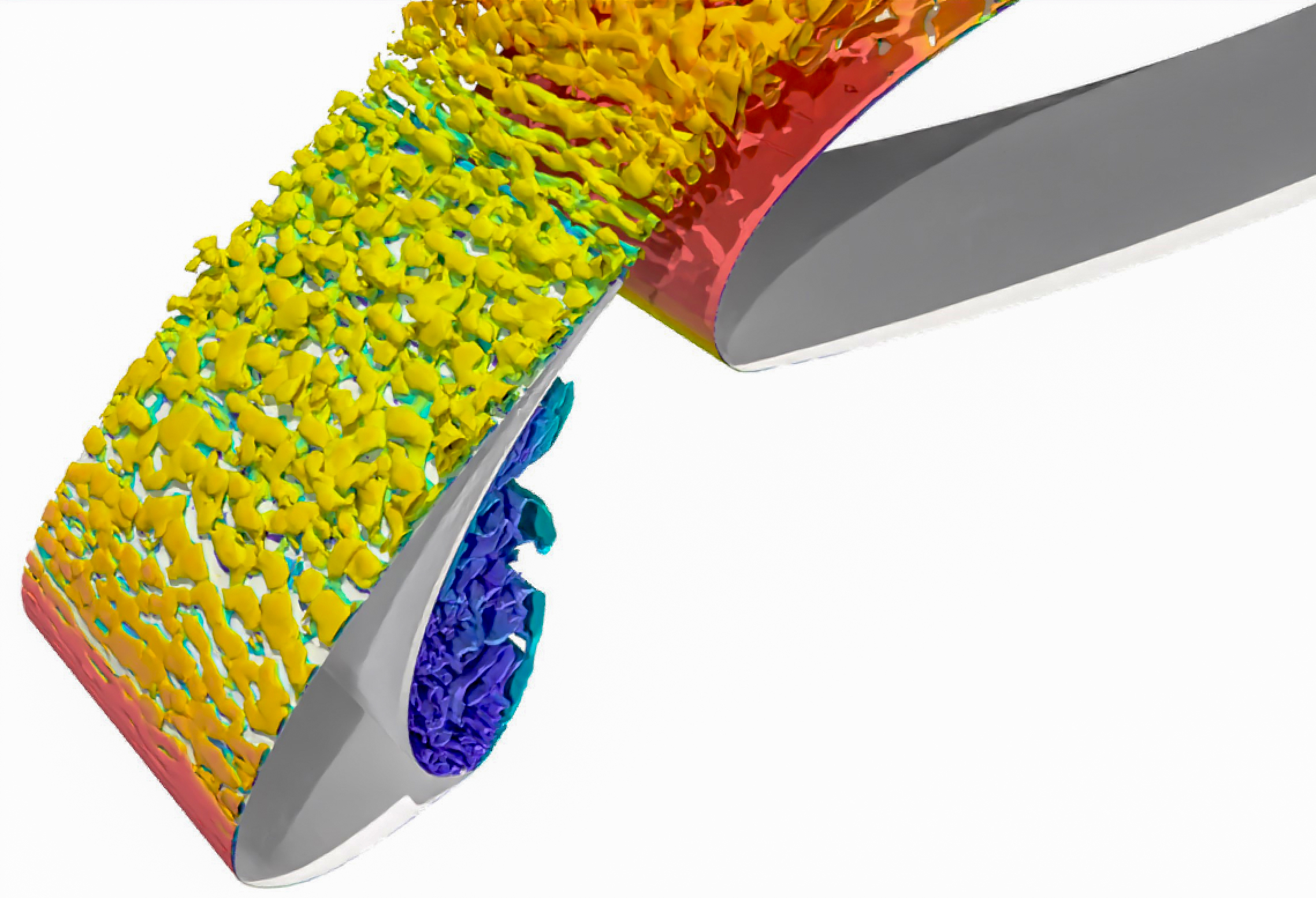}}
        \quad
    \caption{\label{fig:AoA_Q} Instantaneous iso-surface of $Q=2e6$ on the slat upper surface colored by velocity magnitude at 0.17 Ma under different AoAs.}
\end{figure}

\begin{figure}[hbt!]
    \centering
        \captionsetup{justification=raggedright, singlelinecheck=false}
        \subcaptionbox{\label{fig:OWSPL_83}Overall wall pressure level over 0.8 to 3 kHz}{
        \includegraphics[width = 0.5\textwidth]{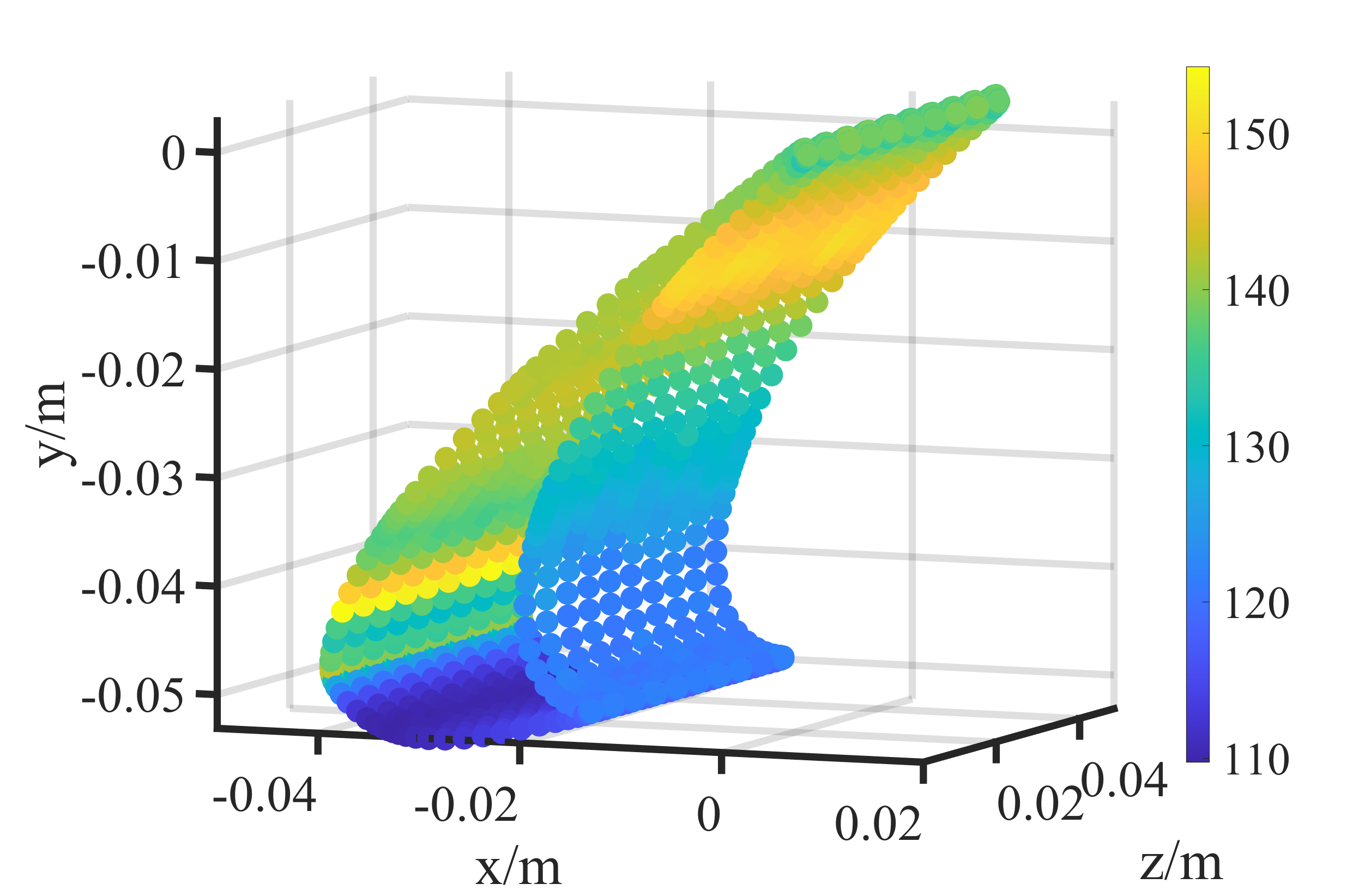}}
        \quad
        \subcaptionbox{\label{fig:OWSPL_38}Overall wall pressure level over 3 to 8 kHz}{\includegraphics[width = 0.5\textwidth]{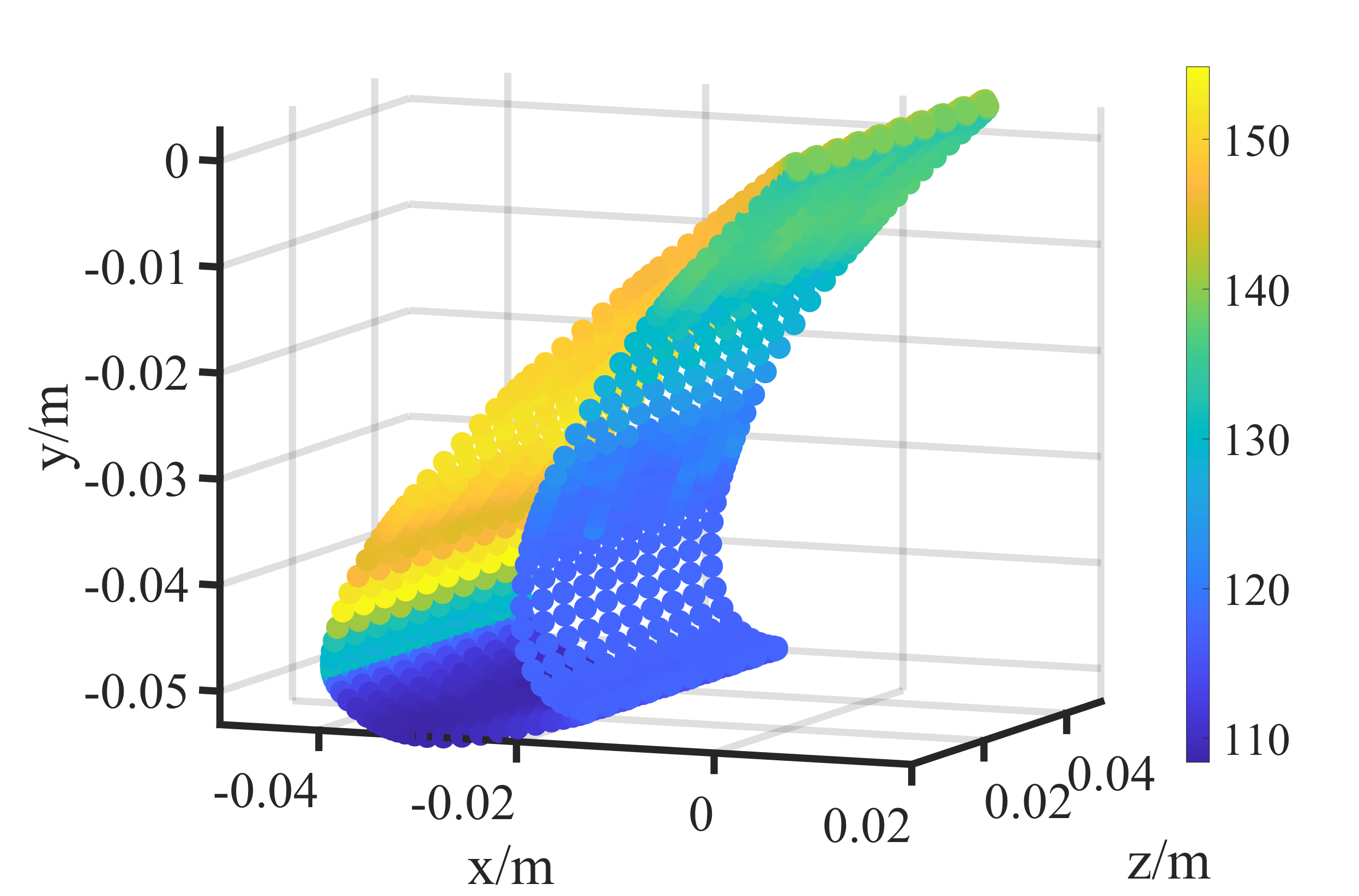}} 
        \subcaptionbox{\label{fig:OWSPL_8}Overall wall pressure level over 8 kHz}{\includegraphics[width = 0.5\textwidth]{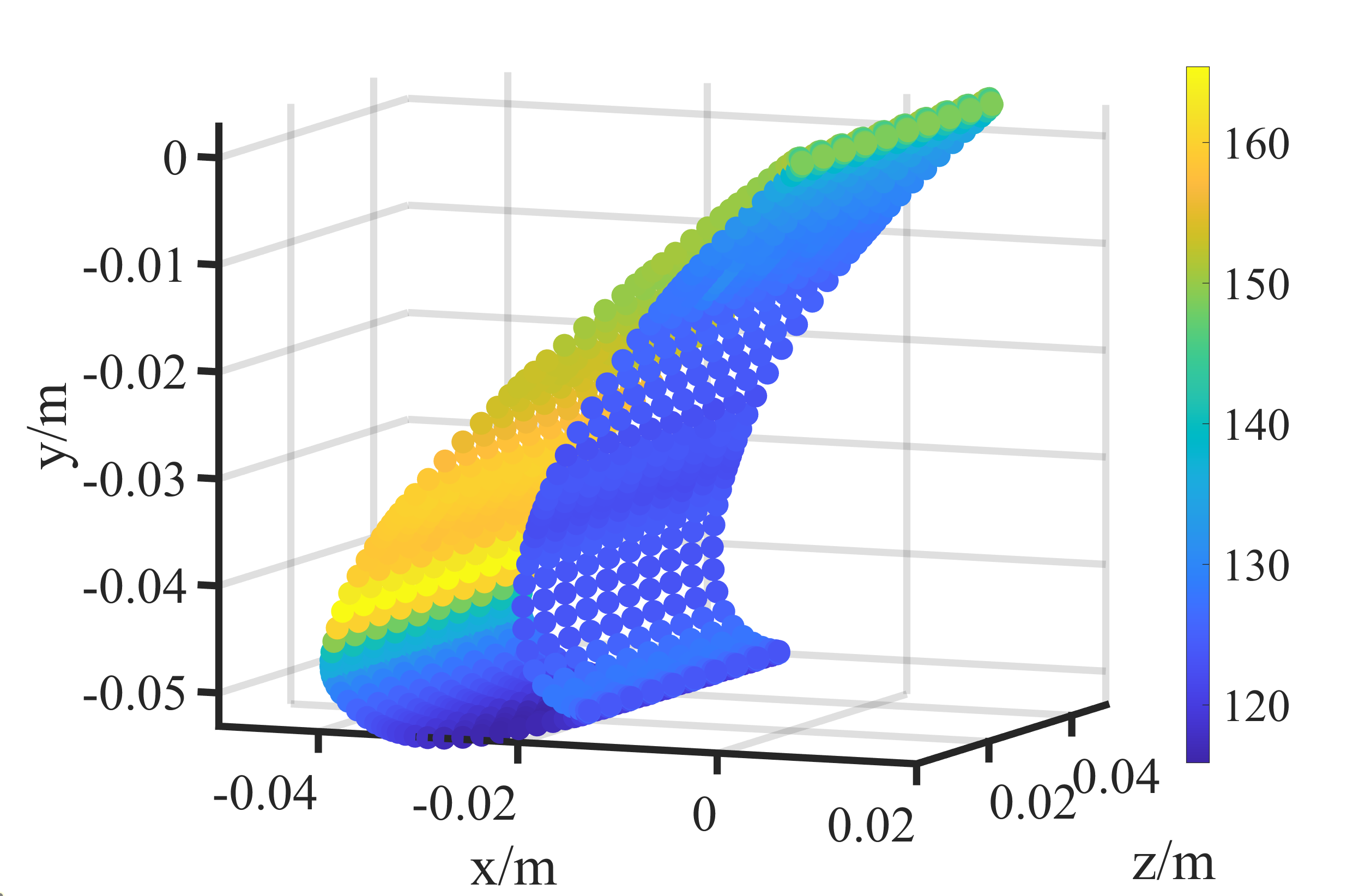}} 
        \quad
    \captionsetup{justification=centering}
    \caption{\label{fig:14_OWSPL} Overall wall pressure level on the slat surface at $14^\circ$ AoA and 0.17Ma.}
\end{figure}

Figure~\ref{fig:AoA_Q} presents the instantaneous iso-surface of the $Q$ criterion on the upper surface of the slat under different AoAs. It can be clearly seen that under $14^\circ$, transition occurs on the slat upper surface, while under lower AoAs, the slat upper surface remains laminar. Figure~\ref{fig:14_OWSPL} demonstrates the overall wall pressure level between 0.8 to 3 kHz, 3 to 8 kHz, and above 8 kHz on the slat surface under $14^\circ$. Notably, significant pressure fluctuations occur at the specified transition position across all examined frequency ranges, with the most substantial contribution from the transition occurring above 8 kHz, as indicated by the high-frequency hump in Fig.~\ref{fig:ff-14}. In addition, there is a strong contribution from the transition location in the 3 to 8 kHz range, as shown in Fig.~\ref{fig:OWSPL_38}. It is also worth noting that Fig.~\ref{fig:OWSPL_83} clearly indicates the tonal noise is associated with the impact of the shear layer on the slat surface. When the slat surface where the transition occurs is removed, the resulting far-field noise spectrum is plotted in Fig.~\ref{fig:ff-14} as part slat. This modification yields improved consistency with experimental result in the mid and high frequency range. 

A new hump appears at 13 kHz in the part slat spectrum. This is attributed to the vortices in the turbulent boundary layer merging with the high-frequency vortex shedding at the slat TE, as shown in Fig.~\ref{fig:Q_14}. These vortices are larger in size, resulting in a lower shedding frequency. The experimental spectrum exhibits a flat broadband profile above 10 kHz, without any discernible humps. However, the spectral amplitude of the experiment in this range closely matches the amplitude of the two humps from the part slat spectrum. It can be inferred that this discrepancy may be due to insufficient mesh density on the upper surface of the slat, which fails to capture smaller turbulent vortices. These smaller vortices could lead to multiple overlapping humps, thereby creating the observed flat broadband profile in the experimental spectrum.

Comparing the far-field noise spectrum of different AoAs in Fig.~\ref{fig:AoA_ff}. It is observed that as the AoA increases, the tonal noise frequencies decrease. This observation aligns with the feedback loop model proposed by \citet{terracol2016investigation} and \citet{souza2019dynamics}, which identifies the associated tonal noise frequencies as follows:
\begin{align} 
    St_{n,Terracol}&=n\frac{c_{slat}}{U_0}\left(\frac{L_v}{U_v}+\frac{L_a}{c_0}\right)^{-1}\label{eq:St_Terracol}
\end{align}
\begin{align} 
    St_{n,Souza}&=n\frac{c_{slat}}{U_0}\left(T_h+\frac{L_a^{\prime}}{c_0}\right)^{-1}\label{eq:St_Souza}
\end{align}
where $T_h$ is equal to $L_v^{\prime}/U_v^{\prime}$ in this paper. The corresponding tonal noise frequency results are listed in Table.~\ref{tab:TN}. Both models predict that the trend of tonal noise frequencies decreases as the AoA increases. This suggests that higher AoAs lead to a weaker feedback loop, indicating the shear layer disturbances take more time from generation to produce acoustic waves.

\begin{figure}[hbt!]
    \centering
    \includegraphics[width=.48\textwidth]{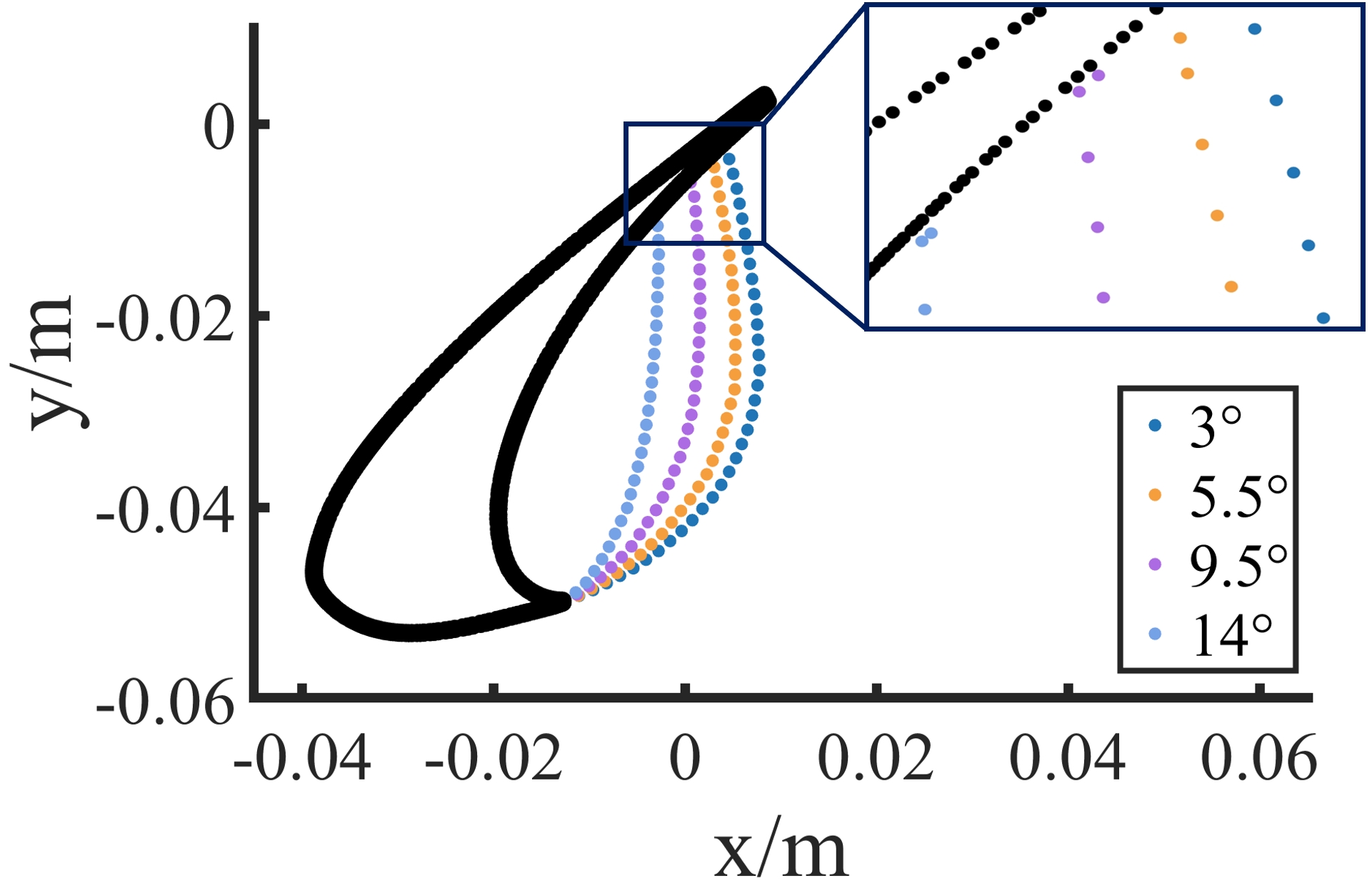}
    \caption{\label{fig:SLL} Time-averaged shear layer path with schematic of acoustic wave generation points.} 
\end{figure}

\begin{figure}[hbt!]
    \centering
        \captionsetup{justification=raggedright, singlelinecheck=false}
        \subcaptionbox{\label{fig:Cwz_3}$3^\circ$}{
        \includegraphics[width = 0.22\textwidth]{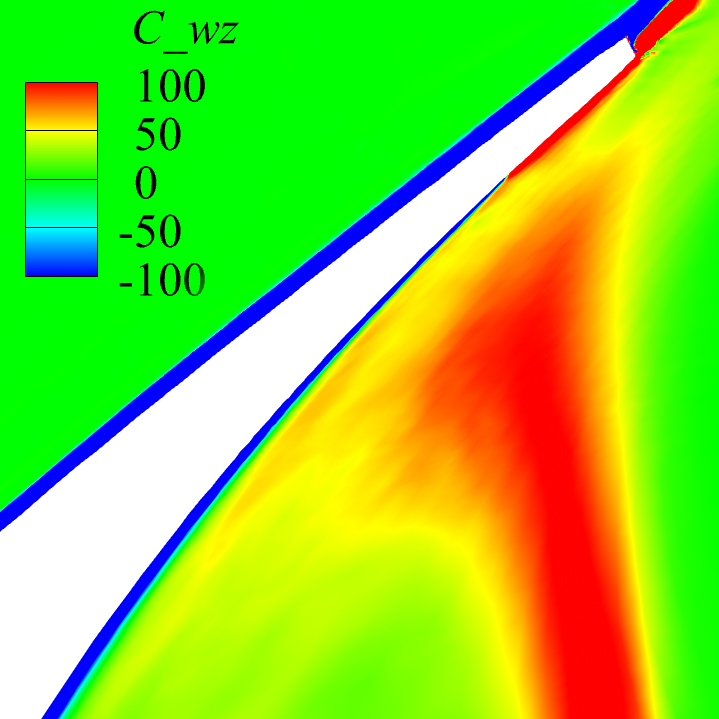}}
        \quad
        \subcaptionbox{\label{fig:Cwz_5.5}$5.5^\circ$}{\includegraphics[width = 0.22\textwidth]{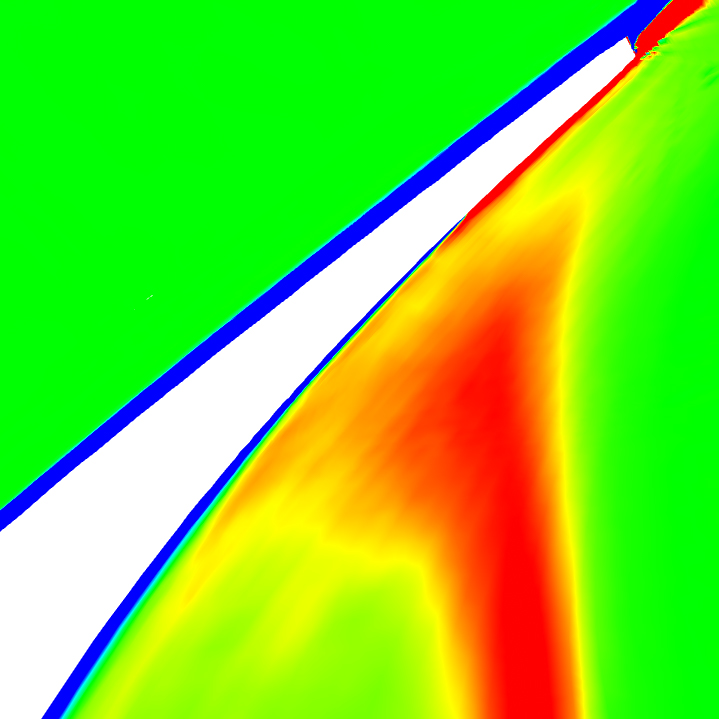}} 
        \quad
        \subcaptionbox{\label{fig:Cwz_9.5}$9.5^\circ$}{\includegraphics[width = 0.22\textwidth]{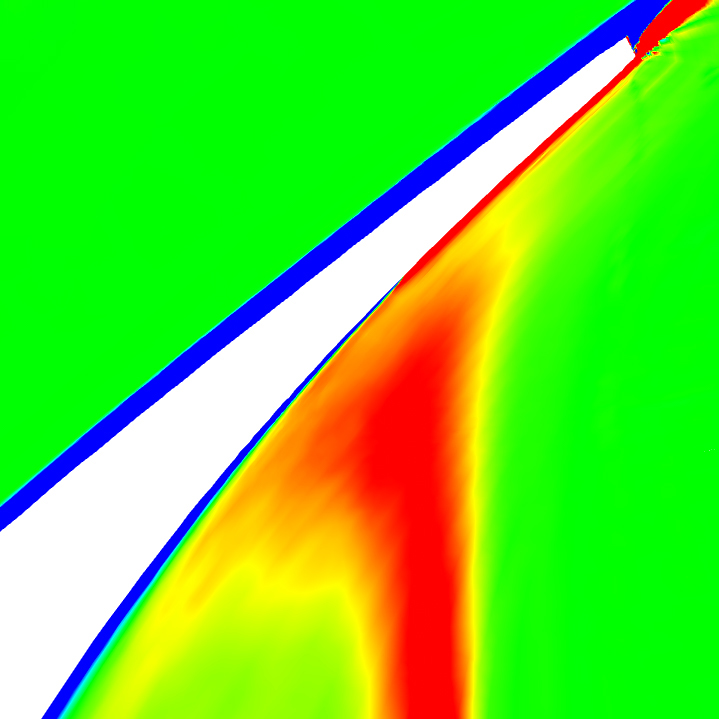}} 
        \quad
        \subcaptionbox{\label{fig:Cwz_14}$14^\circ$}{\includegraphics[width = 0.22\textwidth]{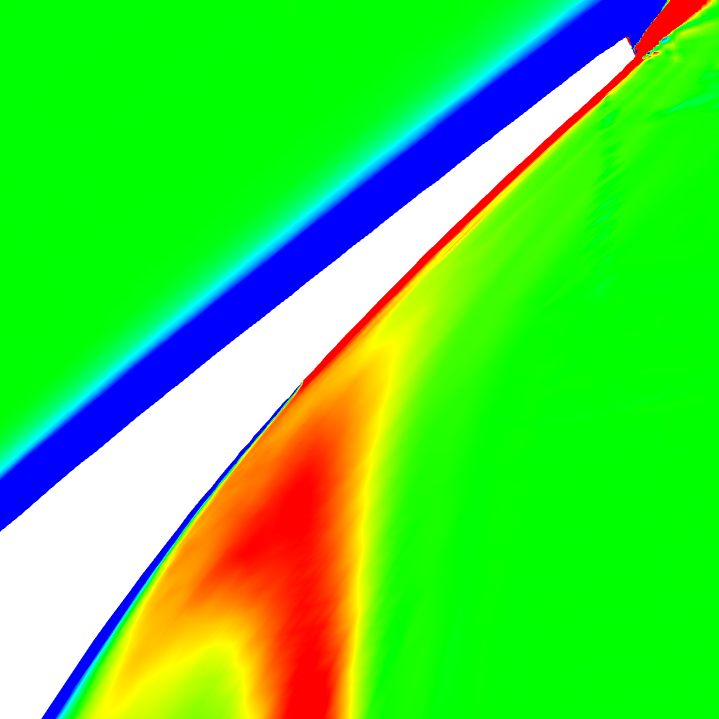}}
        \quad
    \caption{\label{fig:AoA_Cwz} Contours of time-averaged normalized spanwise vorticity ($C_{\omega z} = \omega_zc_s/U_0$) on the airfoil midplane plane around impingement point for different AoAs at 0.17 Ma.}
\end{figure}

Moreover, Eq.~(\ref{eq:St_Terracol}) well predict the tonal noise frequencies at $3^\circ$, $5.5^\circ$, and $14^\circ$, whereas Eq.~(\ref{eq:St_Souza}) predict the tonal noise frequencies more accurately at $9.5^\circ$. This implies that the point of acoustic wave generation in the feedback loop varies with different AoAs. Therefore, we suggest an improved formula for predicting tonal noise frequencies:
\begin{align} 
    St_{n}&=n\frac{c_{slat}}{U_0}\left(\frac{L_v^{\prime\prime}}{U_v^{\prime\prime}}+\frac{L_a^{\prime\prime}}{c_0}\right)^{-1}=n\left(\frac{L_v^{\prime\prime}/c_{slat}}{U_v^{\prime\prime}/U_0}+M_0\frac{L_a^{\prime\prime}}{c_{slat}}\right)^{-1}\label{eq:St_n}
\end{align}
the acoustic wave generation point at different AoAs are illustrated in Fig.~\ref{fig:SLL}. The corresponding tonal noise frequency prediction results are listed Table.~\ref{tab:TN}. Improved predictions are achieved across all AoAs, with a maximum error of less than $3.9\%$ compared to K3 simulations. As depicted in Fig.~\ref{fig:AoA_Cwz}, the high vorticity region of the shear layer gradually extends toward the impingement point as the AoA increases, which corresponds with Fig.~\ref{fig:SLL}. This phenomenon occurs because the angle at which the shear layer vortices strike the slat surface increases with a higher AoA, as also shown in Fig.~\ref{fig:SLL}. Consequently, the mechanism for acoustic wave generation shifts from the impingement of shear layer vortices to the interaction between these vortices and the slat surface as the AoA increases.

\begin{table}[hbt!]
    \caption{\label{tab:TN} Tonal noise Strouhal numbers}
    \centering
    \begin{tabular}{clccccccc}
    \hline
                &                   & $n=2$  & $3$  & $4$  & $5$  & $6$  & $7$  \\\hline
    $3^\circ$   &K3 WRLES + FW-H  & 1.66   & 2.50 & 3.37 & 4.25 & 5.12 & 5.89 \\
                &$St_{n,Terracol}$  & 1.60   & 2.41 & 3.21 & 4.01 & 4.81 & 5.62 \\
                &$St_{n,Souza}$     & 1.52   & 2.28 & 3.04 & 3.80 & 4.56 & 5.32 \\
                &Eq.~(\ref{eq:St_n}) & 1.67   & 2.51 & 3.34 & 4.18 & 5.02 & 5.85 \\
                &                   &        &      &      &      &      &      \\
    $5.5^\circ$ &K3 WRLES + FW-H  & 1.62   & 2.39 & 3.20 & 4.00 & 4.82 & 5.60 \\
                &$St_{n,Terracol}$  & 1.60   & 2.40 & 3.21 & 4.01 & 4.81 & 5.61 \\
                &$St_{n,Souza}$     & 1.42   & 2.12 & 2.83 & 3.54 & 4.25 & 4.95 \\
                &Eq.~(\ref{eq:St_n}) & 1.60   & 2.40 & 3.21 & 4.01 & 4.81 & 5.61 \\
                &                   &        &      &      &      &      &      \\
    $9.5^\circ$ &K3 WRLES + FW-H  & 1.46   & 2.13 & 2.78 & 3.48 & 4.27 & 5.01 \\
                &$St_{n,Terracol}$  & 1.50   & 2.25 & 3.00 & 3.76 & 4.51 & 5.26 \\
                &$St_{n,Souza}$     & 1.41   & 2.11 & 2.82 & 3.52 & 4.23 & 4.93 \\
                &Eq.~(\ref{eq:St_n}) & 1.44   & 2.16 & 2.88 & 3.60 & 4.32 & 5.04 \\
                &                   &        &      &      &      &      &      \\                
    $14^\circ$  &K3 WRLES + FW-H  & 1.25   & 1.82 & 2.37 & 2.93 & 3.63 & 4.39 \\
                &$St_{n,Terracol}$  & 1.24   & 1.85 & 2.47 & 3.09 & 3.71 & 4.33 \\
                &$St_{n,Souza}$     & 1.29   & 1.94 & 2.59 & 3.12 & 3.88 & 4.53 \\
                &Eq.~(\ref{eq:St_n}) & 1.21   & 1.81 & 2.41 & 3.02 & 3.62 & 4.22 \\
    \hline
\end{tabular}
\end{table}

\begin{figure}[hbt!]
    \centering
        \captionsetup{justification=raggedright, singlelinecheck=false}
        \subcaptionbox{\label{fig:Div_3}$3^\circ$}{
        \includegraphics[width = 0.22\textwidth]{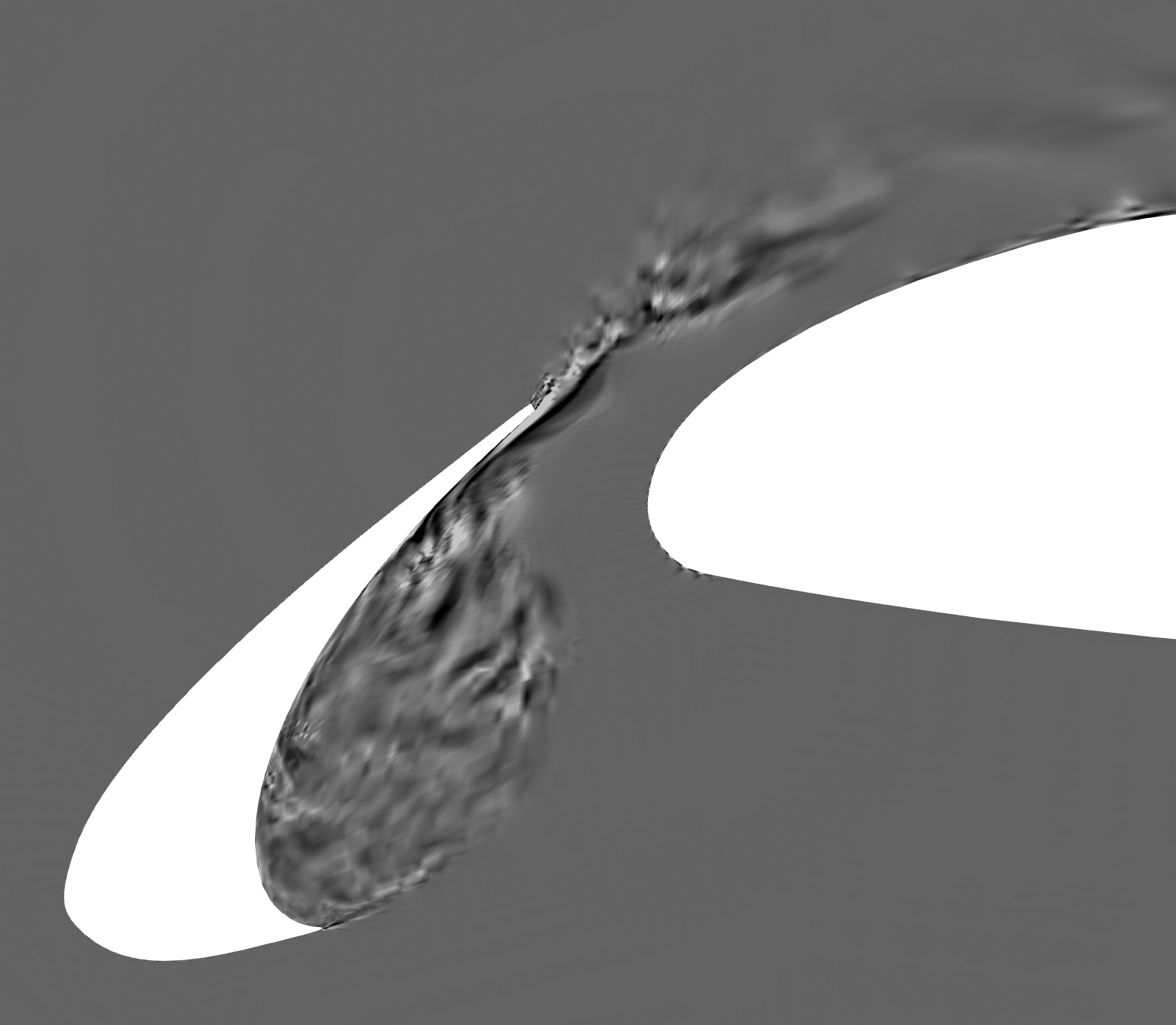}}
        \quad
        \subcaptionbox{\label{fig:Div_5.5}$5.5^\circ$}{\includegraphics[width = 0.22\textwidth]{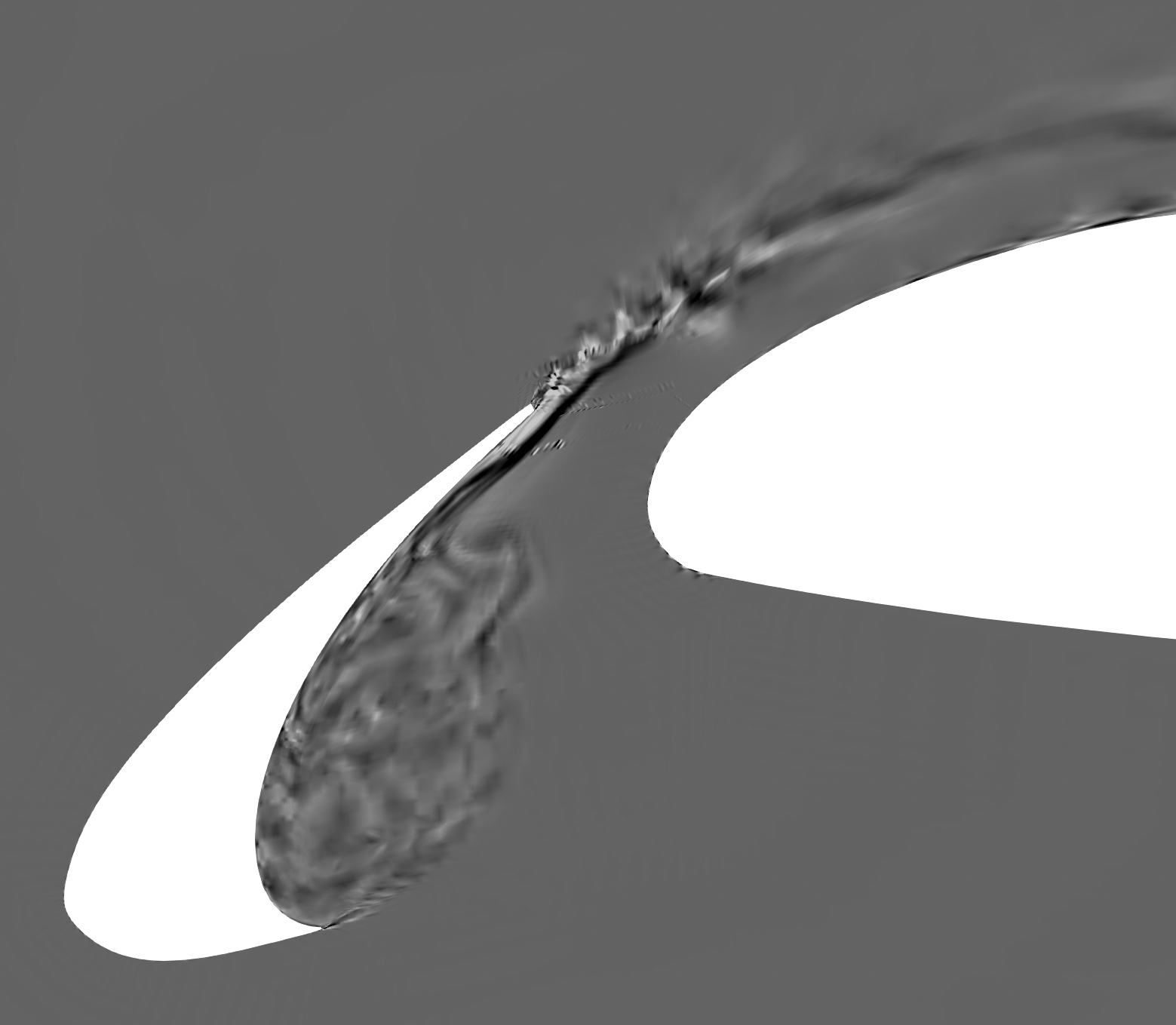}}
        \quad 
        \subcaptionbox{\label{fig:Div_9.5}$9.5^\circ$}{\includegraphics[width = 0.22\textwidth]{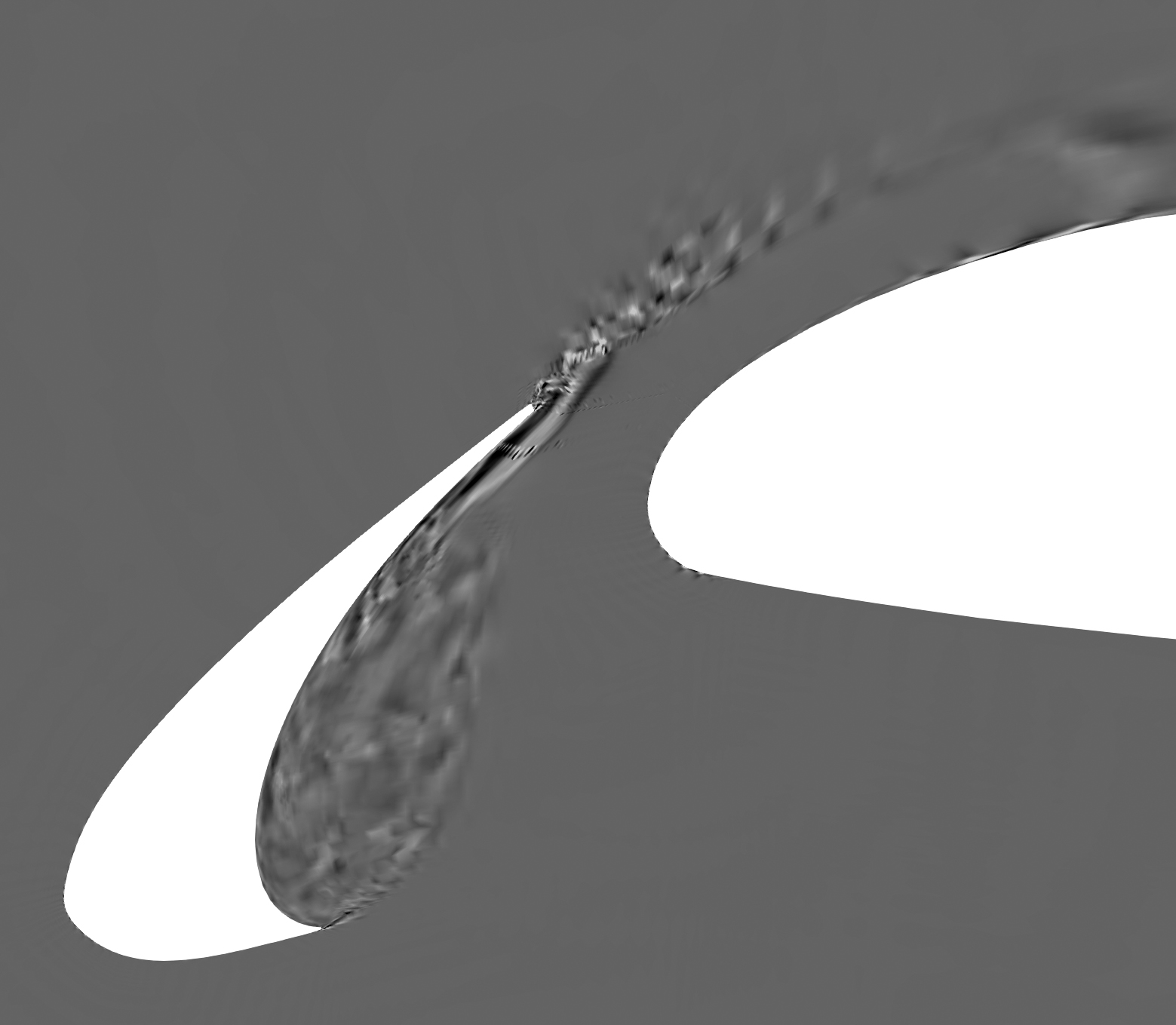}} 
        \quad
        \subcaptionbox{\label{fig:Div_14}$14^\circ$}{\includegraphics[width = 0.22\textwidth]{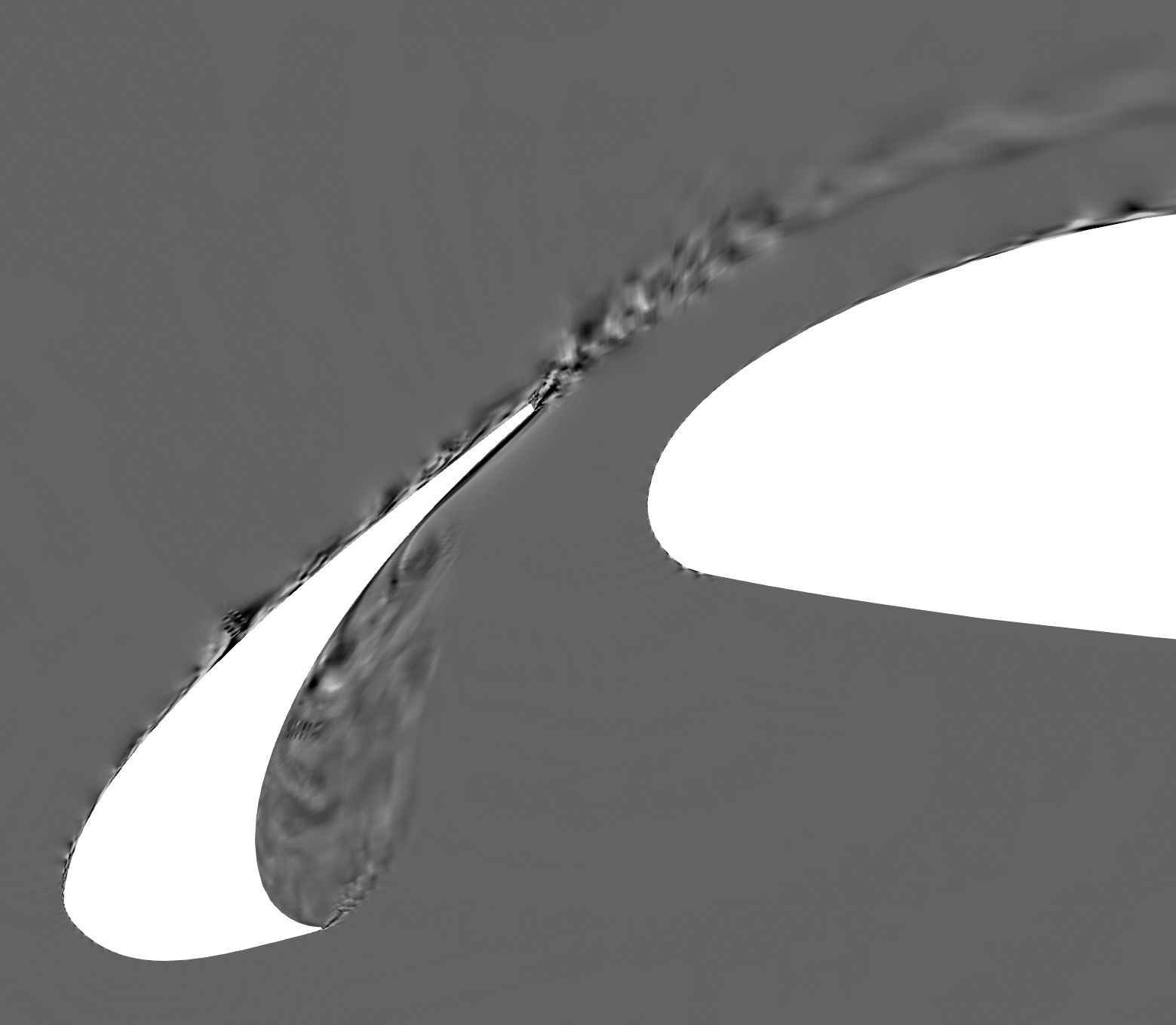}}
        \quad
    \caption{\label{fig:AoA_Div} Instantaneous contour of $Div(\rho \boldsymbol{u})$ in the midspan plane around slat region at 0.17 Ma under different AoAs (-1e4<$Div(\rho \boldsymbol{u})$<1e4).}
\end{figure}

\begin{figure}[hbt!]
    \centering
        \captionsetup{justification=raggedright, singlelinecheck=false}
        \subcaptionbox{\label{fig:Q2_3}$3^\circ$}{
        \includegraphics[width = 0.37\textwidth]{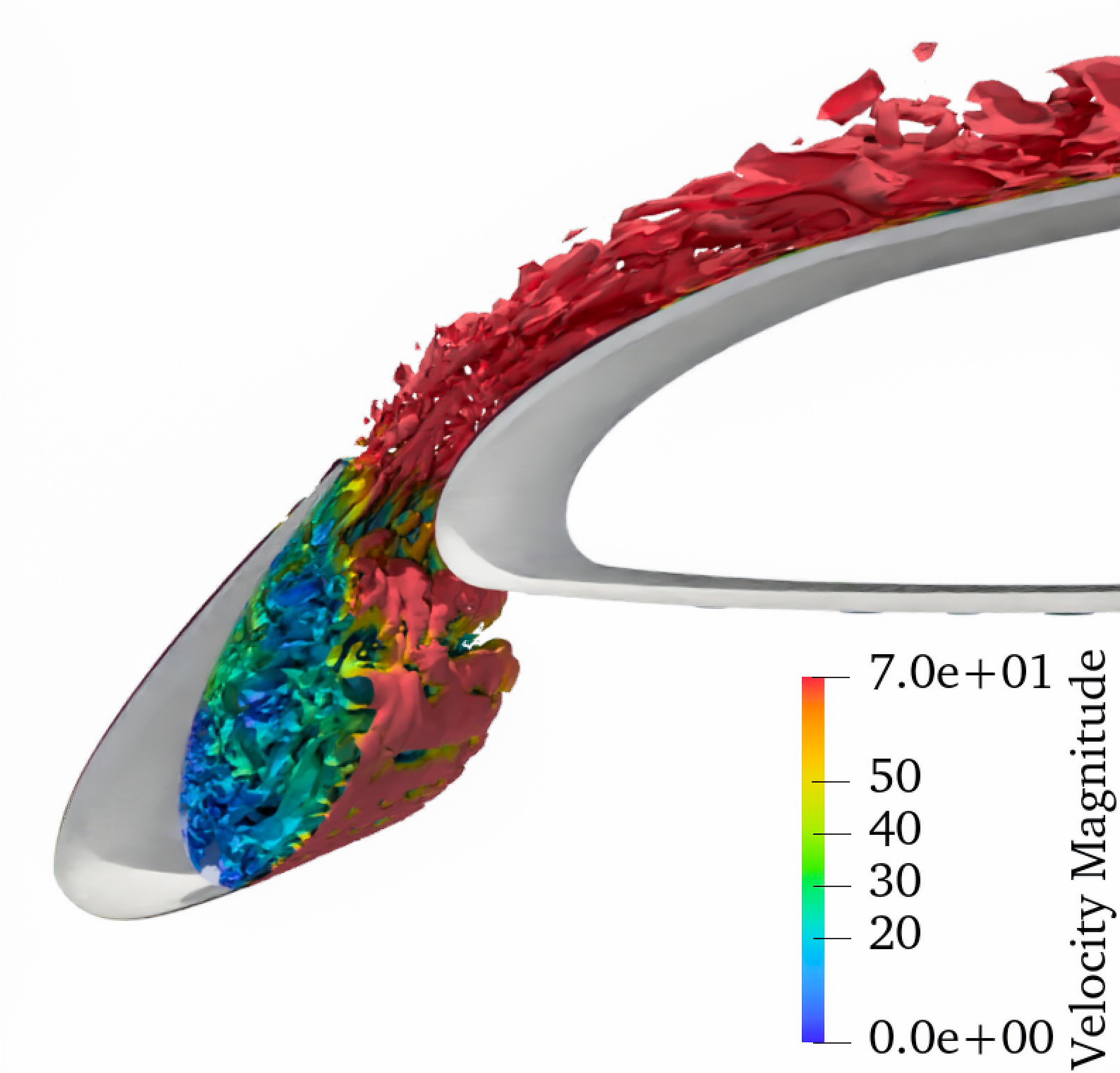}}
        \quad
        \subcaptionbox{\label{fig:Q2_5.5}$5.5^\circ$}{\includegraphics[width = 0.37\textwidth]{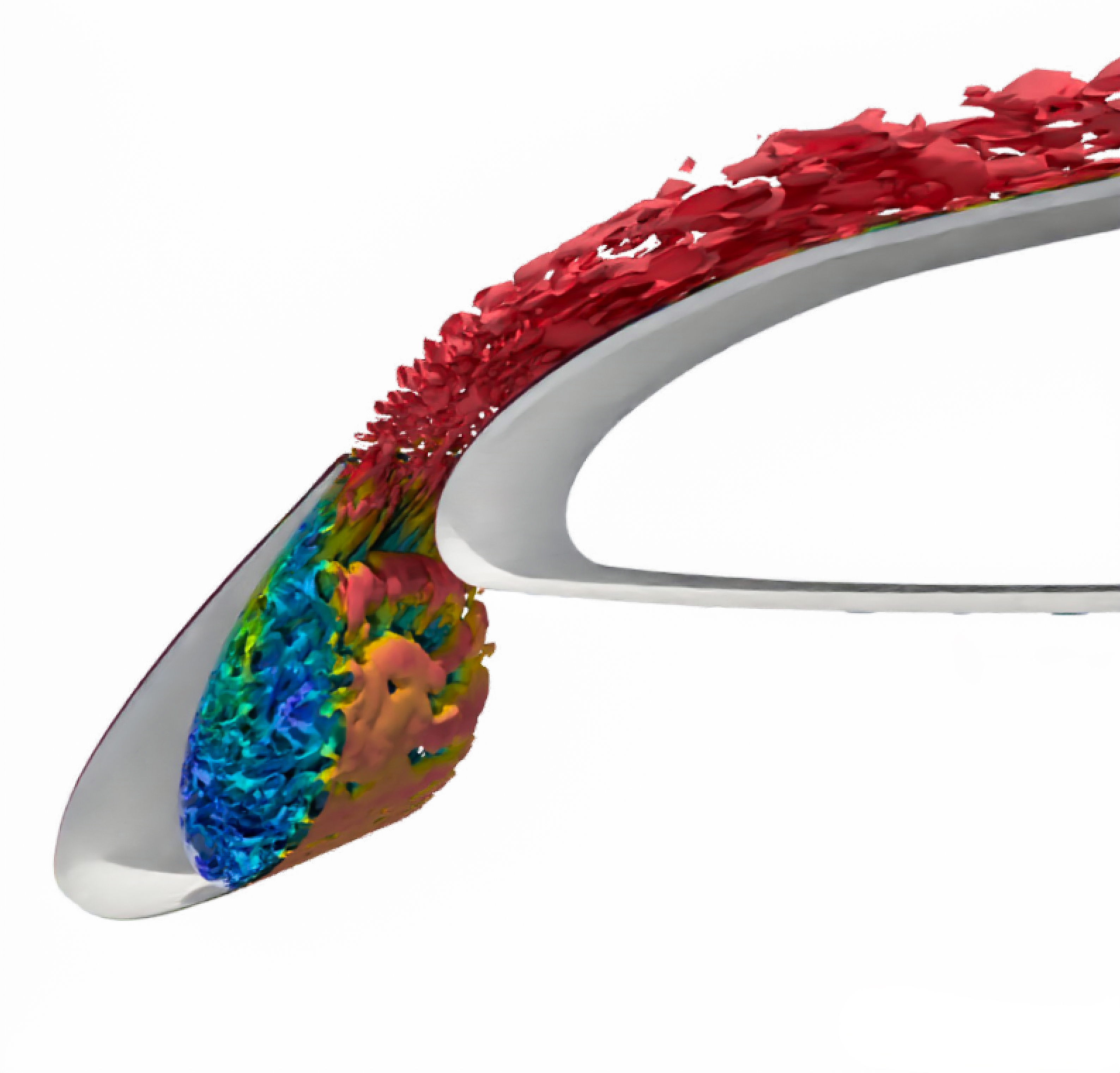}} 
        \subcaptionbox{\label{fig:Q2_9.5}$9.5^\circ$}{\includegraphics[width = 0.37\textwidth]{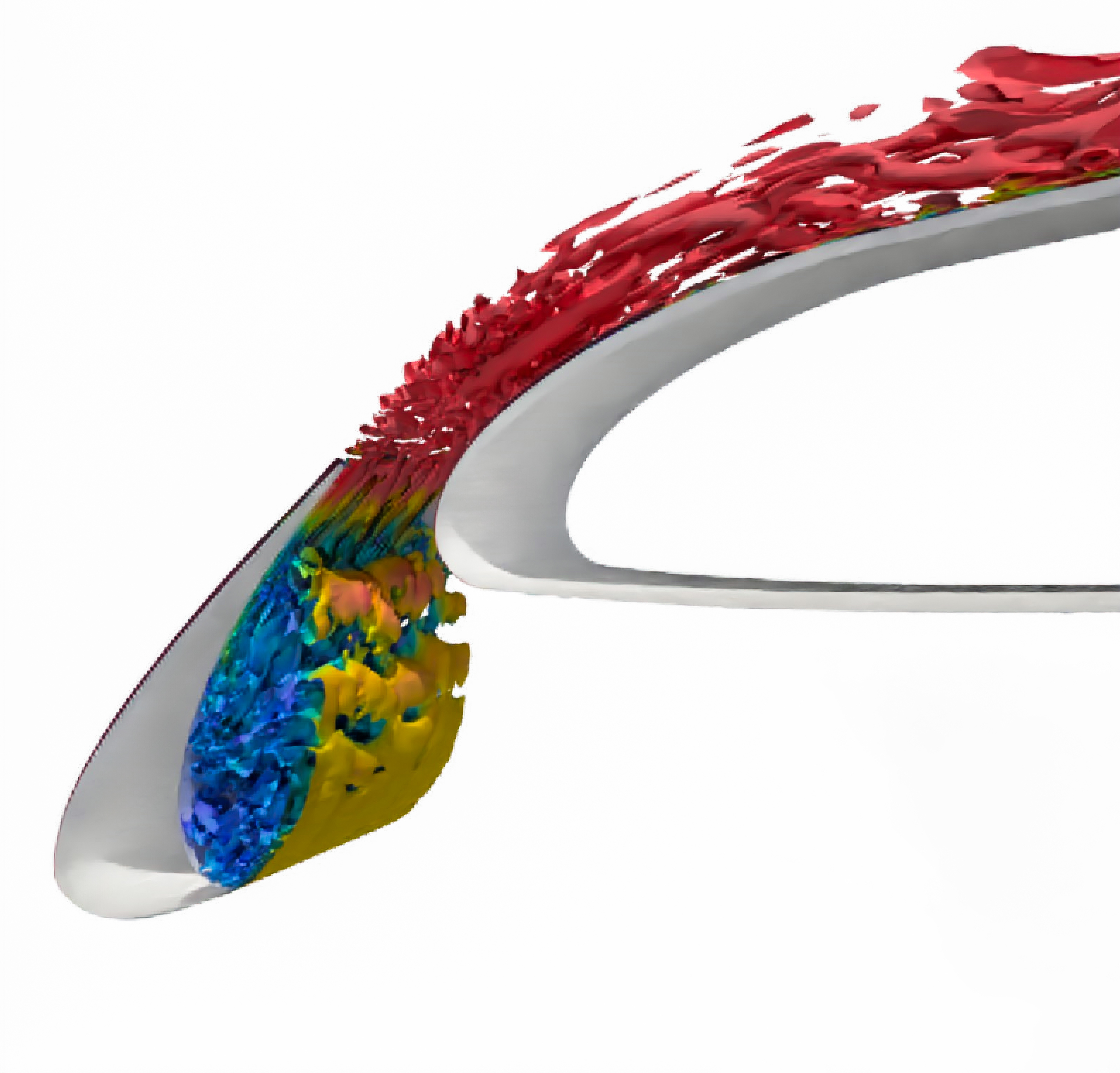}} 
        \quad
        \subcaptionbox{\label{fig:Q2_14}$14^\circ$}{\includegraphics[width = 0.37\textwidth]{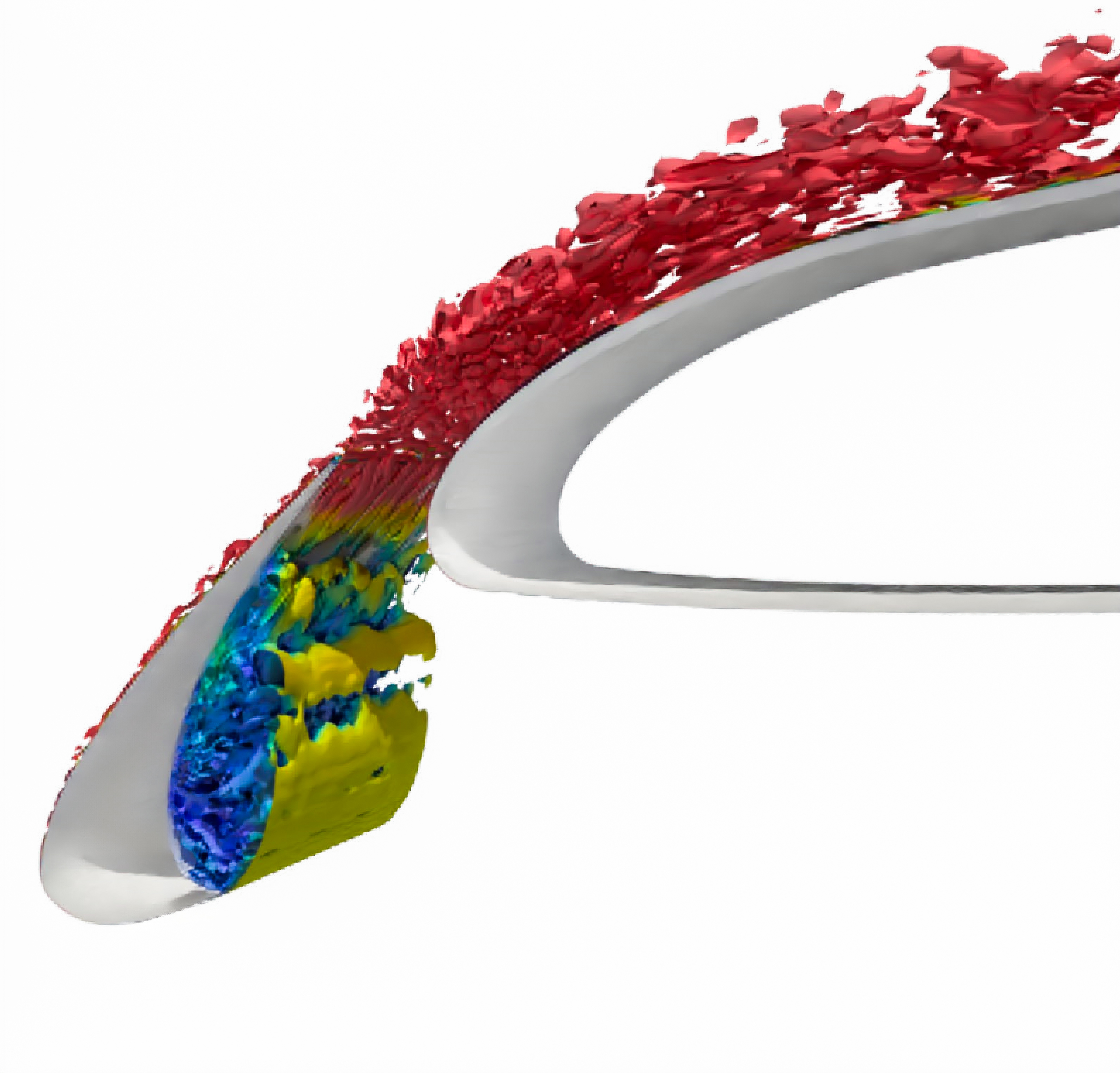}}
        \quad
    \caption{\label{fig:AoA_Q2} Instantaneous iso-surface of $Q=2e6$ in the slat cove region colored by velocity magnitude at 0.17 Ma under different AoAs.}
\end{figure}

In terms of noise amplitude, as the AoA increases, both the amplitude of the tonal noise and broadband noise decrease, as shown in Fig.~\ref{fig:AoA_ff}. This observation is consistent with the result from \citet{souza2019dynamics}. Regarding the dilatation field in the midplane around the slat region, as shown in Fig.~\ref{fig:AoA_Div}, it can be observed that with a higher AoA, both the disturbances in the shear layer and the intensity of the impingement diminish, resulting in a less intense feedback loop, as previously mentioned. Figure~\ref{fig:AoA_Q2} demonstrates that the convection velocity of the shear layer decreases with increasing AoA, which in turn leads to a reduction in the intensity of disturbances within the shear layer and a decrease in the impingement strength. Thus, it can be concluded that the amplitude of the slat noise, whether tonal or broadband, is directly related to the convection velocity of the shear layer.

\begin{figure}[hbt!]
    \centering
    \includegraphics[width=.48\textwidth]{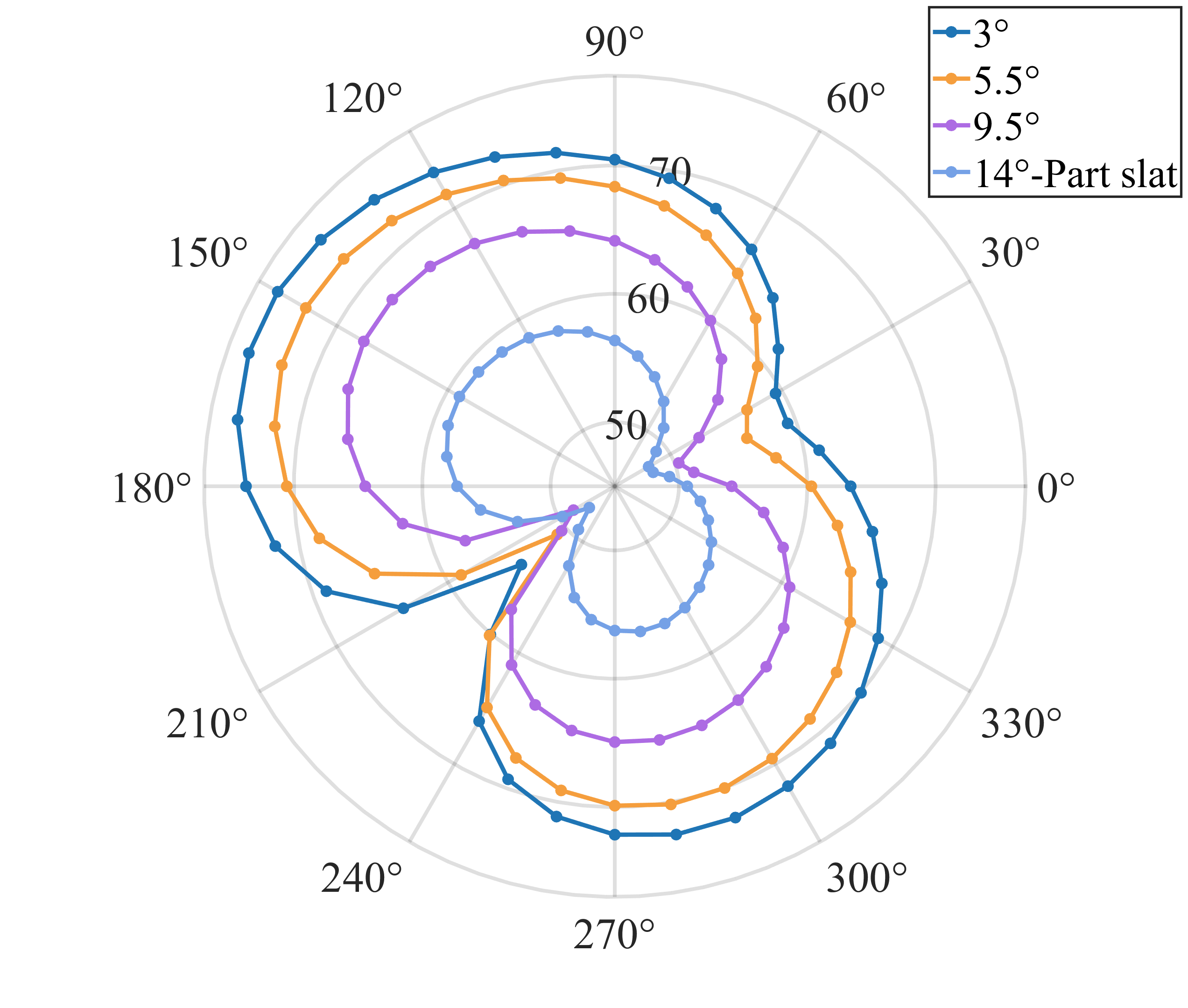}
    \caption{\label{fig:AoA_OASPL} Far-field noise directivity patterns under 0.17 Ma with different AoAs at a distance of $10c_s$ in dB.} 
\end{figure}

The far-field noise OASPL for different AoAs integrated within 10 kHz is shown in Fig.~\ref{fig:AoA_OASPL}. It can be seen that an increase in AoA results in a lower slat noise level across nearly all directions. When comparing the results for $3^\circ$ and $14^\circ$, the average difference is 14.30 dB, with a maximum noise level differs by 16.55 dB at the $170^\circ$ direction.

\section{\label{sec:C}Conclusion}
This paper presents high-order WRLES of the flow over the JAXA modified 30P30N high-lift configuration, focusing on exploring best practice toward the rapid and accurate prediction of slat noise, investigating the effect of Reynolds number on slat noise, and pushing a step further on understanding the AoA effect on the mechanisms of slat tonal noise generation.

Initially, various independence tests are conducted, including different FW-H sampling surfaces, three meshes with varying densities, K3 and K4 simulations, and three different spanwise lengths. The results are compared with JAXA wind tunnel experiments and other simulations to evaluate the best practice of the STE-KEP-FR method for predicting the slat noise of high-lift airfoil. A mesh consisting of 235,750 elements with a spanwise length of $1/9c_s$ and utilizing K3 simulation (with 15.1 million DoF) is sufficient to capture all flow details related to the slat noise while achieving full spanwise decorrelation. In addition, a noise spectrum that coincides with the experimental data is obtained by using the slat surface as the FW-H sampling surface. More importantly, this setup only requires a single Nvidia Ada Lovelace GPU running for one and a half days to produce 10 FPTs of the flow field, significantly reducing computational expenses compared to traditional second-order methods.

Subsequently, Reynolds number effect is investigated through directly scaling the airfoil model with Reynolds numbers ranging from $8.55\times 10^5$ to a real aircraft level of $1.71\times 10^7$. By taking into account the Strouhal number, spanwise length, and distance from the receiver, the far-field noise spectra under different Reynolds numbers can be scaled and matched together, at least for the 30P30N 2.5D high-lift airfoil within the study Reynolds number range. This is extremely beneficial for reduced-scale model slat noise study. Meanwhile, a direct comparison of different slat TE thicknesses at the highest Reynolds number is performed, demonstrating that the high-frequency hump is still an important noise source under the Reynolds number of real aircraft.

Finally, a comparison of simulations for four different AoAs has been conducted. All simulations exhibit good consistency with the JAXA Kevlar wall experiment within the tonal noise frequency range. An improved formula for predicting slat tonal noise frequencies has been developed based on theoretical models from previous research. This improved formula indicates that as the AoA increases, the point at which acoustic waves are generated in the feedback loop shifts from near the impingement point to a downstream location. This change signifies a transition in the acoustic wave generation mechanism from the impact of shear layer vortices to their interaction with the slat surface. Furthermore, it has been found that the amplitude of the slat noise is correlated with the intensity of the feedback loop, which is directly related to the convection velocity of the shear layer.



\section*{Acknowledgments}
This work is supported by the National Natural Science Foundation of China (NSFC) under grant number 12272024. The first author would like to express gratitude to Professor Xinliang Li and his team at LHD, Institute of Mechanics, Chinese Academy of Sciences, for their assistance in the development of OpenCFD-FWH.
\bibliography{sample}

\end{document}